\documentclass[superscriptaddress,twocolumn,aps,showpacs,prc]{revtex4-1}
\usepackage{epsfig}
\usepackage{amsmath}
\usepackage{graphicx}
\usepackage{amssymb}
\usepackage{longtable}
\begin{document}

\title{The electric dipole response of $^{76}$Se above 4 MeV}

\author{P.~M.~Goddard}
\affiliation{Department of Physics, University of Surrey, Guildford, GU2 7XH, UK}
\affiliation{A. W. Wright Nuclear Structure Laboratory, Yale University, New Haven, Connecticut 06520, USA }
\author{N. Cooper}
\affiliation{A. W. Wright Nuclear Structure Laboratory, Yale University, New Haven, Connecticut 06520, USA }
\author{V. Werner}
\affiliation{A. W. Wright Nuclear Structure Laboratory, Yale University, New Haven, Connecticut 06520, USA }
\author{G. Rusev}
\altaffiliation{Present address: Chemistry Division, Los Alamos National Laboratory, Los Alamos, New Mexico 87545, USA.}
\affiliation{Duke University, Durham, North Carolina 27708-0308, USA}
\affiliation{Triangle Universities Nuclear Laboratory, Durham, North Carolina 27708, USA}
\author{P.~D.~Stevenson}
\affiliation{Department of Physics, University of Surrey, Guildford, GU2 7XH, UK}
\author{A. Rios}
\affiliation{Department of Physics, University of Surrey, Guildford, GU2 7XH, UK}
\author{C.~Bernards}
\affiliation{A. W. Wright Nuclear Structure Laboratory, Yale University, New Haven, Connecticut 06520, USA }
\author{A. Chakraborty}
\affiliation{Departments of Chemistry and Physics \& Astronomy, University of Kentucky, Lexington, Kentucky, 40506, USA}
\author{B. P. Crider}
\affiliation{Departments of Chemistry and Physics \& Astronomy, University of Kentucky, Lexington, Kentucky, 40506, USA}
\author{J. H. Kelley}
\affiliation{Triangle Universities Nuclear Laboratory, Durham, North Carolina 27708, USA}
\affiliation{Department of Physics, North Carolina State University, Raleigh, North Carolina 27695, USA}
\author{E. Kwan}
\altaffiliation{Present address: National Superconducting Cyclotron Laboratory, Michigan State University, East Lansing, Michigan 48824, USA.}
\affiliation{Duke University, Durham, North Carolina 27708-0308, USA}
\affiliation{Triangle Universities Nuclear Laboratory, Durham, North Carolina 27708, USA}
\author{J. Glorius}
\affiliation{Institut f\"ur Kernphysik, TU Darmstadt, Schlossgartenstra{\ss}e 9, D-64289 Darmstadt, Germany }
\author{E. E. Peters}
\affiliation{Departments of Chemistry and Physics \& Astronomy, University of Kentucky, Lexington, Kentucky, 40506, USA}
\author{N.~Pietralla}
\affiliation{Institut f\"ur Kernphysik, TU Darmstadt, Schlossgartenstra{\ss}e 9, D-64289 Darmstadt, Germany }
\author{R. Raut}
\altaffiliation{Present address: UGC-DAE Consortium for Scientific Research, Kolkata Centre LB-8 Sector-III Bidhannagar, Kolkata 700098, India.}
\affiliation{Duke University, Durham, North Carolina 27708-0308, USA}
\affiliation{Triangle Universities Nuclear Laboratory, Durham, North Carolina 27708, USA}
\author{C. Romig}
\affiliation{Institut f\"ur Kernphysik, TU Darmstadt, Schlossgartenstra{\ss}e 9, D-64289 Darmstadt, Germany }
\author{D. Savran}
\affiliation{ExtreMe Matter Institute and Research Devision, GSI, Planckstr. 1, 64291 Darmstadt, Germany }
\affiliation{Frankfurt Institute for Advanced Studies, Ruth-Moufang-Str. 1, 60438 Frankfurt, Germany}
\author{L. Schnorrenberger}
\affiliation{Institut f\"ur Kernphysik, TU Darmstadt, Schlossgartenstra{\ss}e 9, D-64289 Darmstadt, Germany }
\author{M. K. Smith}
\affiliation{A. W. Wright Nuclear Structure Laboratory, Yale University, New Haven, Connecticut 06520, USA }
\author{K. Sonnabend}
\affiliation{Institut f\"ur Kernphysik, TU Darmstadt, Schlossgartenstra{\ss}e 9, D-64289 Darmstadt, Germany }
\affiliation{Goethe UniversitŠt Frankfurt, 60438 Frankfurt am Main, Germany}
\author{A.~P.~Tonchev}
\altaffiliation{Present address: Physics Division, Lawrence Livermore National Laboratory, Livermore, California 94550, USA.}
\affiliation{Duke University, Durham, North Carolina 27708-0308, USA}
\affiliation{Triangle Universities Nuclear Laboratory, Durham, North Carolina 27708, USA}
\author{W. Tornow}
\affiliation{Duke University, Durham, North Carolina 27708-0308, USA}
\affiliation{Triangle Universities Nuclear Laboratory, Durham, North Carolina 27708, USA}
\author{S. W. Yates} 
\affiliation{Departments of Chemistry and Physics \& Astronomy, University of Kentucky, Lexington, Kentucky, 40506, USA}

\date{\today}

\begin{abstract}
The dipole response of  $^{76}_{34}$Se in the energy range from 4 to 9 MeV has been analyzed using a $(\vec\gamma,{\gamma}')$ polarized photon scattering technique, performed at the High Intensity $\gamma$-Ray Source facility, to complement previous work performed using unpolarized photons. The results of this work offer both an enhanced sensitivity scan of the dipole response and an unambiguous determination of the parities of the observed $J=1$ states. The dipole response is found to be dominated by $E1$ excitations, and can reasonably be attributed to a pygmy dipole resonance. Evidence is presented to suggest that a significant amount of directly unobserved excitation strength is present in the region, due to unobserved branching transitions in the decays of resonantly excited states. The dipole response of the region is underestimated when considering only ground state decay branches. 
We investigate the electric dipole response theoretically, performing calculations in a 3D cartesian-basis time-dependent Skyrme-Hartree-Fock framework.

 \end{abstract}


\maketitle
\section{Introduction}

With the advent of high resolution nuclear resonance fluorescence (NRF) experiments \cite{Kne96}, interest in low-lying collective dipole resonances of the nucleus has intensified. The so-called pygmy dipole resonance (PDR) is an electric resonance situated upon the low-energy tail of the giant dipole resonance (GDR) \cite{Herz97,Herz99,Sav2013}. It is found typically at energies between 5 and 8 MeV, and its strength contributes less than 1-2\% of the energy-weighted sum rule of the $E1$ excitation strength in the nucleus \cite{Adr05,Vol06,Tso08}. A common interpretation of the PDR is a proton-neutron core vibrating against a neutron skin in nuclei with an excess of neutrons \cite{Moh71,Van92,Paa07,Tso08}. The neutron excess of the nucleus is thought to be correlated to the magnitude of the excitation strength of the PDR, this reasoning stemming from the assumption that a greater neutron excess will result in a thicker neutron skin. The relationship between the excitation strength of the PDR and neutron excess may not be simply correlated, however, as away from spherical nuclei it has been suggested that the low-lying $E1$ strength can be hindered by deformation effects, even in neutron-rich nuclei \cite{Art09}.
It has also been suggested that in proton-rich nuclei a PDR-type resonance can occur, which associates a proton skin oscillating against a proton-neutron core \cite{Paa09}.

The PDR might have significant implications in nuclear astrophysics regarding the synthesis of certain heavy elements via rapid neutron capture \cite{Gor98}. Experimentally, it has been studied extensively in several semi- or doubly-magic nuclei \cite{Vol06,Tso08,Sav08,Pol12,Sav06,Ton10,Gov98,Sch07,Rusev08,Wie09,Sav2013,Adr05,End10}. The case of $^{76}$Se offers an examination of the PDR in a medium mass, deformed nucleus ($\beta$~=~0.309(4)~\cite{Ram01}), with a relatively small neutron excess. The GDR in $^{76}$Se has been observed to split into predominant regions due to its axial deformation \cite{Carl76}, intuitively corresponding to vibrations of the nucleus perpendicular and parallel to the axis of symmetry, $K$. More precisely, this is due to the resonance splitting into a $K$ = 0 and a twofold $K=\pm1$ mode {\cite{Mor80}. Therefore, it is of interest to analyze the fine structure of the PDR in deformed nuclei to determine correlations, if any exist.

The nucleus $^{76}$Se also has relevance in the topic of $0\nu2\beta$ decay \cite{Ell02,Sch08,Fre07}; signatures have been observed in $^{76}$Ge \cite{Klap04}, and $^{76}$Se is the daughter nucleus for this decay mode. Data presented in this paper can offer a challenge for theoretical models capable of describing the dipole response of nuclei. Methods such as time-dependent Hartree-Fock and the (quasiparticle) random phase approximation are two such techniques suitable for describing collective excitations of the nucleus. Both (and variations thereof) have been employed to describe dipole resonances in finite nuclei \cite{Bri08,Stev10,Vol06,Ina09,Avo11}. The matrix elements which describe $0\nu2\beta$ decay \cite{Suh97,Sch08} can only be extracted from theoretical models, therefore tests for whether they can correctly describe the structure of involved nuclei over broad energy ranges are crucial.

In this paper, results from a ($\vec\gamma,\gamma'$) photon scattering experiment performed at the High Intensity $\gamma$-Ray Source (HI$\gamma$S) facility at Triangle Universities Nuclear Laboratory are presented, complementing our previous work \cite{Coo10}, to obtain a more complete picture of the nature of the dipole response in $^{76}$Se. 
In addition to the parity determination of the dipole excited states, the high photon fluxes allow observation of many new states, the majority of them at energies exceeding 7 MeV. The previous bremsstrahlung data yielded absolute cross sections for many states, which can be used to normalize data in the present work. The near monoenergetic beams (with a FWHM of $\approx$ 3\% of the centroid beam energy) allow firm assignments of transitions either to the ground state, or lower-lying excited states. 

The use of monoenergetic beams also allow the contribution to the photon scattering cross section from decays to excited states to be deduced, even if they are not observed directly \cite{Ton10}. This can be done by considering the decays of low lying states which are populated entirely by feeding transitions from branching decays of excited states. This point will be discussed further in Section~\ref{six}. 

This paper will be structured as follows. Section \ref{two} will summarise our previous analysis. Section \ref{three} discusses relevant theory of $\gamma$-ray angular distributions for parity determination at the HI$\gamma$S facility. In Section \ref{four}, the experiment at the HI$\gamma$S facility is described, and the results are presented in Section \ref{five}. Section \ref{six} contains a discussion of the results obtained from this work. The dipole response of $^{76}$Se described in the time-dependent Hartree Fock framework is investigated in Section \ref{seven}, and we conclude this paper in Section \ref{eight}.

\section{Summary of Previous Analysis}
\label{two}
In our previous work performed at the Darmstadt High Intensity Photon Setup (DHIPS) facility \cite{Son11} at TU Darmstadt, excitation energies up to 9 MeV were investigated \cite{Coo10}. There was, however, no means available to distinguish the parities of the observed states. Ref.~\cite{Gen13}  discusses evidence of a Giant $M1$ resonance present in the same energy region as one might expect a PDR; therefore before any theoretical models can be compared to experiment, knowledge of the nature of the dipole response is of vital importance. Polarized incident photons used in conjunction with an appropriate polarimetry setup is an ideal method for distinguishing electric from magnetic spin $J=1$ excited states \cite{Pie02}.

The energy and angle integrated differential cross section $I_{i}^{S}$ for photon scattering for an angular momentum state $J_{x}$ at energy $E_{x}$, excited from an initial state of angular momentum $J_{0}$, given by
 \begin{equation}
I^{S}_{i} =  \left(\pi\frac{\hbar c}{E_{x}}\right)^{2} \times g \times\frac{\Gamma_{0}\Gamma_{i}}{\Gamma} \,\,\,\,\,,
\label{intcrosssec}
\end{equation}
where $\Gamma_{0}$ is the width of the transition from $J_{x}$ to $J_{0}$, and $\Gamma_{i}$ the width of the transition from $J_{x}$ to $J_{i}$. The branching ratios of transitions to states $J_{i}$, relative to the transition to $J_{0}$, are defined by
 \begin{equation}
\frac{\Gamma_{i}}{\Gamma_{0}} =\frac{I^{S}_{i}}{I^{S}_{0}} =  \frac{A_{i}  W^{0}(\theta)}{A_{0}  W^{i}(\theta)} \,\,\,\,\,,
\label{branch}
\end{equation} 
where $A_{i}$ and $A_{0}$ are the observed counts corresponding to a de-excitation to state $J_{i}$ and $J_{0}$, corrected for the detector efficiency. $W^{i}(\theta)$ and $W^{0}(\theta)$ are the effective angular correlation functions of the transitions to the corresponding state. These $W(\theta)$ will be defined further below.
The statistical factor $g$ is defined as
\begin{equation}
g = \frac{2J_{0} + 1}{2J_{i} + 1} \,\,\,\,\,.
\label{spinfac}
\end{equation}
Therefore, use of Eq. \ref{intcrosssec} allows the width of the transition to be deduced from the observed scattering cross section.
The full width $\Gamma$ is defined as the sum of the partial widths. $\Gamma$ is related to the lifetime of the individual state $\tau$ via
\begin{equation}
\Gamma = \sum_{i=0}^{N}\Gamma_{i} = \frac{\hbar}{\tau} \,\,\,\,\,.
\label{lifetime}
\end{equation}

In our previous work, the experiment allowed cross sections of resonantly excited states to be extracted directly from photon scattering data. 
From $\Gamma_{i}$, the excitation strength $B(\Pi\lambda$)$\uparrow$ ($\Pi$ defining an electric or magnetic transition, $\lambda$ the multipolarity) of a state can be obtained. The following are the explicit forms for the transitions of interest for an excitation from ground state $J_{0}$ to $J_{x}$ in a photon scattering experiment, as only low multipole transitions are likely to occur through the absorption of real photons:
\begin{equation}
\frac{B(E1) \uparrow}{[\text{e$^{2}$fm$^{2}$}]} = 9.554\times10^{-4}\, \times g\times \,\frac{\Gamma_{0}}{[\text{meV}]} \times\left(\frac{ [\text{MeV}]} {E_{x}}  \right)^{3}
\label{be1}
\end{equation}
\begin{equation}
\frac{B(M1) \uparrow}{[\mu_{N}^{2}]} = 8.641\times10^{-2}\, \times g \times\,\frac{\Gamma_{0}}{[\text{meV}]} \times\left(\frac{[\text{MeV}]}{E_{x}}\right)^{3} \,\,\,\,\,.
\label{bm1}
\end{equation}

\section{Angular Distributions for Parity Determination}
\label{three}
In its most general form, the angular distribution of emitted $\gamma$-rays from an initial state $J_{0}$, through intermediate state $J_{1}$, to final state $J_{2}$, is given by \cite{Kra73}
\begin{eqnarray}
  \label{w}
  \lefteqn{W(\theta_{1},\theta_{2},{\phi}) =} \\ \nonumber
  & &   \sum_{\lambda_{1}\lambda\lambda_{2}} B_{\lambda_{1}}(J_{0})A_{\lambda}^{\lambda_{2}\lambda_{1}}(\gamma_{1})A_{\lambda_{2}}(\gamma_{2})H_{\lambda_{1}\lambda\lambda_{2}}(\theta_{1},\theta_{2},\phi)\,\,\,\,\, ,
  \end{eqnarray}
where $B_{\lambda_{1}}(J_{0})$ is the orientation parameter of the initial state, defined with respect to the orientation axis, $A_{\lambda}^{\lambda_{2}\lambda_{1}}(\gamma_{1})$ and $A_{\lambda_{2}}(\gamma_{2})$ are the radiation distribution coefficients, and $H_{\lambda_{1}\lambda\lambda_{2}}(\theta_{1},\theta_{2},\phi)$ is the angular function. All are defined in Ref.~\cite{Kra73} using the Krane, Steffen, and Wheeler phase convention. The angles $\theta_{1}$ and $\theta_{2}$ are the polar angles of emission of $\gamma_{1}$ and $\gamma_{2}$, respectively, measured in the polarization plane. $\phi$ describes the azimuthal rotation of the emission. 
The indices $\lambda_{1}$ and $\lambda_{2}$ are the ranks of the statistical tensors that describe the orientation of states $J_{0}$ and $J_{1}$. For the multipole expansion, they take values of even integers. $\lambda$ is the tensor rank of the radiation field. 

For an excitation from a $J^\pi= 0^{+}$ ground state, which is the only case when performing NRF on an even even nucleus, the orientation of the ground state $J_{0}$ is arbitrary. Therefore $\lambda_{1}$ may be set to 0. The $H_{0\lambda\lambda_{2}}(\theta_{1},\theta_{2},\phi)$ will reduce to the ordinary Legendre polynomial $P_\lambda(\cos\theta)$.  The orientation of the nucleus in the excited state $J_{1}$ is therefore defined by $B_{\lambda}(\gamma_{1})$.

In NRF, the formalism assumes that the first transition, $\gamma_{1}$, is responsible for the orientation of state $J_{1}$ as it is excited from $J_0$. Only the second transition, $\gamma_{2}$, is detected as the state $J_{1}$ de-excites to state $J_{2}$. It is for $\gamma_{2}$ that we calculate the angular distribution. 

Without any knowledge of the polarization of the incident $\gamma$-ray, Eq.~(\ref{w}) reduces to 
\begin{equation}
\label{unpol_gen}
W(\theta) = \sum_{\lambda=0,2,4}B_{\lambda}(\gamma_{1})A_{\lambda}(\gamma_{2})P_{\lambda}(\cos\theta)\,\,\,\,\,.
\end{equation}
If $\gamma_{1}$ is linearly polarized (the case of the current work at the HI$\gamma$S facility), the terms $B_\lambda(\gamma_1) P_\lambda \cos(\theta)$ must be replaced with the modified orientation coefficient $\tilde{B}P_\lambda(\theta,\phi,\gamma_1)$, as described in Refs.~\cite{Fag59,Pie03,Wer04} by:
\begin{eqnarray}
  \label{orientation}
 \lefteqn {\tilde{B}P_\lambda(\theta,\phi,\gamma_{1})=} \\ \nonumber
 &&\qquad B_{\lambda}(\gamma_{1})P_{\lambda}(\cos\theta)  + \frac{1}{1+\delta_{1}^{2}}(\cos2\phi) P_{\lambda}^{(2)}(\cos\theta)    \\ \nonumber
  &&\qquad \times\Big{[}(\pm_{L_{1}}\kappa_{\lambda}(L_{1}L_{1})F_{\lambda}(L_{1}L_{1}J_{0}J_1)\\ \nonumber
  &&\qquad +(-1)^{L_{1}+L'_1}(\pm)_{L_1'}\kappa_{\lambda}(L_1L_1')2\delta_1F_\lambda(L_1L_1'J_0J_1) \\ \nonumber
  &&\qquad + (\pm)_{L_1'}\kappa_\lambda(L'_1L'_1)F_\lambda(L'_1L_1'J_0J_1)\Big{]}\,\,\,\,\,.
 \end{eqnarray}
The $P_\lambda^{(\mu)}(\cos\theta)$ is the unnormalized associated Legendre polynomial of order $\mu$. The ordinary $F_{\lambda}(LL'J_nJ)$ coefficients can be found in, e.g., Ref. \cite{Kra73}. The term $(\pm)_L$ gives a positive sign if the multipole radiation $L_1$ is electric, and negative if it is magnetic. The multipole radiation can be in principle of mixed multipole orders; $L_1'$ is the competing field to $L_1$, and the relative contributions are given by the mixing ratio $\delta_1$. However, for an excitation or decay to a $J^\pi=0^+$ state, the multipole field will be pure $L_1$. The coefficients $\kappa$ describe the vector coupling of the multipole fields $L_1$ and $L'_1$, and are given explicitly in Ref. \cite{Fag59}.  We comment here that Ref. \cite{Fag59} uses the convention of Biedenharn and Rose \cite{Bie53} to define the multipole mixing ratios $\delta_n$, whereas we use the convention of Krane, Steffen, and Wheeler. The formalism contained in this paper is fully consistent with that of Ref. \cite{Kne96}, other than a slight difference of notation.

Therefore, when using fully polarized incident photons, Eq.~(\ref{unpol_gen}) can be written:
\begin{equation}
W(\theta,\phi) = \sum_{\lambda=0,2,4}\tilde{B}P_\lambda(\theta,\phi,\gamma_{1})A_{\lambda}(\gamma_{2}) \,\,\,\,\, .
\label{polw}
\end{equation}

 \begin{figure}[h!]
\begin{center} 
\includegraphics[width=9.0cm]{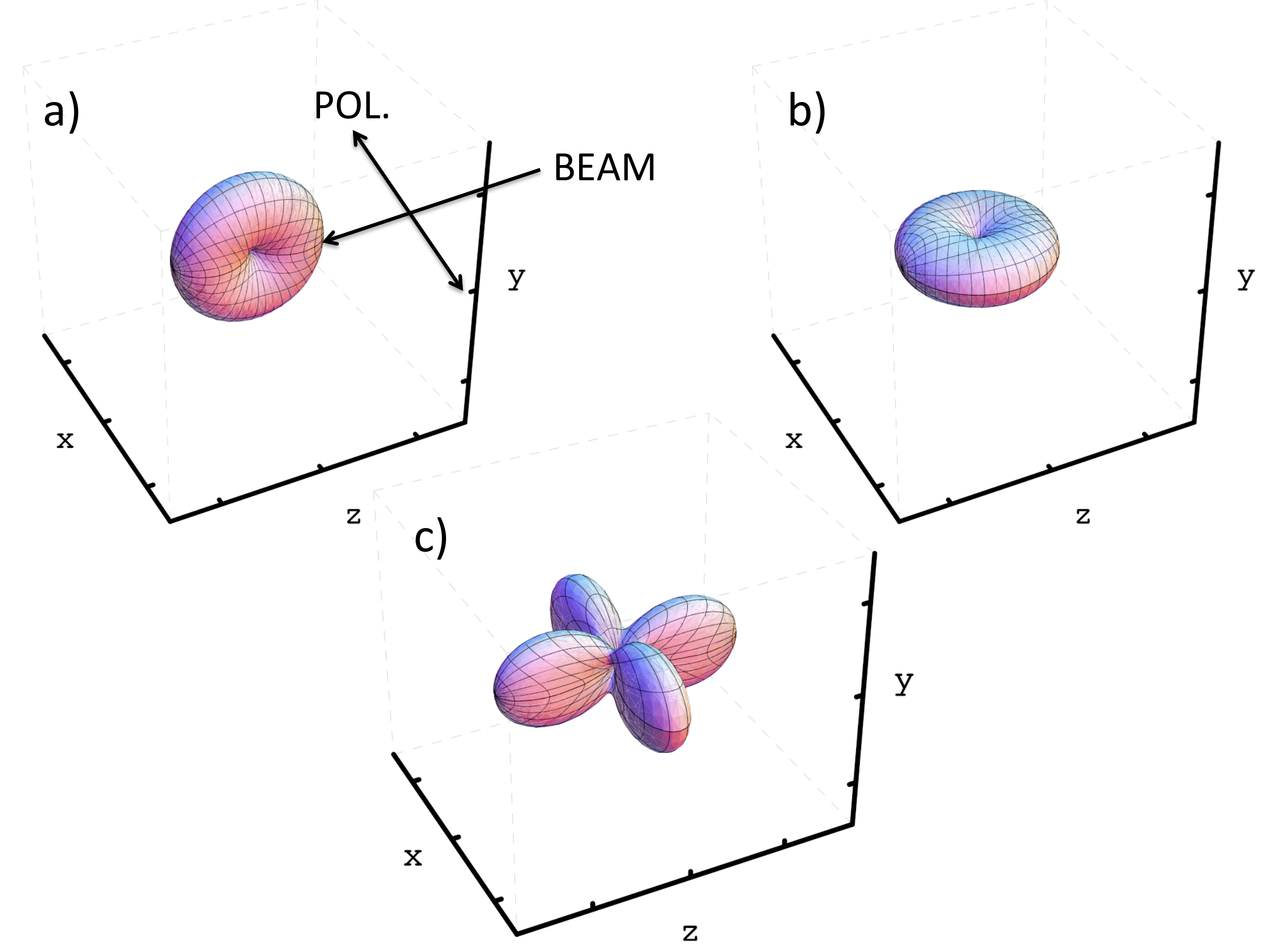} 
\caption{(color online). $W(\theta,\phi)$ for a) a $0^+ \to 1^- \to 0^+$ sequence, b) a $0^+ \to 1^+ \to 0^+$, and c) a  $0^+ \to 2^+ \to 0^+$. The direction of the incident beam and polarization is labeled. The azimuthal asymmetry for a $0^+ \to 1^+ \to 0^+$ and a $0^+ \to 2^+ \to 0^+$ sequence will be the same, therefore an extra detector placed at $135^\circ$ in the $x-z$ plane is necessary to distinguish the two types of transition.}
\label{fig:angdist}
\end{center}
\end{figure} 

In Fig. \ref{fig:angdist}, we show the $W(\theta,\phi)$ for ground-state decays from a resonantly excited $J^{\pi} = 1^\pm$ or $J^{\pi} = 2^+$ state. For our main case of interest, the $W(\theta,\phi)$ for a $0^+\to1^{\pi}\to0^{+}$ sequence is given explicitly by \cite{Pie03,Pie02}
\begin{equation}
W(\theta,\phi) = 1 + \frac{1}{2}\left[P_2(\cos\theta)+\frac{1}{2}\pi_1\cos(2\phi))P_2^{(2)}(\cos\theta)\right] \,\,\,\,\, ,
\end{equation}
where $\pi_1$ represents the parity of the resonantly excited state $J_1$.

The analyzing power $\Sigma$ is defined by
\begin{equation}
\label{w_stuff}
 \Sigma = \frac{W(90^{\circ},0^{\circ}) - W(90^{\circ},90^{\circ})}{W(90^{\circ},0^{\circ})+ W(90^{\circ},90^{\circ})} \,\,\,\,\,,
\end{equation}
and is equal to $+1$ for the ground state decay of a $J^{\pi} = 1^+$ excited state, and $-1$ for a $J^\pi = 1^-$ state.
We therefore define the observed azimuthal count rate asymmetry of scattered photons by
\begin{equation}
\label{assym}
\epsilon = \frac{A_{h} - A_{v}}{A_{h} + A_{v}} = P_{\gamma}\Sigma\,\,\,\,\,,
\end{equation}
where $A_{h}$ and $A_{v}$ are the corresponding efficiency corrected count rates observed for the $\gamma$-rays by detectors horizontal and vertical to the scattering target.
$P_{\gamma}$ is the polarization of the photon beam, which is assumed to be 1 for all energies at the HI$\gamma$S facility. Therefore, the count rate asymmetry is equivalent to the analyzing power $\Sigma$.
The asymmetry $\epsilon$ will be equal to $+1$ for a $J^\pi_1 = 1^+$ state decaying by an $M1$ emission to the ground state, and $-1$ for a $J^\pi_1 = 1^-$ state decaying by an $E1$ emission to the ground state. Experimental observations will deviate slightly from this, as expressions given for $\epsilon$ have not accounted for the finite solid angles of the detectors, and statistical uncertainties in the data. 

\section{Experiment}
\label{four}
At the HI$\gamma$S facility \cite{Car96,Lit97,Wel08}, nearly monoenergetic  photon beams were produced by the intra-cavity Compton backscattering of linearly polarized Free-Electron Laser photons with a high energy electron beam. Polarization is conserved in this Compton scattering process, so intense, fully polarized photon beams can be produced with this technique. A typical schematic of the setup for parity measurements is shown in, e.g., Ref. \cite{Ham12}.

The photon beam was collimated by a lead collimator of length 30.5 cm with a cylindrical hole of diameter 2.54~cm before passing through to the target. The energy distribution of the photon flux was measured with a large volume high-purity germanium (HPGe) detector, of efficiency 123\% relative to a 3" $\times$ 3" NaI scintillator, placed in the incident beam. For this measurement, the beam was attenuated by copper absorbers mounted upstream. The large distance neglects the probability of the detector to measure the small angle Compton-scattered beam photons from the absorbers.

The scattered $\gamma$-rays from the target were measured by four HPGe detectors, each of 60\% relative efficiency, positioned around the Se target at $(\theta,\phi)~=~(90^{\circ},0^{\circ})$, $(90^{\circ},90^{\circ})$, $(90^{\circ},180^{\circ})$, and $(90^{\circ},270^{\circ})$, where $\theta$ is the polar angle with respect to the horizontally polarized incoming photon beam (this is defined as the polarization plane), and $\phi$ the azimuthal angle measured from the polarization plane. A fifth detector, of relative efficiency 25\%, was placed at  $(\theta,\phi)\,=\,(135^{\circ},0^{\circ})$ to distinguish the spins of positive-parity states. The distance between the center of the target to the front surface of the 90$^\circ$ detectors was 10 cm. All detectors had passive shielding consisting of 3 mm copper and 2~cm thick lead cylinders. Lead and copper absorbers of thickness 5 and 3 mm, respectively, covered the openings of the detectors to reduce the low-energy part of the scattered spectrum. The target used consisted of 11.96 g of Se powder with an enrichment of  97\% in $^{76}$Se. The powder was held in a cylindrical polypropylene container of density 2.99~g/cm$^3$, with an inner diameter and height of 1.4 cm and 2.6 cm, respectively.

The energy range of interest was scanned, with beam centroid energies incrementing up in steps of approximately the FWHM of the beam, from 4.2 MeV up to 8.8 MeV. The target was exposed for two to three hours for each beam energy.
The efficiency response of the detectors was measured using a $^{56}$Co source placed in the target position for energies below 3.2 MeV, and simulated with a GEANT4 Monte Carlo simulation \cite{Ago03} for energies exceeding this.

The HPGe detector placed at $(\theta,\phi)\,=\,(135^{\circ},0^{\circ})$ did not yield sufficient statistics for spin determination. However, Ref.~\cite{Pie94} suggests that little $E2$ strength from a $2^{+}_{i} \to 0^{+}_{g}$ transition is likely to be observed outside the $2^{+}_{1} \to 0^{+}_{g}$ transition at high energies in even-even vibrational nuclei. Therefore, all resonantly excited positive parity states are reasonably assumed to be $M1$ excited states.

 The only significant contaminant observed in the energy range between 4 and 9 MeV was from $^{12}$C, due to the composition of the target container, which has a $2^{+}_{1}$ state at 4.439 MeV \cite{ndsC}.

\section{Results}
\label{five}

The measured azimuthal asymmetries of ground-state decays observed over the energy range are shown in Fig.~\ref{fig:ass}. The mean value of the asymmetry was fitted separately for positive and negative parity states. For negative parity states $\epsilon=-0.77(2)$, and for positive parity states $\epsilon= 0.94(6)$. The deviations from the expected values of $\pm$1 are due to transitions bordering the sensitivity limit of our experiment, which may deviate from the mean observed values of $\epsilon$ as they are not well resolved above the background. None the less, the parities of these states may be firmly deduced from the plane in which the scattered $\gamma$-rays are observed.

\begin{figure}[h!]
\begin{center} 
\includegraphics[width=9cm]{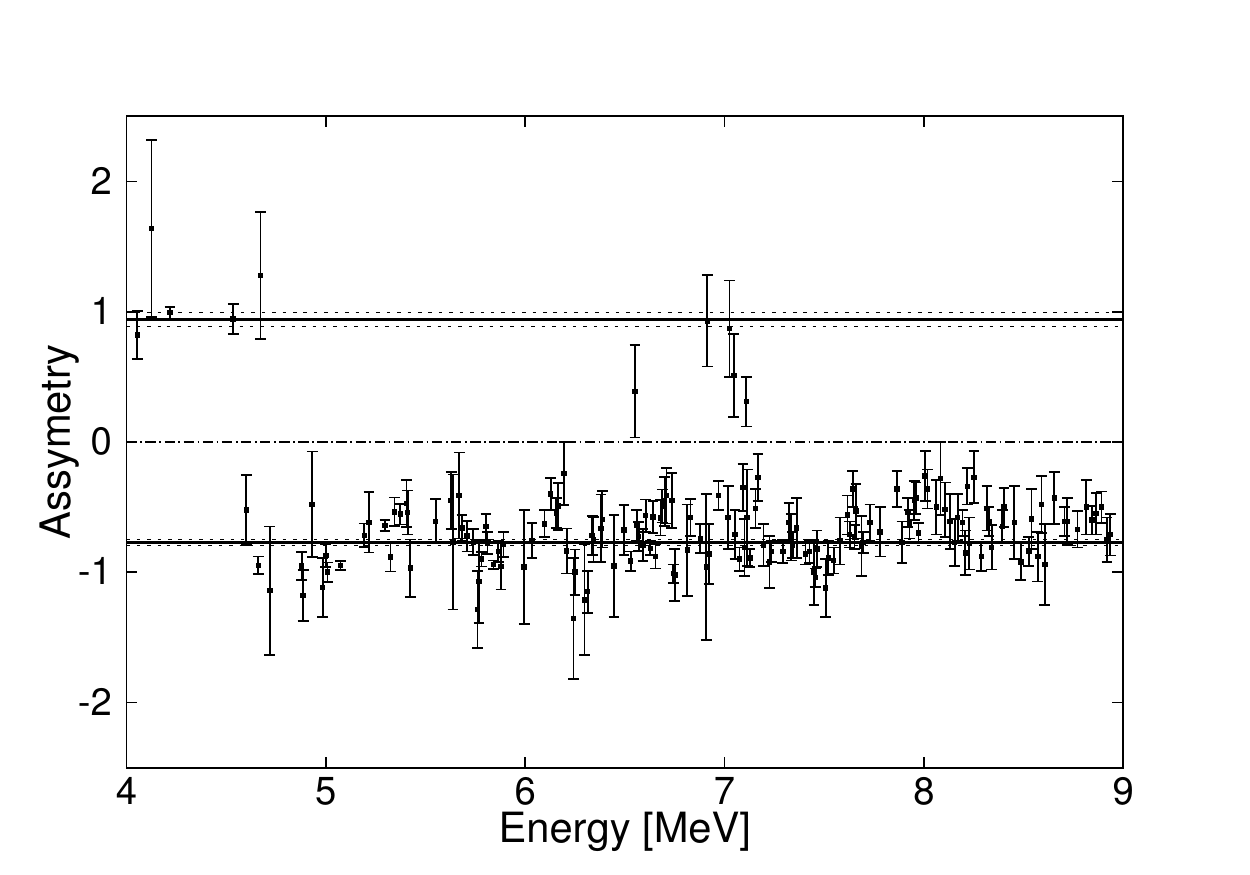} 
\caption{Azimuthal efficiency corrected count rate asymmetries for all states in $^{76}$Se observed at the HI$\gamma$S facility. A positive asymmetry corresponds to a positive parity state, and a negative asymmetry to a negative parity state. The deviations from the expected values of $\pm1$ are due to the finite opening angles of the detectors and limited statistics.}
\label{fig:ass}
\end{center}
\end{figure}

With the high photon flux available at the HI$\gamma$S facility over the entire energy range, many states of interest previously unresolved from the DHIPS  data were observed. As discrete beam energies were used, techniques to calibrate the photon flux differ from bremsstrahlung experiments. Methods to accurately deduce the photon flux at the HI$\gamma$S facility are discussed in, e.g., Refs. \cite{Sun09,Sun09i,Kwa11,Ham12,Gen13}. However, as many cross sections had been deduced from the DHIPS data, only knowledge of the energy distribution of the beams was required to infer cross sections for any newly observed state in the HI$\gamma$S data.

To calculate the photon scattering cross section, $I^{S}_{i\,\,X}$, for a newly observed state by comparison to the cross section of a known state, $I^S_{i\,\,Y}$, which was observed at the DHIPS facility, the relation
\begin{equation}
I_{i\,\, X}^{S} = \frac{n_{Y}A_{X}}{n_{X}A_{Y}}  \times  I_{i\,\, Y}^{S}
\label{cross}
\end{equation}
was considered. The normalisation factor $n_{X(Y)}$ depends on the state's position in the beam energy distribution (corresponding to the relative flux), and $A_{X(Y)}$ is the observed efficiency corrected counts in a peak. 
The energy distribution of the photon flux $n_{X(Y)}$ was measured with the large volume HPGe detector placed in the incident beam, and the full energy peak can be extracted from the measured spectrum using the methods outlined in Ref. \cite{Sun09}. The top panel of Fig. \ref{fig:tudhigs} shows an example of the measured energy distribution. Cross sections of newly observed states were then determined relative to known ones using Eq.~(\ref{cross}). This method was validated as it provided consistent results for those states observed at the DHIPS facility. 

 \begin{figure}[h!]
\begin{center} 
\includegraphics[width=9.cm]{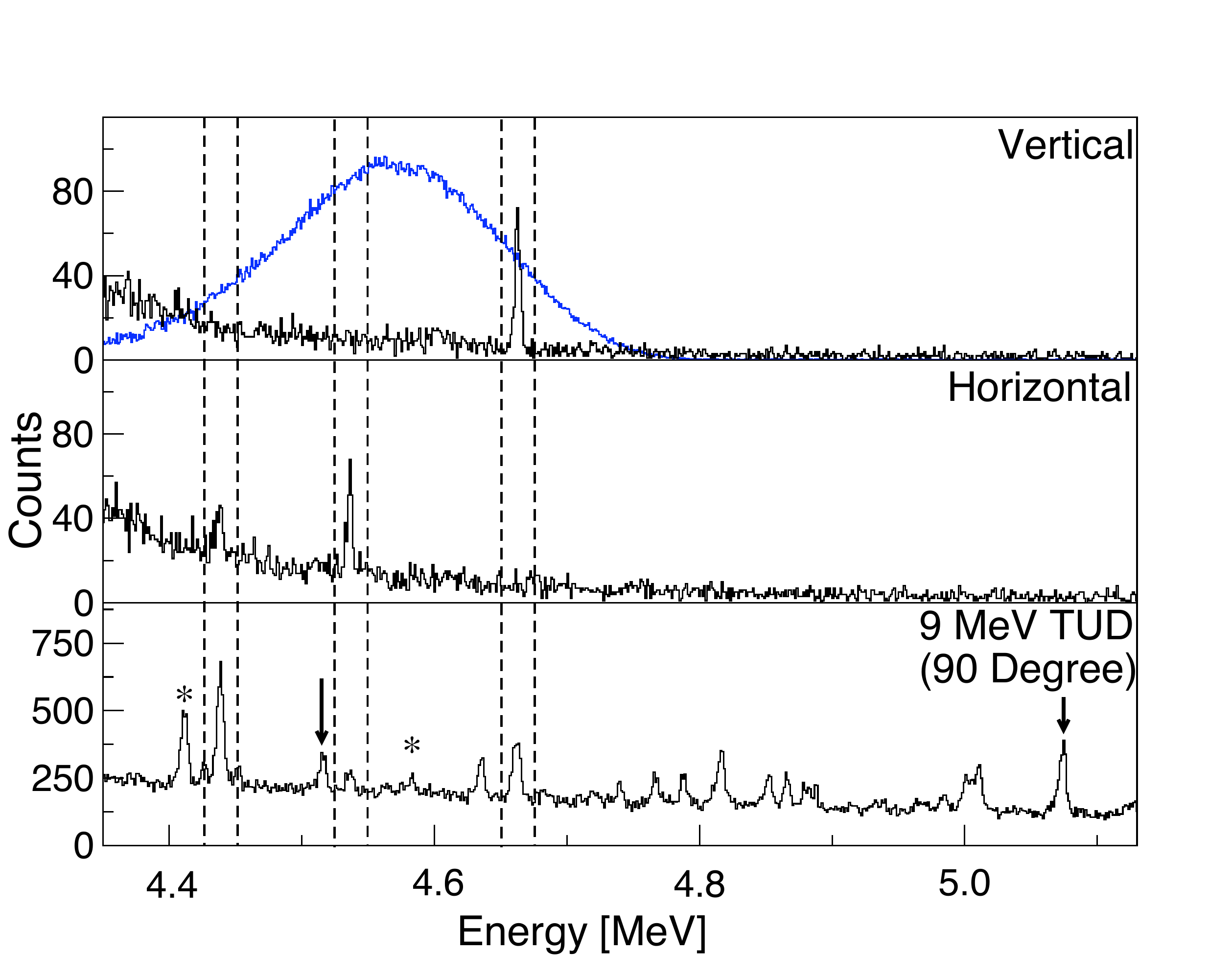} 
\caption{(color online). Comparison of the summed spectrum of vertical and horizontal detectors obtained at the HI$\gamma$S facility, to the spectrum obtained from the 90$^\circ$ detector at the DHIPS facility using the 9~MeV bremsstrahlung beam. The photon beam distribution from the 4.6 MeV beam at the HI$\gamma$S facility is superimposed (blue) upon the spectrum in the top panel for visualization of the energy range excited. Only levels decaying to the ground state are visible in the HI$\gamma$S spectrum (highlighted), and the plane in which they are observed corresponds to the parity. In the DHIPS data, $\gamma$-rays corresponding to decays to states other than the ground state are visible; an example of a decay to an excited state and a decay to the ground state from the same initially excited state is marked with arrows. Peaks marked with an $\ast$ are from the $^{27}$Al  or the $^{11}$B photon flux calibration sources. }
\label{fig:tudhigs}
\end{center}
\end{figure} 

A parity doublet of a close lying $E1$ and $M1$ excited state was observed at 5297.7(3) and 5298.4(2) keV. Shown in Fig.~\ref{fig:doub} is a peak from a ground state decay, which appears in both horizontal and vertical spectrum. From the available data, the energy difference in the fitted peak position is 0.7(3) keV.  At the DHIPS facility, a total cross section of  66.6(42) eVb was deduced, and by comparing the relative number of counts in the peak in each plane at the HI$\gamma$S facility, two states with respective cross sections of 53(3)~eVb and 13.7(8)~eVb were distinguished. 

\begin{figure}[h!]
\begin{center} 
\includegraphics[width=8.95cm]{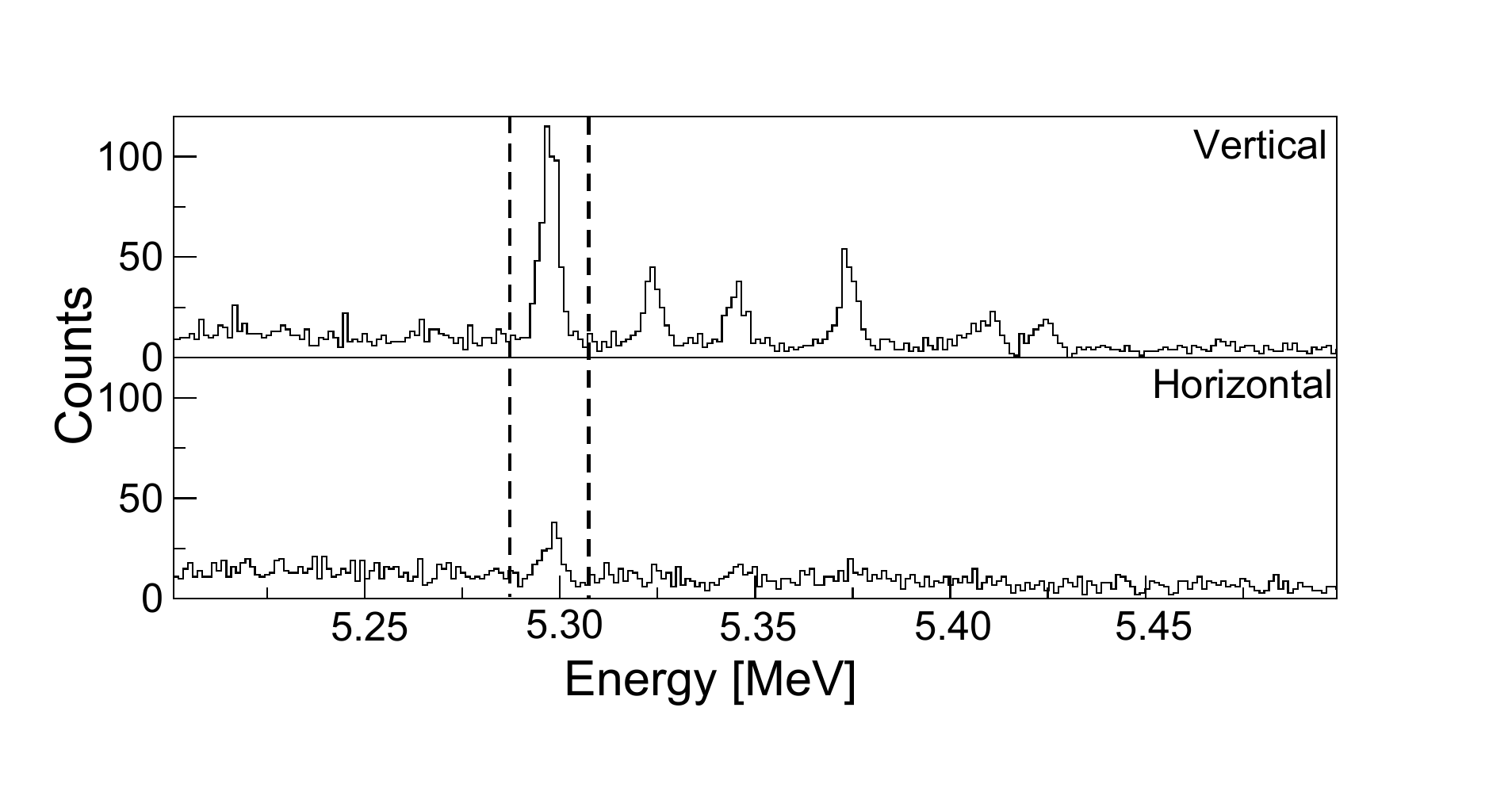} 
\caption{A doublet of positive and negative-parity $J=1$ states has been highlighted. This doublet would have been unresolvable without the use of polarized $\gamma$-rays. }
\label{fig:doub}
\end{center}
\end{figure}

The experiment at the HI$\gamma$S facility also allowed for unambiguous determination of ground-state transitions. Referring to Fig.~\ref{fig:tudhigs}, peaks corresponding to resonantly excited states decaying to the ground state, and those decaying to lower-lying excited states, were intermingled in the DHIPS spectrum. In addition, escape peaks from electron-positron annihilation further contaminated the spectrum; these can lie underneath or very close to peaks corresponding to excited states decaying. At the HI$\gamma$S facility, the narrow width of the beam profiles removed any ambiguity when distinguishing between ground state decays and either electron-positron annihilation peaks or branching decays to excited states. 

The branching ratios of the decays from excited states were of interest as they are necessary for obtaining an accurate value for $\Gamma_{0}$, and therefore the $B(\Pi\lambda)$$\uparrow$ strength. Due to relatively short exposure times at each energy window to accumulate statistics, only a few decays branching to the $2^{+}_1$ state at 559.1 keV \cite{ndsSe} could be resolved in the HI$\gamma$S spectrum. Table \ref{results} contains a compilation of the results from the experiments at the DHIPS and HI$\gamma$S facilities, using the DHIPS data where available.

\section{Discussion}
\label{six}
The distribution of the electric dipole excited states for the covered energy range is shown in Fig.~\ref{fig:con}. An increased density of $1^{-}$ states, beginning at approximately 4.5 MeV, is apparent. 

\begin{figure}[h!]
\begin{center} 
\includegraphics[width=9cm]{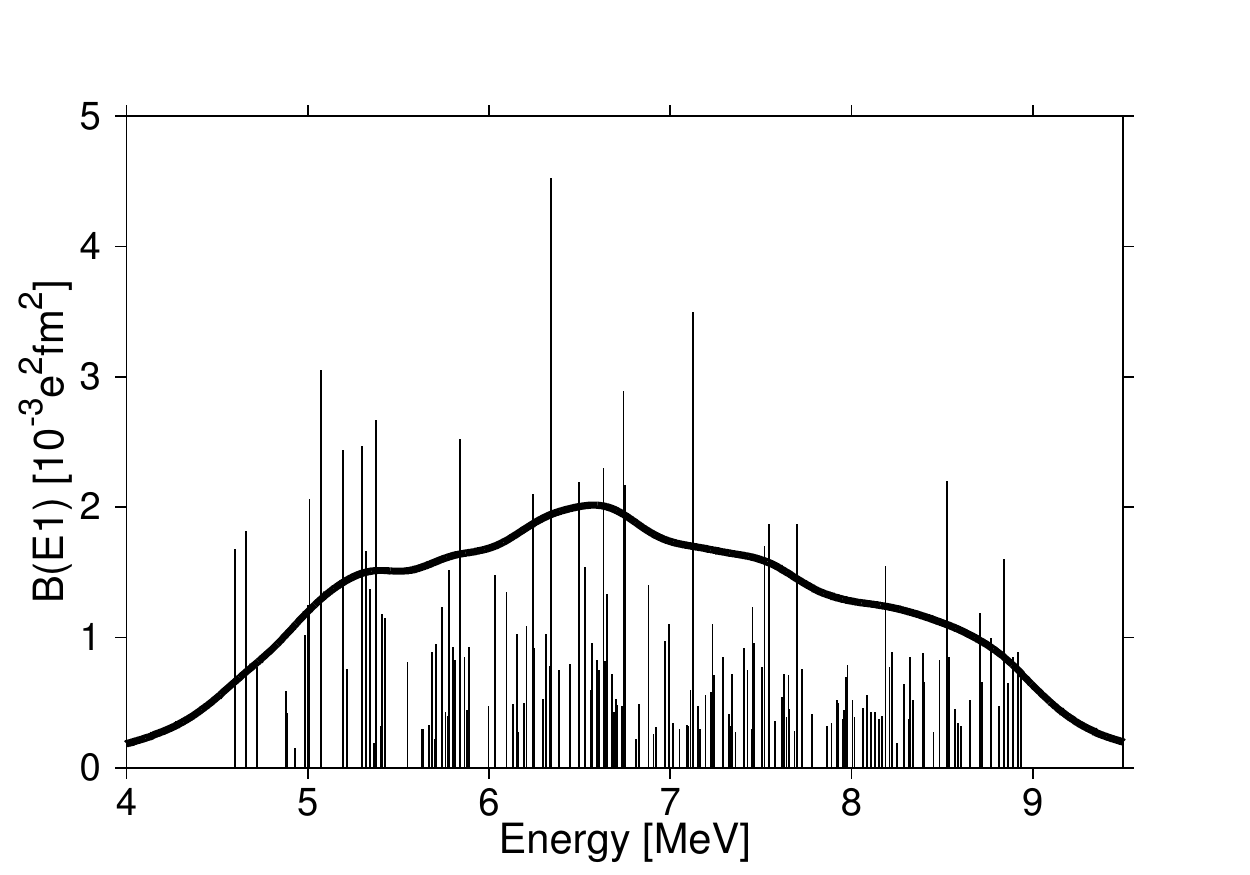} 
\caption{$B(E1)$$\uparrow$ strength distribution for resonantly excited states between 4 and 9 MeV, with a Lorentzian convoluted over the distribution for visualisation purposes. The convoluted Lorentzian does not correspond to the vertical scale of the figure, and the width was chosen so that no individual state dominated the shape. Individual contributions and uncertainties are tabulated in Table \ref{results}.} 
\label{fig:con}
\end{center}
\end{figure}

To analyse the gross features of the strength distribution of the $E1$ response, we convolute a standard Lorentzian over the $B(E1)$$\uparrow$ strengths of the resonantly excited states, and scale it for visibility. The width can be chosen to ensure that no individual state dominates the distribution. It proves useful for locating regions of concentrated strength, although other than slight enhancements around 5.2 and 6.5 MeV, no regions are seen to be prominently enhanced. Further, a pronounced splitting in the distribution, as seen for the GDR \cite{Carl76}, is not obvious.

Several $M1$ excited states were also observed, as shown in Fig.~\ref{fig:BM1}. However, it is clear that the dipole response in the energy region is predominantly electric. The dominant $M1$ strength around 4 MeV could reasonably be attributed to the Scissors Mode, which the semi-empirical formula derived by Pietralla \emph{et al.} \cite{Pie98} predicts it to be at an energy of approximately 3.9 MeV. An additional concentration of $M1$ excited states is seen around 7 MeV, which may be attributed to an $M1$ spin-flip resonance \cite{Hey10}.

Further $(\vec\gamma,{\gamma}')$ data has been acquired at the HI$\gamma$S facility, and an analysis of the nature of the dipole response of $^{76}$Se below 5 MeV will be published in a forthcoming paper. 

\begin{figure}[h!]
\begin{center} 
\includegraphics[width=9cm]{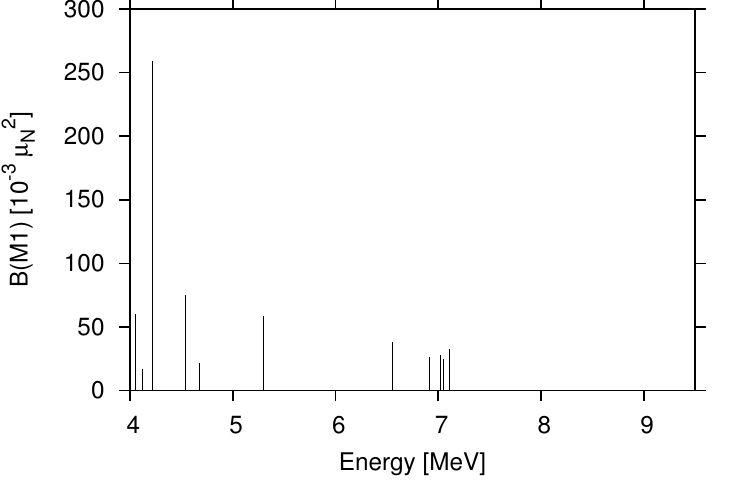} 
\caption{Distribution of observed $B(M1)$$\uparrow$ strength in the energy range covered. Individual contributions and uncertainties are tabulated in Table \ref{results}.}
\label{fig:BM1}
\end{center}
\end{figure}

\newpage

\begin{widetext}
\begin{center}
\setlongtables
\begin{longtable}[htbp]{c c c c c c c c c }
\caption{Spin $J=1$ states of $^{76}$Se observed in the energy range 4 to 9 MeV. Data has been compiled using the results from the experiments at the DHIPS and HI$\gamma$S facilities. The final state $J^{\pi}_f$ is $0^{+}_{g}$ unless otherwise mentioned. For consistency, decays to excited states are listed as $B(\Pi\lambda)\uparrow$; these can be converted to $B(\Pi\lambda)\downarrow$ by considering the statistical $g$ factor (Eq. \ref{spinfac}).  Errors are statistical only. If a $\gamma$-ray transition is observed in the DHIPS data, it is marked with a $\dagger$.} 
\label{results}
\\
$E_x$ 		&J$^{\pi}_x$	&	J$^{\pi}_f$	&	$E_\gamma$	&	$\Gamma$ 	&$\tau$ 			&	$\frac{\Gamma_i}{\Gamma}$	&	$B(E1)$$\uparrow$		&	$B(M1)$$\uparrow$ 	\\

[keV] &&&[keV]&[meV]&[fs]&&[10$^{-3}$e$^2$fm$^2$]&[10$^{-3}\mu_N^2$] \\		\hline\hline
4055.1	(3)&	1$^{+}$	&		&	4055.1	(3)$\dagger$&	15.6	(14)&	42.3	(38)&			&			-		&	60.5	(77)	\\
4125.4	(10)&	1$^{+}$	&		&	4125.4	(10)$\dagger$&	4.6	(18)&142	(55)&	&			-		&	17	(7)	\\
			4218.8	(3)&			1$^{+}$		&	&	4218.8	(3)$\dagger$&			154	(17)&		4.3	(5)&0.49	(6)&			-	&	259	(30)	\\
				&&		2$^{+}_1$	&	3659.6	(2)$\dagger$	&			&			&0.51	(6)&-&83.5 (92)												\\
4535.4	(6)&			1$^{+}$	&		&		4535.4	(6)$\dagger$&	45	(10)	&14.6	(34)&	0.60	(11)	&					-&		75 (8)				\\
			&		&	2$^{+}_1$	&	3977.2	(11)$\dagger$&				&							&0.40	(9)	&-&		14.5 (45)				\\
4601.5	(11)&			1$^{-}$	&		&	4601.5	(11)&			57	(17)&					11.6	(34)&	&					1.68	(50)&		-				\\
4662.7	(4)&			1$^{-}$	&		&	4662.7	(4)$\dagger$&			85	(14)&7.8	(13)&			0.76	(10)&							1.82	(25)&		-	\\
			&		&	2$^{+}_1$	&	4104.2	(5)$\dagger$&						&									&0.24	(4)&0.17	(3)&		-	\\
4673.5	(14)&			1$^{+}$	&		&	4673.5	(14)&			8.5	(29)&						78	(26)&&					-		&	21.6	(73)\\
4720.5	(7)&	1$^{+}$	&				&	4720.5	(7)$\dagger$&	71	(11)&	9.3	(15)&	0.4	(7)&	0.77	(13)&	-	\\
	&				&	2$^{+}_1$	&	4160.7	(4)$\dagger$&		&			&0.6	(9)&	0.34	(5)&	-	\\
4879.8	(4)&	1$^{-}$	&		&	4879.8	(4)$\dagger$&	24	(5)&	27.3	(59)&		&	0.59	(13)&-	\\
4886.9	(3)&	1$^{-}$	&		&	4886.9	(3)$\dagger$&	17	(6)&	39	(13)&		&	0.42	(14)&-	\\
4931.4	(17)&	1$^{-}$	&		&	4931.4	(17)$\dagger$&	5.8	(15)&	114	(30)&		&	0.15	(5)&-	\\
4984.3	(5)&	1$^{-}$	&		&	4984.3	(5)$\dagger$&	76	(14)&	8.7	(16)&	0.58	(8)&	1.02	(14)&-	\\
	&		&	2$^{+}_1$	&	4426.1	(5)$\dagger$&		&		&	0.42	(8)&	0.21	(4)&-	\\
5001.3	(3)&	1$^{-}$	&		&	5001.3	(3)&	54.5	(40)&	12.1	(9)&		&	1.25	(13)&-	\\
5010.3	(3)&	1$^{-}$	&		&	5010.3	(3)$\dagger$&	121	(26)&	5.4	(10)&	0.75	(7)&	2.06	(20)&-	\\
	&		&	2$^{+}_1$	&	4451.8	(3)$\dagger$&		&		&	0.25	(5)&	0.20	(4)&-	\\
5073.7	(2)&	1$^{-}$	&		&	5073.7	(2)$\dagger$&	187	(20)&	3.5	(4)&	0.74	(7)&	3.05	(28)&-	\\
	&		&	2$^{+}_1$	&	4515.8	(3)$\dagger$&		&		&	0.26	(3)&	0.30	(3)&-	\\
5194.5	(3)&	1$^{-}$	&		&	5194.5	(3)$\dagger$&	200	(22)&	3.3	(4)&	0.60	(6)&	2.44	(25)& -	\\
	&		&	2$^{+}_1$	&	4635.1	(3)$\dagger$&		&		&	0.40	(4)&	0.47	(5)&-	\\
5217.6	(11)&	1$^{-}$	&		&	5217.6	(11)&	37.6	(81)&	17.5	(38)&		&	0.76	(16)&-	\\
5297.7	(3)&	1$^{+}$	&		&	5298.4	(2)&	33.4	(20)&	19.7	(12)&		&	-&	58.2	(34)\\
5298.4	(2)&	1$^{-}$	&		&	5298.4	(2)$\dagger$&	128	(8)&	5.13	(33)&		&	2.47	(16)&-	\\
5323.8	(4)&	1$^{-}$	&		&	5323.8	(4)$\dagger$&	147	(17)&	4.5	(5)&	0.60	(6)&	1.66	(23)&-	\\
	&		&	2$^{+}_1$	&	4766.9	(10)&		&		&	0.40	(6)&	0.31	(5)&-	\\
5346.0	(4)&	1$^{-}$	&		&	5346.0	(4)$\dagger$&	133	(30)&	5.0	(11)&	0.55	(7)&	1.37	(17)&-	\\
	&		&	2$^{+}_1$	&	4788.0	(3)$\dagger$&		&		&	0.24	(4)&	0.17	(3)&-	\\
	&		&	2$^{+}_2$	&	4131.5	(9)$\dagger$&		&		&	0.21	(4)&	0.23	(4)&-	\\

5375.6	(4)&	1$^{-}$	&		&	5375.6	(4)$\dagger$&	319	(35)&	2.1	(2)&	0.45	(5)&	2.67	(29)&-	\\
	&		&	2$^{+}_1$	&	4816.1	(2)$\dagger$&		&		&	0.55	(6)&	0.89	(10)&-	\\
5405	(18)&	1$^{-}$	&		&	5405	(18)&	17.7	(55)&	37	(12)&		&	0.32	(10)&-	\\
5412.4	(14)&	1$^{-}$	&		&	5412.4	(14)$\dagger$&	14.3	(83)&	2.2	(6)&	0.22	(8)&	1.18	(42)&-	\\
	&		&	2$^{+}_1$	&	4852.0	(3)$\dagger$&		&		&	0.78	(21)&	1.16	(31)&-	\\
5425.1	(6)&	1$^{-}$	&		&	5425.1	(6)$\dagger$&	127	(18)&	5.2	(7)&	0.5	(7)&	1.15	(16)&-	\\
	&		&	2$^{+}_1$	&	4865.9	(3)$\dagger$&		&		&	0.5	(7)&	0.31	(4)&-	\\
5551.6	(15)&	1$^{-}$	&		&	5551.6 (15)	&	48	(12)&	13.6	(34)&		&	0.81	(21)&-	\\
5629.6	(15)&	1$^{-}$	&		&	5629.6	(15)&	18.7	(58)&	35	(11)&		&	0.30	(9)&-	\\
5637.5	(15)&	1$^{-}$	&		&	5637.5	(15)&	19	(6)&	35	(11)&		&	0.30	(10)&-	\\
5669.0	(15)&	1$^{-}$	&		&	5669.0	(15)&	20.9	(74)&	32	(11)&		&	0.33	(12)&-	\\
5685.3	(4)&	1$^{-}$	&		&	5685.3	(4)$\dagger$&	57.4	(51)&	11.5	(10)&		&	0.89	(11)&-	\\

5709.6	(5)&	1$^{-}$	&		&	5709.6	(5)$\dagger$&	61.9	(57)&	10.6	(10)&		&	0.95	(12)&-	\\
5740.5	(5)&	1$^{-}$	&		&	5740.5	(5)$\dagger$&	81.1	(66)&	8.1	(7)&		&	1.23	(14)&-	\\
5761.8	(10)&	1$^{-}$	&		&	5761.8	(10)&	29	(6)&	22.7	(49)&		&	0.43	(9)&-	\\
5773.1	(20)&	1$^{-}$	&		&	5773.1	(10)$\dagger$&	26.6	(40)&	24.7	(37)&		&	0.40	(8)&-	\\
5781.0	(2)&	1$^{-}$	&		&	5781.0	(2)$\dagger$&	102	(22)&	6.4	(14)&		&	1.52	(33)&-	\\
5803.4	(7)&	1$^{-}$	&		&	5803.4	(7)$\dagger$&	178	(43)&	3.7	(9)&	0.36	(9)&	0.93	(22)&-	\\
	&		&	2$^{+}_1$	&	5246.1	(14)$\dagger$&		&		&	0.64	(16)&	0.46	(11)&-	\\
	5813.7	(5)&	1$^{-}$	&		&	5813.7	(5)$\dagger$&	57.2	(54)&	11.5	(11)&		&	0.83	(11)&-	\\
5842.0	(3)&	1$^{-}$	&		&	5842.0	(3)$\dagger$&	221	(65)&	3.0	(9)&	0.80	(9)&	2.54	(90)&-	\\
	&		&	2$^{+}_1$	&	5283.8	(10)$\dagger$&		&		&	0.20	(6)&	0.17	(5)&-	\\
5865.1	(7)&	1$^{-}$	&		&	5865.1	(7)&	59.8	(85)&	11.0	(16)&		&	0.85	(12)&-	\\
5879.4	(7)&	1$^{-}$	&		&	5879.4	(7)$\dagger$&	31	(4)&	21.3	(28)&		&	0.44	(8)&-	\\
5891.9	(6)&	1$^{-}$	&		&	5891.9	(6)$\dagger$&	136	(24)&	4.9	(9)&	0.56	(9)&	0.93	(22)&-	\\
	&		&	2$^{+}_1$	&	5333.1	(5)$\dagger$&		&		&	0.44	(8)&	0.46	(11)&-	\\
5998.4	(14)&	1$^{-}$	&		&	5998.4	(14)$\dagger$&	85	(19)&	7.7	(18)&	0.41	(11)&	0.47	(13)&-	\\
	&		&	2$^{+}_1$	&	5435.2	(11)$\dagger$&		&		&	0.59	(13)&	0.18	(4)&-	\\
6035.4	(5)&	1$^{-}$	&		&	6035.4	(5)$\dagger$&	174	(26)&	3.8	(6)&	0.66	(8)&	1.48	(27)&-	\\
	&		&	2$^{+}_1$	&	5474.6	(13)&		&		&	0.34	(7)&	0.21	(4)&-	\\
6098.9	(6)&	1$^{-}$	&		&	6098.9	(6)$\dagger$&	164	(27)&	4.00	(7)&	0.65	(10)&	1.35	(20)&-	\\
	&		&	2$^{+}_1$	&	5540.2	(7)$\dagger$&		&		&	0.35	(6)&	0.19	(3)&-	\\
6131.2	(6)&	1$^{-}$	&		&	6131.2	(6)$\dagger$&	39.6	(61)&	16.6	(26)&		&	0.49	(11)&-	\\
6156.3	(14)&	1$^{-}$	&		&	6156.3	(14)&	84	(15)&	79	(14)&		&	1.03	(18)&-	\\
6164.8	(11)&	1$^{-}$	&		&	6164.8	(11)&	22.3	(66)&	29.6	(87)&		&	0.27	(8)&-	\\
6195.9	(11)&	1$^{-}$	&		&	6195.9	(11)&	45.7	(61)&	14.4	(19)&		&	0.55	(10)&-	\\
6208.4	(15)&	1$^{-}$	&		&	6208.4	(15)&	91	(18)&	7.2	(14)&		&	1.09	(21)&-	\\
6242.4	(6)&	1$^{-}$	&		&	6242.4	(6)$\dagger$&	175	(76)&	3.8	(16)&		&	2.1	(9)&-	\\
6250.4	(5)&	1$^{-}$	&		&	6250.4	(5)$\dagger$&	79	(20)&	8.4	(22)&		&	0.92	(24)&- 	\\
6297.6	(14)&	1$^{-}$	&		&	6297.6	(14)$\dagger$&	45.8	(66)&	14.4	(21)&		&	0.53	(11)&-	\\
6315.6	(4)&	1$^{-}$	&		&	6315.6	(4)$\dagger$&	91	(23)&	7.3	(18)&		&	1.03	(26)&-	\\
6336.5	(20)&	1$^{-}$	&		&	6336.5	(20)$\dagger$&	69	(35)&	3.0	(15)&		&	0.78	(15)&-	\\
6342.3	(11)&	1$^{-}$	&		&	6342.3	(11)$\dagger$&	1440 	(350)&	0.4	(1)&	0.28	(5)&	4.53	(96)&	-\\
	&		&	2$^{+}_1$	&	5783.3	(3)&		&		&	0.72	(10)&	3.33	(66)&	-\\
6387.2	(14)&	1$^{-}$	&		&	6387.2	(14)$\dagger$&	68	(11)&	9.6	(15)&		&	0.75	(16)&-	\\
6448.7	(20)&	1$^{-}$	&		&	6448.7	(20)$\dagger$&	75	(12)&	8.8	(15)&		&	0.8	(2)&-	\\
6497.4	(6)&	1$^{-}$	&		&	6497.4	(6)$\dagger$&	210	(65)&	3.13	(97)&		&	2.19	(68)&-	\\
6532.4	(4)&	1$^{-}$	&		&	6532.4	(4)$\dagger$&	150	(14)&	4.4	(4)&		&	1.54	(21)&-	\\
6550.7	(3)&	1$^{+}$	&		&	6550.7	(3)$\dagger$&	41.6	(74)&	15.8	(28)&		&	-&	38.4	(96)\\
6562.6	(9)&	1$^{-}$	&		&	6562.6	(9)$\dagger$&	59	(3)&	11.1	(4)&		&	0.60	(7)&-	\\
6570.1	(9)&	1$^{-}$	&		&	6570.1	(9)$\dagger$&	95	(13)&	7.0	(9)&		&	0.96	(18)&-	\\
6595.9	(7)&	1$^{-}$	&		&	6595.9	(7)$\dagger$&	83	(10)&	7.9	(10)&		&	0.83	(15)&-	\\
6608.2	(9)&	1$^{-}$	&		&	6608.2	(9)$\dagger$&	76	(10)&	8.7	(12)&		&	0.75	(14)&-	\\
6632.9	(12)&	1$^{-}$	&		&	6632.9	(12)$\dagger$&	327	(50)&	2.0	(4)&	0.71	(22)&	2.3	(4)&-	\\
	&		&	2$^{+}_1$	&	6071.8	(8)$\dagger$&		&		&	0.28	(14)&	0.24	(11)&-	\\
6641.0	(17)&	1$^{-}$	&		&	6641.0	(17)&	84	(18)&	7.9	(17)&		&	0.82	(18)&-	\\
6653.4	(14)&	1$^{-}$	&		&	6653.4	(14)&	136	(27)&	4.8	(10)&		&	1.33	(26)&-	\\
6679.7	(18)&	1$^{-}$	&		&	6679.7	(18)&	75	(17)&	8.8	(10)&		&	0.72	(16)&-	\\
6691.2	(8)&	1$^{-}$	&		&	6691.2	(8)$\dagger$&	44.7	(74)&	14.7	(24)&		&	0.43	(7)&-	\\
6700.0	(20)&	1$^{-}$	&		&	6700.0	(20)&	56	(14)&	11.8	(30)&		&	0.53	(13)&-	\\
6708.7	(21)&	1$^{-}$	&		&	6708.7	(21)&	51	(14)&	13.1	(36)&		&	0.48	(13)&-	\\
6735.9	(15)&	1$^{-}$	&		&	6735.9	(15)&	50	(14)&	13.1	(36)&		&	0.47	(13)&-	\\
6743.2	(3)&	1$^{-}$	&		&	6743.2	(3)$\dagger$&	401	(39)&	1.6	(2)&	0.77	(10)&	2.89	(27)&-	\\
	&		&	2$^{+}_1$	&	6182.8	(7)$\dagger$&		&		&	0.23	(4)&	0.22	(5)&-	\\
6750.9	(9)&	1$^{-}$	&		&	6748.7	(5)$\dagger$&	532	(51)&	1.9	(3)&	0.66	(13)&	2.17	(28)&-	\\
	&		&	2$^{+}_1$	&	6190.0	(6)$\dagger$&		&		&	0.34	(9)&	0.29	(9)&-	\\
6813.6	(20)&	1$^{-}$	&		&	6813.6	(20)&	24.1	(71)&	23.7	(81)&		&	0.22	(6)&-	\\
6829.9	(15)&	1$^{-}$	&		&	6829.9	(15)&	55	(12)&	12.0	(26)&		&	0.49	(11)&-	\\
6881.9	(14)&	1$^{-}$	&		&	6881.9	(14)$\dagger$&	296	(59)&	2.2	(4)&	0.54	(14)&	1.40	(24)&- 	\\
	&		&	2$^{+}_1$	&	6323.4	(6)$\dagger$&		&	&		0.46	(16)&	0.31	(12)&-	\\
6908.0	(20)&	1$^{-}$	&		&	6908.0	(20)&	29.9	(78)&	22.0	(58)&		&	0.26	(7)&-	\\
6913.0 (17)&	1$^{+}$	&		&	6913	(17)&	33	(11)&	19.7	(63)&		&	-&	26.2	(84)\\
6921.9	(18)&	1$^{-}$	&		&	6921.9	(18)&	36.1	(94)&	18.2	(47)&		&	0.31	(8)&-	\\
6970.0	(5)&	1$^{-}$	&		&	6970.0	(5)$\dagger$&	115	(26)&	5.7	(13)&		&	0.97	(22)&-	\\
6992.5	(5)&	1$^{-}$	&		&	6992.5	(5)$\dagger$&	130	(18)&	4.7	(7)&		&	1.10	(15)&-	\\
7017.7	(18)&	1$^{-}$	&		&	7017.7	(18)&	41	(17)&	16.1	(66)&		&	0.34	(14)&-	\\
7024.7	(20)&	1$^{+}$	&		&	7024.7	(20)&	37	(13)&	17.7	(60)&		&	-&	27.8	(95)\\
7047.0 (15)&	1$^{+}$	&		&	7047	(15)&	33	(11)&	19.9	(68)&		&	-&	24.5	(84)\\
7052.7	(19)&	1$^{-}$	&		&	7052.7	(19)&	36	(11)&	18.1	(54)&		&	0.30	(9)&-	\\
7092.7	(20)&	1$^{-}$	&		&	7092.7	(20)&	41	(11)&	16.2	(44)&		&	0.33	(9)&-	\\
7100.7	(19)&	1$^{-}$	&		&	7100.7	(19)&	40	(12)&	16.5	(51)&		&	0.32	(10)&-	\\
7109.7	(19)&	1$^{+}$	&		&	7109.7	(19)&	46	(13)&	14.4	(42)&		&	-&	32.9	(97)\\
7113.6	(19)&	1$^{-}$	&		&	7113.6	(19)&	115	(51)&	4.2	(14)&	0.49	(18)&	0.60	(24)&- 	\\
	&		&	2$^{+}_1$	&	6557.2	(16)&		&		&	0.51	(19)&	0.13	(6)&-	\\
7127.3	(13)&	1$^{-}$	&		&	7127.3	(13)&	570	(150)&	2.9	(2)&	0.77	(23)&	3.5	(14)&-	\\
	&		&	2$^{+}_1$	&	6570.6	(19)&		&		&	0.23	(17)&	0.26	(20)&-	\\
7155.6	(17)&	1$^{-}$	&		&	7155.6	(17)&	61	(17)&	11	(3)&		&	0.47	(13)&-	\\
7167.7	(18)&	1$^{-}$	&		&	7167.7	(18)&	39	(11)&	17	(5)&		&	0.30	(9)&-	\\
7195.2	(14)&	1$^{-}$	&		&	7195.2	(14)&	72	(21)&	9.1	(26)&		&	0.56	(16)&-	\\
7225.2	(20)&	1$^{-}$	&		&	7225.2	(20)&	77	(19)&	8.6	(21)&		&	0.58	(14)&-	\\
7241.2	(7)&	1$^{-}$	&		&	7241.2	(7)$\dagger$&	94	(19)&	7.0	(14)&		&	0.71	(14)&-	\\
7292.4	(15)&	1$^{-}$	&		&	7292.4	(15)&	115	(31)&	5.7	(15)&		&	0.85	(23)&-	\\
7324.2	(18)&	1$^{-}$	&		&	7324.2	(18)&	56	(16)&	12.0	(34)&		&	0.41	(12)&-	\\
7334.6	(20)&	1$^{-}$	&		&	7334.6	(20)&	44	(14)&	14.9	(47)&		&	0.32	(10)&-	\\
7341.8	(14)&	1$^{-}$	&		&	7341.8	(14)&	99	(26)&	6.6	(18)&		&	0.72	(19)&-	\\
7361.8	(21)&	1$^{-}$	&		&	7361.8	(21)&	37	(12)&	17.8	(57)&		&	0.27	(9)&-	\\
7392.2	(8)&	1$^{-}$	&		&	7392.2	(8)&	35	(11)&	19	(6)&		&	0.25	(8)&-	\\
7406.0	(15)&	1$^{-}$	&		&	7406.0	(15)&	188	(99)&	3.5	(18)&	0.69	(26)&	0.92	(61)&	-\\
	&		&	2$^{+}_1$	&	6846.0	(17)&		&		&	0.31	(20)&	0.08	(7)&-	\\
7426.7	(14)&	1$^{-}$	&		&	7426.7	(14)&	108	(28)&	6.1	(16)&		&	0.75	(20)&-	\\
7455.1	(13)&	1$^{-}$	&		&	7455.1	(13)$\dagger$&	178	(46)&	3.7	(9)&		&	1.23	(32)&-	\\
7464.3	(18)&	1$^{-}$	&		&	7464.3	(18)&	252	(88)&	2.6	(9)&	0.55	(20)&	0.96	(41)&-	\\
	&		&	2$^{+}_1$	&	6905.8	(21)&		&		&	0.45	(19)&	0.20	(9)&-	\\
7508.0	(8)&	1$^{-}$	&		&	7508.0	(8)$\dagger$&	114	(24)&	5.8	(7)&		&	0.77	(9)&-	\\
7521.7	(7)&	1$^{-}$	&		&	7521.7	(7)$\dagger$&	396	(71)&	1.7	(3)&	0.64	(16)&	1.70	(29)&-	\\
	&		&	2$^{+}_1$	&	6963.9	(7)$\dagger$&		&		&	0.36	(11)&	0.24	(8)&-	\\
7546.5	(6)&	1$^{-}$	&		&	7546.5	(6)$\dagger$&	280	(29)&	2.4	(2)&		&	1.87	(19)&-	\\
7580.1	(16)&	1$^{-}$	&		&	7580.1	(16)&	55	(16)&	11.9	(33)&		&	0.36	(10)&-	\\
7616.8	(17)&	1$^{-}$	&		&	7616.8	(17)&	83	(17)&	7.9	(16)&		&	0.54	(11)&-	\\
7627.4	(15)&	1$^{-}$	&		&	7627.4	(15)&	111	(20)&	5.9	(11)&		&	0.72	(13)&-	\\
7642.9	(17)&	1$^{-}$	&		&	7642.9	(17)&	61	(15)&	10.8	(27)&		&	0.39	(10)&-	\\
7652.5	(17)&	1$^{-}$	&		&	7652.5	(17)&	110	(22)&	5.9	(12)&		&	0.71	(14)&-	\\
7658.3	(2)&	1$^{-}$	&		&	7658.3	(2)$\dagger$&	71	(12)&	9.3	(15)&		&	0.45	(7)&-	\\
7698.2	(9)&	1$^{-}$	&		&	7698.2	(9)$\dagger$&	460	(140)&	1.4	(4)&	0.65	(16)&	1.87	(70)&-	\\
	&		&	2$^{+}_1$	&	7137.0	(20)&		&		&	0.35	(14)&	0.26	(10)&-	\\
7729.3	(16)&	1$^{-}$	&		&	7729.3	(16)&	122	(25)&	5.4	(11)&		&	0.76	(16)&-	\\
7781.2	(18)&	1$^{-}$	&		&	7781.2	(18)&	67	(22)&	9.9	(32)&		&	0.41	(13)&-	\\
7817	.1	(10)&	1$^{-}$	&		&	7817	.1	(10)&	47	(17)&	14	(5)&		&	0.28	(10)&-	\\
7829.6	(9)&	1$^{-}$	&		&	7829.6	(9)&	50	(20)&	13	(5)&		&	0.30	(10)&-	\\
7865.7	(17)&	1$^{-}$	&		&	7865.7	(17)&	55	(18)&	12.0	(39)&		&	0.32	(10)&-	\\
7890.5	(18)&	1$^{-}$	&		&	7890.5	(18)&	59	(19)&	11.2	(36)&		&	0.34	(11)&-	\\
7919.7	(17)&	1$^{-}$	&		&	7919.7	(17)&	90	(28)&	7.3	(23)&		&	0.52	(16)&-	\\
7927.2	(17)&	1$^{-}$	&		&	7927.2	(17)&	87	(27)&	7.6	(24)&		&	0.50	(16)&-	\\
7951.6	(21)&	1$^{-}$	&		&	7951.6	(21)&	64	(21)&	10.3	(34)&		&	0.37	(12)&-	\\
7959.9	(18)&	1$^{-}$	&		&	7959.9	(18)&	77	(24)&	8.5	(27)&		&	0.44	(14)&-	\\
7978.5	(8)&	1$^{-}$	&		&	7978.5	(8)$\dagger$&	139	(34)&	4.7	(11)&		&	0.79	(19)&-	\\
8017.4	(23)&	1$^{-}$	&		&	8017.4	(23)&	69	(23)&	9.5	(31)&		&	0.39	(13)&-	\\
8062.0	(22)&	1$^{-}$	&		&	8062.0	(22)&	85	(27)&	7.8	(25)&		&	0.46	(15)&-	\\
8084.2	(26)&	1$^{-}$	&		&	8084.2	(26)&	220	(100)&	3.3	(12)&	0.46	(25)&	0.56	(31)&-	\\
	&		&	2$^{+}_1$	&	7521.3	(25)&		&		&	0.54	(26)&	0.16	(9)&-	\\
8106.8	(22)&	1$^{-}$	&		&	8106.8	(22)&	80	(25)&	8.2	(25)&		&	0.43	(13)&-	\\
8131.6	(22)&	1$^{-}$	&		&	8131.6	(22)&	79	(24)&	8.4	(25)&		&	0.43	(13)&-	\\
8154.4	(21)&	1$^{-}$	&		&	8154.4	(21)&	70	(21)&	9.4	(28)&		&	0.37	(11)&-	\\
8169.6	(22)&	1$^{-}$	&		&	8169.6	(22)&	76	(22)&	8.7	(25)&		&	0.40	(11)&-	\\
8197.0	(13)&	1$^{-}$	&		&	8196.5	(13)$\dagger$&	580	(120)&	1.1	(2)&	0.52	(14)&	1.55	(27)&-	\\
	&		&	2$^{+}_2$	&	6982.8	(15)$\dagger$&		&		&	0.48	(16)&	0.47	(18)&-	\\
8210.0	(20)&	1$^{-}$	&		&	8210.0	(20)&	114	(29)&	5.8	(14)&		&	0.77	(19)&-	\\
8222.0	(20)&	1$^{-}$	&		&	8222.0	(20)&	183	(45)&	3.6	(9)&		&	0.89	(22)&-	\\
8251.4	(23)&	1$^{-}$	&		&	8251.4	(23)&	37	(15)&	17.9	(74)&		&	0.19	(8)&-	\\
8288.0	(23)&	1$^{-}$	&		&	8288.0	(23)&	127	(32)&	5.2	(13)&		&	0.64	(16)&-	\\
8316.2	(22)&	1$^{-}$	&		&	8316.2	(22)&	75	(25)&	8.8	(30)&		&	0.37	(13)&-	\\
8340.2	(10)&	1$^{-}$	&		&	8340.2	(10)&	104	(31)&	6.3	(19)&		&	0.52	(15)&-	\\
8394.4	(19)&	1$^{-}$	&		&	8394.4	(19)$\dagger$&	520	(140)&	1.25	(33)&		&	0.88	(12)&-	\\
8453.0	(21)&	1$^{-}$	&		&	8453.0	(21)&	162	(60)&	4	(1)&		&	0.27	(10)&-	\\
8486.0	(18)&	1$^{-}$	&		&	8486.0	(18)&	500	(120)&	1.32	(33)&		&	0.83	(21)&-	\\
8526.0	(5)&	1$^{-}$	&		&	8526.0	(5)$\dagger$&	950	(210)&	0.69	(14)&	0.50	(12)&	2.20	(36)&-	\\
	&		&	2$^{+}_1$	&	7970.8	(6)$\dagger$&		&		&	0.50	(15)&	0.54	(20)&-	\\
8540.4	(20)&	1$^{-}$	&		&	8540.4	(20)&	488	(91)&	1.35	(25)&	0.38	(15)&	0.85	(48)&-	\\
	&		&	2$^{+}_1$	&	7979.7	(13)&		&		&	0.62	(18)&	0.32	(16)&-	\\
8571.2	(19)&	1$^{-}$	&		&	8571.2	(19)&	270	(79)&	2.43	(71)&		&	0.45	(13)&-	\\
8589.6	(20)&	1$^{-}$	&		&	8589.6	(20)&	199	(64)&	3.3	(11)&		&	0.34	(11)&-	\\
8654.4	(19)&	1$^{-}$	&		&	8654.4	(19)&	228	(68)&	2.88	(87)&		&	0.52	(11)&-	\\
8709.4	(13)&	1$^{-}$	&		&	8709.4	(13)$\dagger$&	274	(42)&	2.4	(4)&		&	1.19	(18)&-	\\
8719.0	(21)&	1$^{-}$	&		&	8719.0	(21)&	154	(54)&	4.3	(15)&		&	0.66	(23)&-	\\
8770.4	(23)&	1$^{-}$	&		&	8770.4	(23)&	236	(67)&	2.8	(8)&		&	1.00	(29)&-	\\
8843.2	(18)&	1$^{-}$	&		&	8843.2	(18)&	560	(290)&	1.2	(6)&	0.68	(26)&	1.60	(10)&-	\\
	&		&	2$^{+}_1$	&	8283.3	(20)&		&		&	0.32	(20)&	0.18	(10)&-	\\
8864.2	(20)&	1$^{-}$	&		&	8864.2	(20)&	158	(50)&	4.2	(13)&		&	0.65	(20)&-	\\
8890.2	(19)&	1$^{-}$	&		&	8890.2	(19)&	209	(60)&	3.1	(9)&		&	0.85	(24)&- 	\\
8918.2	(19)&	1$^{-}$	&		&	8918.2	(19)&	221	(64)&	3.0	(9)&		&	0.89	(26)&-	\\
8935.0	(20)&	1$^{-}$	&		&	8935.0	(20)&	178	(56)&	3.7	(12)&		&	0.71	(23)&-	\\

\end{longtable}
\end{center}
\end{widetext}

Unobserved branching decays due to poor statistics and high background were a potential issue when summing the $B(\Pi\lambda)\uparrow$ strengths up to 9~MeV. The vast majority of transitions directly observed in this work and in Ref. \cite{Coo10} are decays to the ground state. If a branching in the decay of a resonantly excited $J^\pi =1^-$ state is unobserved, the corresponding calculated $B(E1)\uparrow$ strength will be underestimated.
 
Resonantly excited $J=1$ states, which branch to states other than the ground state in their decay, are likely to cascade through excited states, eventually collecting in low-lying states before the nucleus de-excites to the ground state. In $^{76}$Se, only the first two low-lying $2^+$ states have significant ground-state decay branches. Therefore, it is likely that any excited state decays will cascade through one or both of these states before decaying to the ground state by $E2$ $\gamma$-ray emission. Due to this, consideration of the intensity of the decays from the low-lying $2^{+}$ states can provide a quantitative measure of the `missing' cross section due to unobserved branching decays \cite{Ton10}. 

For beam energy settings higher than 6 MeV, we begin to observe transitions corresponding to low-lying $2^+$ states with significant statistics. As the beams were nearly monoenergetic, and tuned to energies of several MeV above the $2^{+}$ states, this implies that they were not excited directly from the ground state.  Indeed, Ref. \cite{Hag09} demonstrates that the effect of Compton scattered $\gamma$-rays exciting states below the incident beam energy is negligible. Therefore, it is safe to assume that the low-lying $2^{+}$ states were populated entirely by feeding from the decays of higher-lying resonantly excited states.

The angular distribution of the $E2$ $\gamma$-ray transitions to the ground state from the low-lying $2^{+}$ states will differ from that of a $0^{+}\to2^{+}\to0^{+}$ sequence, as shown in Fig.~\ref{fig:angdist}, if they are populated by feeding from higher lying excited states. For each consecutive intermediate transition between the resonantly excited state and the ground state the nucleus realigns, resulting in an increasingly isotropic angular distribution of the emitted $\gamma$-rays.
The formalism contained in Ref.~\cite{Kra73} was used to calculate the expected angular distributions of the $E2$ $\gamma$-ray transitions from the low-lying quadrupole states to the ground state, assuming they were populated by feeding. 

The expression for the angular distribution of the observed multipole emission $\gamma_n$ is the following:
\begin{eqnarray}
\lefteqn{W(\theta) =} && \\ \nonumber
 &&\sum_{\lambda_1=0,2,4}B_{\lambda_1}(\gamma_{1})\left(\prod^{n-1}_{k=1}U_{\lambda_1\lambda_1 }(\gamma_{k})\right)A_{\lambda_1}(\gamma_{n})P_{\lambda_1}(\cos\theta) \,\,\,\,\,,
\label{ind_casc}
\end{eqnarray}
where we have set $\lambda= 0$ to account for unobserved transitions. We assume in our case that the initial state has been excited and oriented by linearly polarized $\gamma$-rays. The $U_{\lambda\lambda}$ are the deorientation coefficients defining the $n-1$ intermediate unobserved transitions between the initially excited state and the penultimate state; for every intermediate state in the cascade an additional $U_{\lambda\lambda}$ must be included in the expansion. It is only the $\gamma$-ray emission from the penultimate to ultimate state which is observed. Ref. \cite{Kra73} allows us to define the $U_{\lambda_1\lambda_1}$ coefficient by considering the generalised $A^{\lambda_2\lambda_1}_{\lambda}$ coefficient:
\begin{equation}
A_{\lambda=0}^{\lambda_2\lambda_1}(\gamma_1) = \sqrt{(2\lambda_1+1)}U_{\lambda_1\lambda_1}(\gamma_1)\delta_{\lambda_1\lambda_2}\,\,\,\,\,.
\label{a_gen}
\end{equation}

The $A^{\lambda_2\lambda_1}_{\lambda}$, with $\lambda=0$ reduces to
\begin{eqnarray}
\label{alamgen}
\lefteqn {A_{0}^{\lambda_2\lambda_1}= \left(\frac{1}{1+\delta^2}\right) \Big{[}F^{\lambda_2\lambda_1}_{0}(LLJ_fJ_i)} \\ \nonumber
&& + 2\delta F^{\lambda_2\lambda_1}_{0}(LL'J_fJ_i) + \delta^2F^{\lambda_2\lambda_1}_{0}(L'L'J_fJ_i)\Big{]}\,\,\,\,\, ,
\end{eqnarray}
where the intermediate state $J_i$ transitions to state $J_f$. The $F^{\lambda_2\lambda_1}_{0}$ coefficients are given in Ref. \cite{Kra73}.

The calculated azimuthal asymmetries confirm that the angular distribution does become more isotropic as extra intermediate states are involved in the cascade. For a $0^{+}\to1^{-}\to2^{+}\to0^{+}$ sequence the final $E2$ transition has $\epsilon$~=~0.33 (assuming a pure dipole transition feeding the $2^{+}$ state). A $0^{+}\to1^{-}\to2^{+}\to2^{+}\to0^{+}$ sequence yields $\epsilon$~=~$-0.08$ (assuming the $2^{+}$ states decay to one another with a pure $E2$ emission). A $0^{+}\to1^{-}\to2^{+}\to2^{+}\to2^{+}\to0^{+}$ sequence has $\epsilon$~=~0.02. Additional intermediate $2^{+}$ (or other $J^\pi$) states further increase the isotropy. The observed azimuthal asymmetry of the $2^{+}_{1,2}\to0^{+}_{g}$ transitions are shown in Fig \ref{2+ass}. The experimentally observed angular distribution can be a superposition of the results of a collection of excited states, de-exciting through different levels, with different multipole mixing ratios, so the values calculated using Eq. (\ref{ind_casc}) rather serve as guidance.

\begin{figure}[h!]
\begin{center} 
\includegraphics[width=9.5cm]{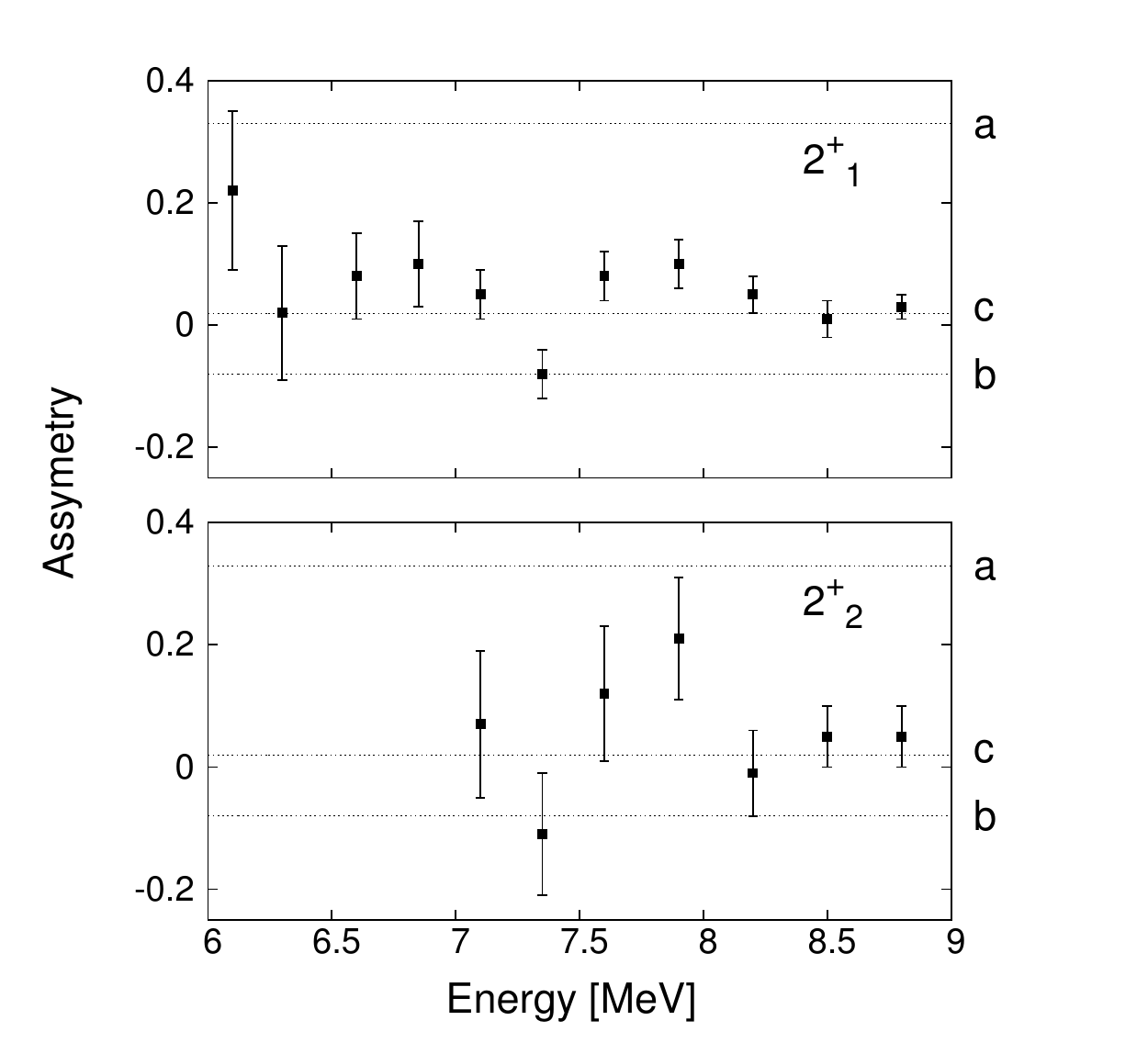} 
\caption{Observed count rate asymmetry of the 2$^+_{1,2}$ to ground state transitions. Dashed lines show the expected asymmetries of the final $E2$ transition of an a) $0^{+}\to1^{-}\to2^{+}\to0^{+}$, b) $0^{+}\to1^{-}\to2^+\to2^{+}\to0^{+}$, and c) $0^{+}\to1^{-}\to2^{+}\to2^+\to2^+\to0^{+}$ cascade, with respective values of 0.33, -0.08, and 0.02.}
\label{2+ass}
\end{center}
\end{figure}

As the beam energy increased, transitions from an increased number of low-lying states became visible. We show in Fig. \ref{hfhj} several of the observed transitions to and from low-lying 2$^{+}$ states for the 8.8~MeV photon beam. Observations of de-excitations from the $2^{+}_{3}$ state, or higher spin states such as the $3^{+}_1$, are additional confirmation that the $2^+_{1,2}$ states were populated by feeding and not directly excited from scattered $\gamma$-rays, as they have an almost negligible width to the ground state \cite{ndsSe}. Observation of higher spin states also confirm that the cascades become longer and more complex for resonantly excited states at higher energies. This is also indicated by the rather large asymmetry of the 2$^+_1$ to ground state transition at 6 MeV, suggesting only one intermediate transition, whereas the negative value at 7.35~MeV suggests two intermediate transitions. At higher energies, more complicated decay patterns may lead to near-isotropy.
 
\begin{figure}[h!]
\begin{center} 
\includegraphics[width=9cm]{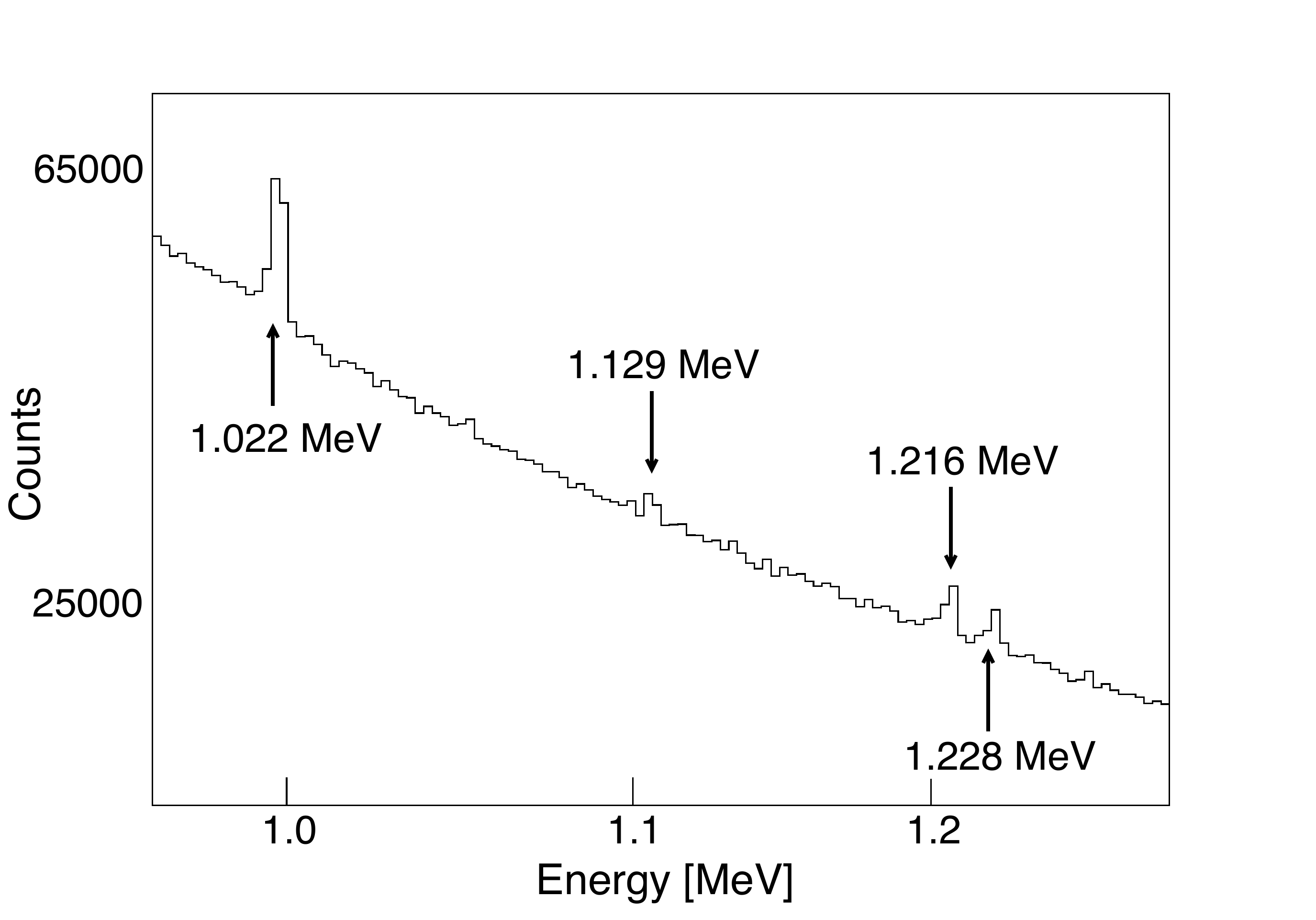} 
\caption{Some visible transitions from low lying states for the 8.8 MeV beam (horizontal and vertical detectors have been summed). The low-lying 2$^{+}$ states are significantly populated by feeding as the incident photon beam energy increases. The visible $\gamma$-ray transitions correspond to \cite{ndsSe}: $2^+_3\to2^+_1$ (1.228 MeV), $2^+_2\to0^+_g$ (1.216 MeV), and $3^+_1\to2^+_1$ (1.129 MeV). The electron-positron annihilation peak at 1.022 MeV is also visible.}
\label{hfhj}
\end{center}
\end{figure}

By considering the decays of the $2^+$ states, the total scattering cross section for all decays from the initially excited $J^\pi=1^-$ states to excited states at each beam energy can be deduced. The $W(\theta)$ for the $2^{+}$ to ground state transitions were chosen to correspond to the cascade that described best the observed asymmetry. As the angular distributions become more isotropic as more intermittent decays are introduced, it can be assumed that the systematic error introduced by the feeding assumptions above are insignificant compared to the uncertainties in the observed peak areas $A$. An average branching ratio for each beam energy can be deduced by considering the observed counts and cross sections for ground state transitions of the resonantly excited states, and the observed counts from the ground state transitions of the $2^{+}$ states \cite{Ton10}. With this branching ratio, an averaged differential cross section for the states within the FWHM of each beam energy window can be calculated. We comment that for higher beam energies, ground state transition peaks were not all well resolved against the background. With better statistics, it is likely that more ground state transitions would be observed, which would influence the deduced average branching ratio for the beam energy intervals.

In Fig. \ref{fig:av_cross}, we show the resulting differential cross section $\sigma(E)$ obtained from this work. The solid and dashed lines correspond to a generalised Lorentzian (GLO) and standard Lorentzian (SLO) fitted to data over 10 MeV as taken by the Saclay group \cite{Carl76}, and extrapolated to low energies. The merging into high energy data  suggests that we are correct to take into account the deduced contribution to the cross section from the branching decays. Without this, the total averaged cross sections would be grossly underestimated in the energy region above 6 MeV.

\begin{figure}[h!]
\begin{center} 
\includegraphics[width=9cm]{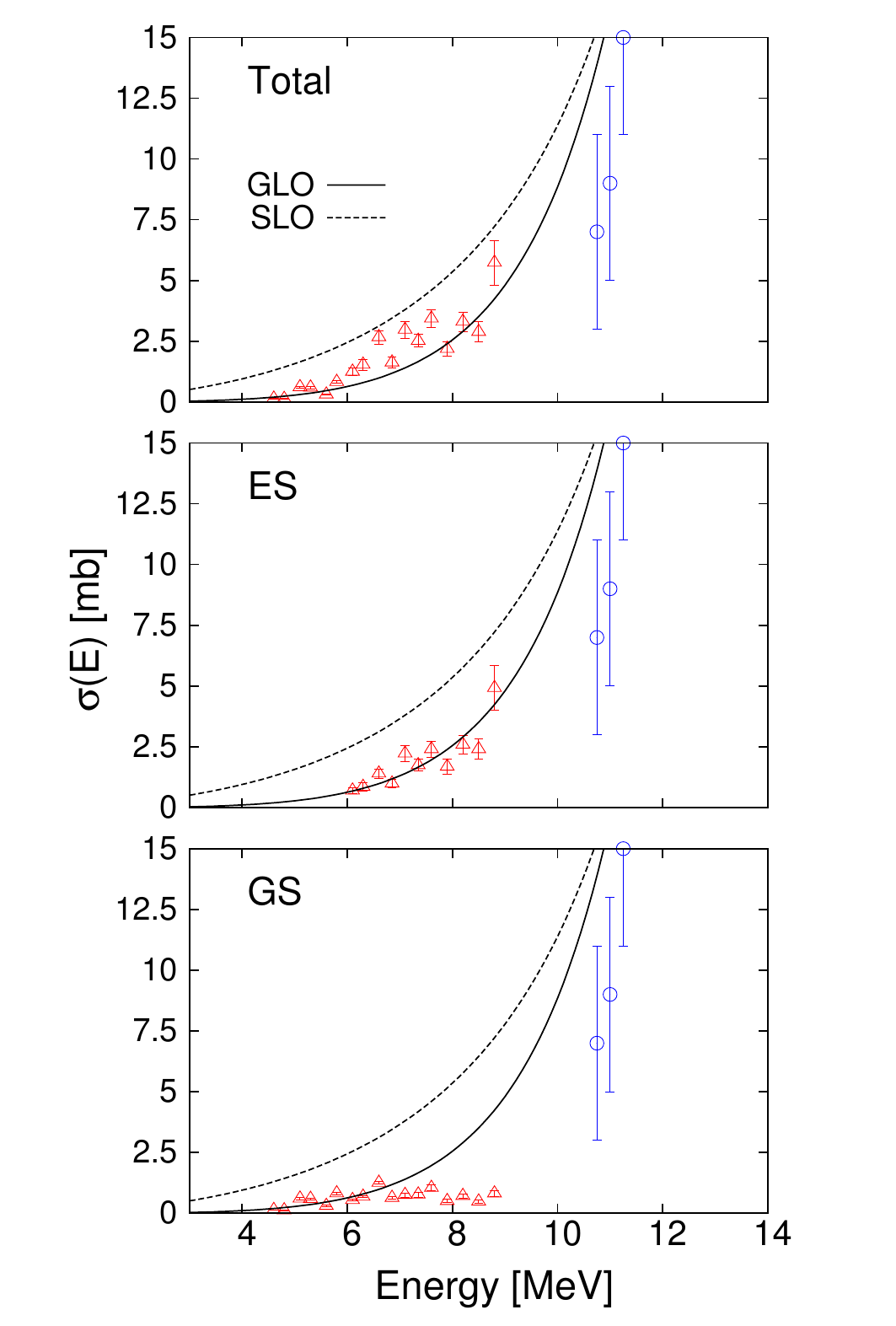} 
\caption{(color online). Averaged cross section for $E1$ excited states. The bottom panel shows the contribution from considering ground state decays only, the middle panel the contribution from excited state decays, and the top panel the total. Data above 10 MeV is $\sigma\left[(\gamma,\text{n})+(\gamma,\text{np}) +(\gamma,\text{2n})\right]$ (circles), taken from Ref. \cite{Carl76}. Results from this work are shown by triangles, and results from Ref. \cite{Carl76} by circles. The data over 10 MeV has been fitted to both a standard Lorentzian (dashed line) and generalized Lorentzian (solid line), and extrapolated to low energies. }
\label{fig:av_cross}
\end{center}
\end{figure}

Shown in Fig. \ref{fig:loren} is all available cross section data, with the same fits to the Saclay data as shown in Fig. \ref{fig:av_cross}. When assuming these fits, upon comparing to data from this work, when using the GLO form an enhancement upon the extrapolated low energy tail of the GDR can be seen. However, when fitting the Saclay data to the SLO form, the data from this work yields no enhancement upon the extrapolated GDR tail.

\begin{figure}[h!]
\begin{center} 
\includegraphics[width=9cm]{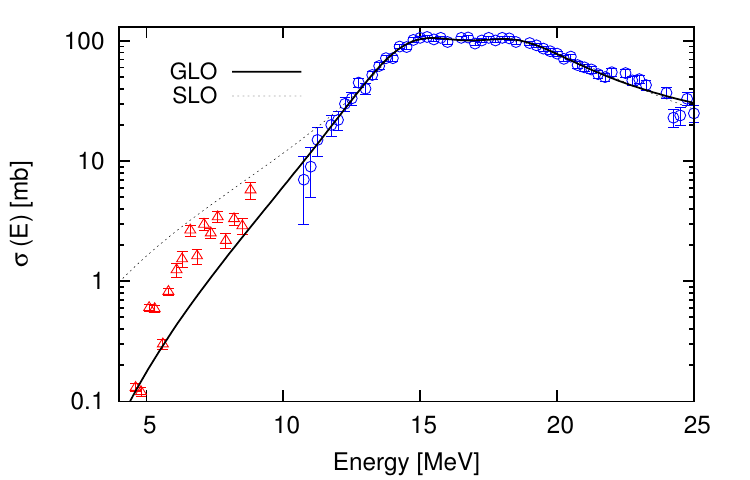} 
\caption{(color online). Deduced total cross sections from this work, and GDR cross section data from above the particle separation threshold \cite{Carl76}, triangles and circles respectively. The GDR data has been fitted to both a standard and generalized Lorentzian form. For the case of the GLO form, an enhancement is observable upon the extrapolated tail of the GDR.}
\label{fig:loren}
\end{center}
\end{figure}

\section{Electric Dipole Response in the Time-Dependent Hartree-Fock Framework}
\label{seven}
The random phase approximation (RPA) is a model often used for describing collective particle-hole excitations of the nucleus \cite{Goe82}. It is in fact the small-amplitude harmonic limit of time-dependent Hartree-Fock (TDHF) \cite{Rin80,Stev10,Simenel12}. An advantage of TDHF lies in the fact that direct access to the time evolution of the nuclear density is available, which allows for an insightful analysis of the dynamics of the system. 

TDHF and RPA are suitable for describing $1p-1h$ excitations and, consequently, they are often employed to analyse Giant Resonances. TDHF describes correlations beyond Hartree-Fock, however it is limited as it cannot describe $p-h$ correlations beyond the lowest order. Therefore, the widths of giant resonances are often underestimated for nuclei where a significant contribution comes from higher order correlations. Several models go beyond the mean-field approximation to include these correlations, such as the quasiparticle phonon model \cite{Sol92}, extended RPA \cite{Lac04}, and extended TDHF \cite{Lac99}. These approaches include coherent (coupling to $p-h$ and phonon) and incoherent (coupling to $2p-2h$) dissipation mechanisms. Other methods such as stochastic TDHF \cite{Rein92,Lac13} go beyond the mean field approximation by considering the time evolution of an ensemble of Slater determinants.

The treatment of pairing is handled in our framework via the BCS approximation \cite{Rin80}. We employ it to calculate the ground-state occupation probabilities, and then freeze these occupations for time-dependent calculations. Time-dependent Hartree-Fock-Bogoliubov \cite{Ave08,Ste11} fully treats pairing in the TDHF framework, but requires severe truncations of the quasiparticle basis to make calculations feasible in nuclei away from shell closures. Other implementations of time-dependent pairing simplify the problem by considering TDHF with time-dependent BCS pairing \cite{Eba10}.

To perform TDHF simulations, the starting point is finding the ground state Slater determinant whose single-particle (s.p.) wavefunctions $\varphi_\alpha$ are given by the solution of the static Hartree-Fock equation \cite{Gro91}:
\begin{equation}
\hat{h}_{HF}\varphi_\alpha = \epsilon_\alpha \varphi_\alpha \,\,\,\, ,
\label{eq:hf}
\end{equation}
where $\hat{h}_{HF}$ is the sum of the kinetic and potential energy operators $\hat{t} + \hat{v}$, and $\alpha$ represents the quantum numbers of the s.p. wavefunctions. The Hartree-Fock formalism is equivalent to the Kohn-Sham equations in density functional theory \cite{Koh65}. The Skyrme-Hartree-Fock (SHF) \cite{Vau72} approximation is often employed in nuclear structure calculations as it allows the inter-particle potential $\hat{v}$ to be expressed in terms of a zero-ranged, two-body, density-dependent interaction \cite{Sky59}. In this framework, the total energy functional of the system can be expressed solely in terms of the particle, kinetic and spin-orbit densities.

Once the ground state wavefunctions have been determined, they are evolved via the TDHF equation \cite{Eng74,Dir30,Rin80}:
\begin{equation}
i\hbar \frac{\partial\rho}{\partial t} = \left[\hat{h}_{HF}[\rho]\,,\,\rho\right] \,\,\,\,\, ,
\label{eq:tdhf}
\end{equation}
where $\rho$ is the density matrix, defined by
\begin{equation}
\rho = \sum_{i,j=1}^{N} |\varphi_i\rangle\langle\varphi_j| \,\,\,\,\, .
\end{equation}
When studying the total electric dipole response of a nucleus, the $E1$ dipole operator \cite{Har01}
\begin{equation}
\mathbf{\hat{D}} = \frac{N}{A}\sum_{p=1}^{Z} \mathbf{r}_p - \frac{Z}{A}\sum_{n=1}^N \mathbf{r}_n \,\,\,\,\, ,
\end{equation}
is applied instantaneously to the s.p. wavefunctions to excite $E1$ modes via the boost $\text{e}^{i{\mathbf{k}}\cdot {\hat{\mathbf D}}}$ , where $\mathbf k$ (in units of fm$^{-1}$) is a vector quantity which scales the initial spatial separation of the protons and neutrons \cite{Mar05}. All of our time-dependent calculations used $|\mathbf k| =$ 0.01~fm$^{-1}$ in the in the $x,y$, and $z$ directions. After the boost is applied at $t=0$, time evolution then begins and the expectation value of $\mathbf{\hat{D}}$ is tracked. 

The strength function associated with $\mathbf{\hat{D}}(t)$ for the case of an instantaneous boost is given by \cite{Rei97,Rei06,Mar05}
\begin{equation}
\mathbf{S_D}(\omega) = -\frac{1}{\pi \hbar k} \,\, \text{Im} \,\, \mathbf{{\tilde{D}}(\omega)}
\end{equation}
where $\mathbf{\tilde{D}}(\omega)$ is the Fourier transform of the expectation value of $\mathbf{\hat{D}}(t)$. From $\mathbf{S_D}(\omega)$ the photon scattering cross section $\sigma(E)$ can be calculated \cite{Boh75}:
\begin{equation}
\label{strength_calc}
\sigma(E) = \frac{4\pi^2e^2E}{\hbar c} \sum_{x,y,z}\mathbf{S_D}(\omega) \,\,\,\,\, .
\end{equation}

To perform the static SHF calculations, a cubic cartesian box with side length spanning from -11.5 to 11.5 fm was defined, with a grid spacing of 1 fm. 65 neutron and 65 proton wavefunctions were considered to allow spreading of the BCS occupations across the valence s.p. levels. Three Skyrme parameterization were used: NRAPR, fitted to an equation of state of nuclear matter \cite{Ste05}, SkI4, fitted to experimental ground state binding energy data \cite{Flo95}, and SLy4, fitted to data for supernovae and neutron-rich nuclei \cite{Chab95}. These various parameterizations were chosen to demonstrate the dependence of the calculated structure of the dipole response upon the chosen Skyrme parameterization. 

The Skyrme parameterization NRAPR was chosen, in addition to the well known forces SkI4 and SLy4, as it has passed a stringent set of tests for nuclear matter calculations \cite{Dut12}. There are well known correlations between nuclear matter properties and Giant Resonances \cite{Roc13}. However, the spin-orbit parameter $b_4$ has been doubled and the $t_1-t_2$ dependence of the spin-orbit term eliminated to ensure it reproduces shell closures in doubly magic finite nuclei \cite{God13}. These modifications should not affect the nuclear matter properties of the force. 

SHF calculations yielded binding energies per nucleon deviating no more than 3\% from the experimental value \cite{Aud03}. The calculated r.m.s. radius $\sqrt{\langle r^2_{tot}\rangle}$ was  4.080~fm for NRAPR, 4.075 fm for SLy4, and 4.043 fm for SkI4. The quadrupole deformation $\beta_2$ is defined in our framework as \cite{Mar05} 
\begin{equation}
\beta_2 = \frac{4\pi}{5A\langle r_{tot}^2\rangle}\sqrt{\left ( Q_{20}\right)^2 +2\left( Q_{22}\right)^2 }
\end{equation}
where $Q_{lm}$ is the spherical quadrupole moment of the nucleus. The deformation parameter $\gamma$, which is a measure of triaxiality, is given by
\begin{equation}
\gamma = \arctan \left(\frac{\sqrt 2 Q_{22}} {Q_{20}}\right) \,\,\, .
\end{equation}
The deformation parameter $\beta_2$ differs from the definition of the deformation parameter $\beta$, often assumed as the quadrupole deformation parameter. Ref. \cite{Ram01}  quotes a value of $\beta= 0.309(4)$ for $^{76}$Se, defined in a model-dependent way from experimentally observed $B(E2)\uparrow$ values. NRAPR yielded a $(\beta_2,\gamma)$ deformation of (0.089,60), SLy4 (0.032,0) and SkI4 (0.061,0). NRAPR reveals an oblate shape for the ground state of the nucleus, whereas SkI4 and SLy4 calculate a prolate shape.

\begin{figure}[h!]
\begin{center} 
\includegraphics[width=9cm]{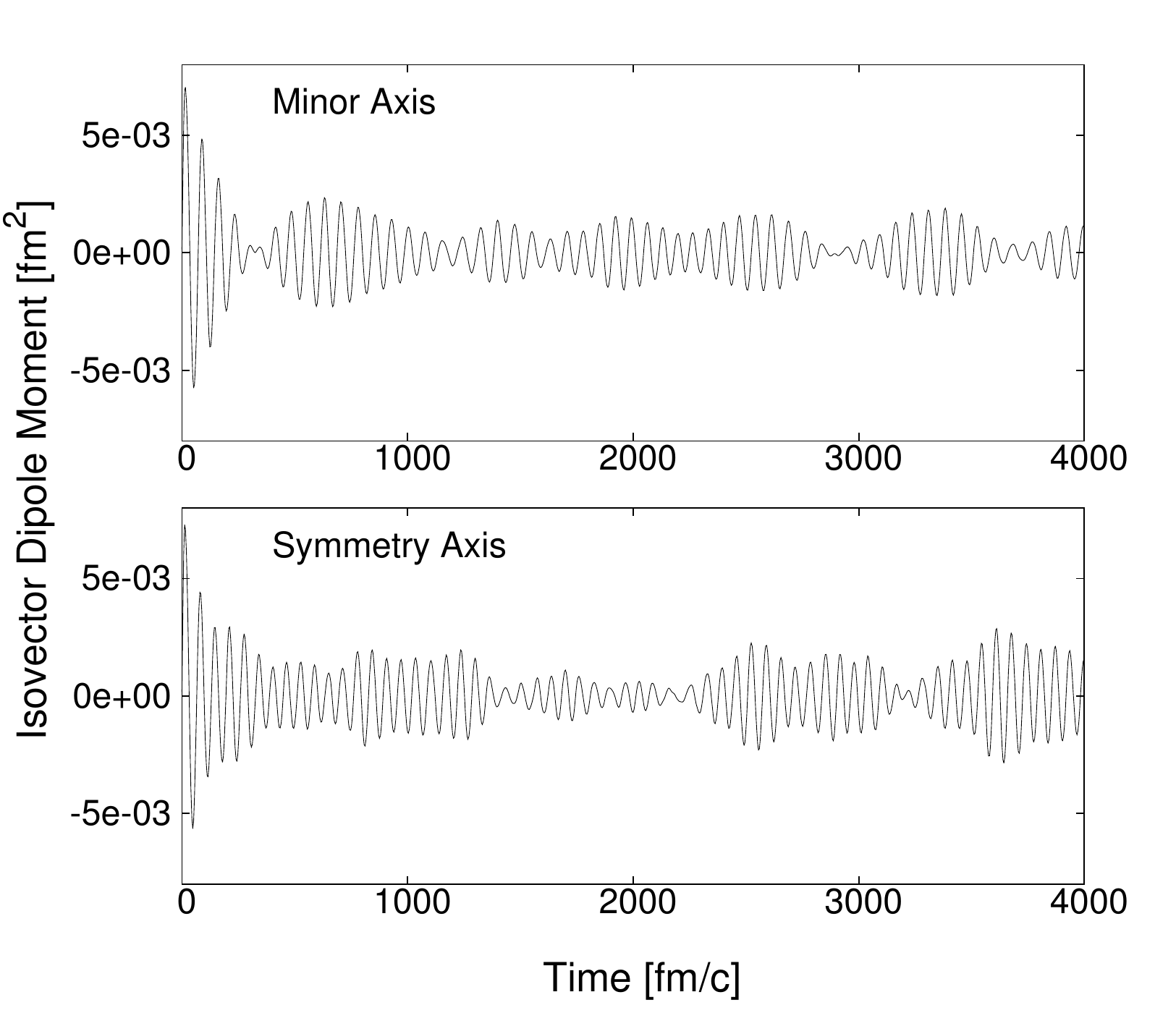} 
\caption{Time evolution of the expectation value of $\mathbf{\hat{D}}$ in the symmetry and minor axis. The Skyrme interaction used was NRAPR.}
\label{fig:D_ev}
\end{center}
\end{figure}

Shown in Fig. \ref{fig:D_ev} is the time evolution of the expectation value of $\mathbf{\hat{D}}$ after the initial boost at $t=0$ was applied, using the Skyrme interaction NRAPR. The Fourier transform into the frequency domain is shown in Fig. \ref{fig:tdhf_resp} for each of the considered Skyrme parameterizations. Due to the finite time $T$ available to run the calculations, the resolution is limited to $\delta E = 2\pi\hbar /T$ in the energy domain. Additionally, the use of reflecting boundary conditions leads to artificial discretization of the strength function \cite{Rei06}. Therefore, the strength function has been smoothed by folding it with a Gaussian of width 500 keV. The effect upon the response function when using absorbing boundaries is discussed in Ref. \cite{Par13}.

\begin{figure}[h!]
\begin{center} 
\includegraphics[width=9cm]{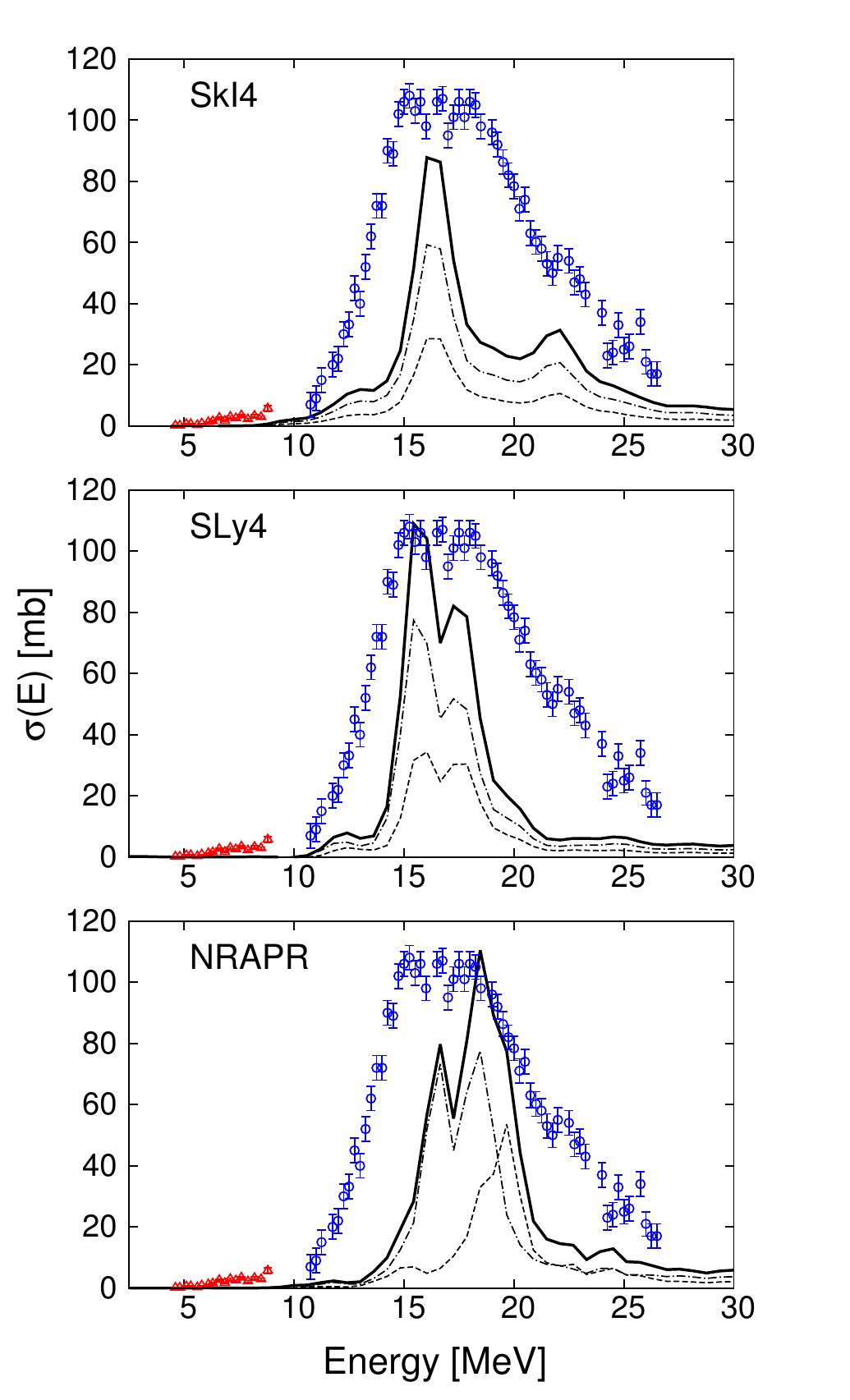} 
\caption{(color online). Photon scattering cross section derived from the $E1$ strength function for various Skyrme forces. The solid line show the total, dot-dash the contribution from minor axis,  and dashes the contribution from the symmetry axis. Circles show GDR data \cite{Carl76}, and triangles the low lying response from this work. In all cases, the TDHF calculations show low-lying enhancement to be in the region of 10 to 13 MeV, in contrast to the region of 6 to 8 MeV inferred by experiment. Low lying enhancements are obvious for SkI4 and SLy4, and although much less apparent for NRAPR, still present. }
\label{fig:tdhf_resp}
\end{center}
\end{figure}

All of the calculated response functions in Fig. \ref{fig:tdhf_resp}  underestimate the total cross section inferred from Ref. \cite{Carl76}. SkI4 reproduces the width, but has a lot of missing strength, whereas SLy4 and NRAPR reproduce the central strength around 15 to 20 MeV with a degree of success, but the calculated widths are far too narrow. We comment that the use of time-dependent pairing may widen the response function and shift the peak positions \cite{Ste11,Scam13}. Additionally, inclusion of beyond mean-field effects may also significantly alter the calculated response \cite{Lac01}. Ref. \cite{Gen13} demonstrates that these effects are required to accurately reproduce experimentally observed dipole responses.

Our calculations yield, for all Skyrme interactions considered, an enhancement upon the low energy tail of the GDR between 10 to 13 MeV. The calculated enhancements lie several MeV above the typical energy range of 5 to 8 MeV of the PDR. This suggests that the PDR may have significant contributions from higher order correlations in addition to its $1p-1h$ part, which our calculations cannot consider.

To investigate the energy dependence of the dipole response within the TDHF framework, we apply an external driving field to the system. With this mechanism, we can force the nucleus to vibrate at fixed frequencies. The TDHF equation is written in this case as:
\begin{equation}
i\hbar \frac{\partial\rho}{\partial t} = \left[\left(\hat{h}_{HF}[\rho] + f(t)\right)\,,\,\rho\right] \,\,\,\,\, ,
\end{equation}
where $f(t)$ is the external driving field. The external field applied is of the form: 
\begin{equation}
f(t) = \cos(\omega(t-\tau_0))\times\cos^2\left(\frac{\pi}{2} \frac{(t-\tau_0)}{\tau_t}\right) \,\,\,\,\, ,
\end{equation}
where $\omega$ is the driving frequency of the system, $\tau_0$ the time at which the external pulse is a maximum, and $\tau_t$ the width of the pulse. This allows us to study the behaviour of the densities and currents of protons and neutrons at different excitation frequencies. For each time step in the TDHF framework, we firstly consider the time derivative of the proton and neutron densities. This quantity can be approximated by $\rho_q(t_n) - \rho_q(t_{n-1})$, where $q$ denotes proton or neutron, and $t_n$ the discrete time step. This gives us direct access to the dynamics of the proton and neutron densities, in contrast to RPA based approaches which have to consider transition densities in order to analyze the dynamics of the nucleus \cite{Tso08}.

\begin{figure}[h!]
\begin{center} 
\includegraphics[width=8.5cm]{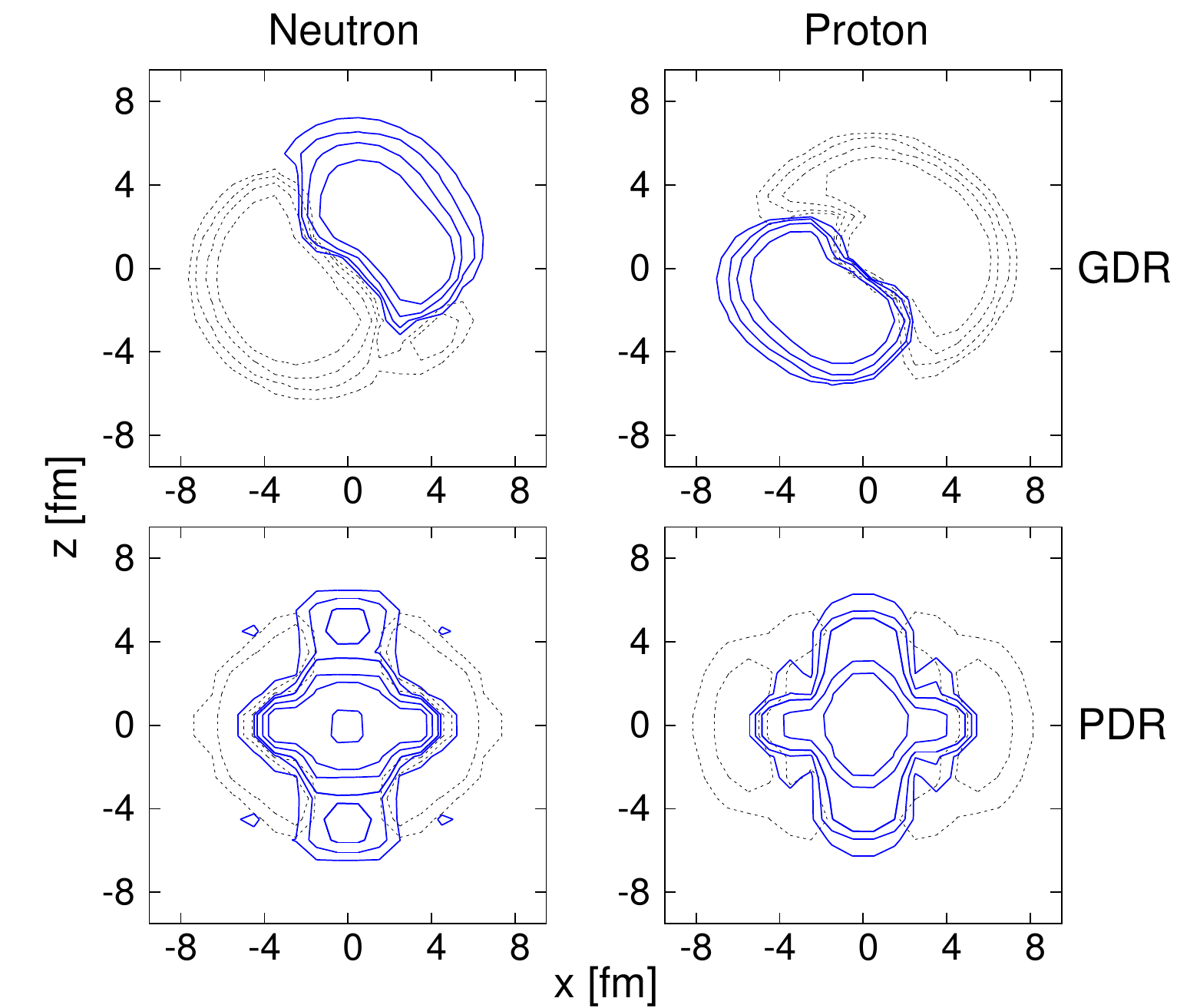} 
\caption{(color online). Slices of the density time derivative, typical of the GDR and PDR. These evolve in time (the behaviour is oscillatory), so we present a snapshot. Shown in Fig. \ref{fig:dens_mag} are the scale of the contours. Positive density is denoted with solid lines, negative with dashed lines.}
\label{dens_slic}
\end{center}
\end{figure}
\begin{figure}[h!]
\begin{center} 
\includegraphics[width=9cm]{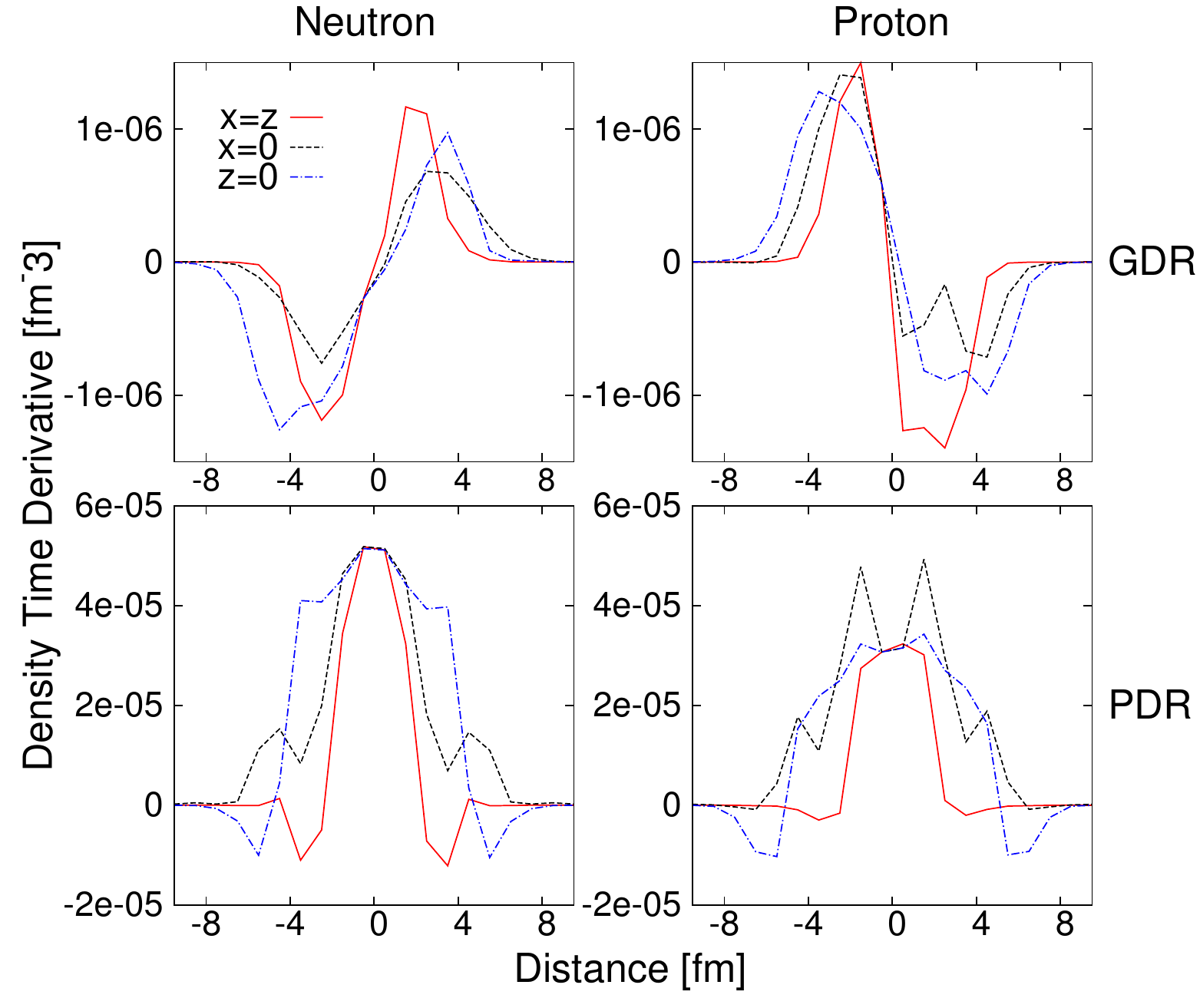} 
\caption{(color online). Snapshot in time of the density time derivative for GDR and PDR, taken from various slices in the $x-z$ plane.}
\label{fig:dens_mag}
\end{center}
\end{figure}

Shown in Fig. \ref{dens_slic} is a snapshot of the time derivative of the proton and neutron densities in the $x-z$ plane for driven frequencies chosen to correspond to the GDR and PDR. For the relative magnitude of the contours shown, we refer to Fig. \ref{fig:dens_mag}, which shows the density time derivative for a selection of slices in the $x-z$ plane. The differences in the vibrational mode are apparent.

When driving at GDR frequencies, the classical hydrodynamic picture described by Goldhaber and Teller is obvious, with the proton and neutron densities oscillating as collective bodies out of phase with one another \cite{Gol48}.  When the system is driven at frequencies corresponding to the low lying enhancement on the GDR tail, a different vibrational mode is apparent. The slice along the line $x=z$ again agrees with the interpretation of the PDR in a neutron rich, spherical nucleus as a proton-neutron core vibrating against a neutron skin. Taking slices over different lines, however, shows more complex behaviour.

These observations are in line with those of Ref. \cite{Tso08}, which investigates (quasiparticle) RPA derived transition densities across the Sn isotopic chain. The reference demonstrates that the classical picture of the PDR is only observed in neutron abundant isotopes, and the behaviour as the neutron excess diminishes deviates from the intuitive picture of a proton-neutron core vibrating out of phase with a neutron skin.

The particle current vectors are also instructive in describing the dynamics of collective resonances of the nucleus.  $\mathbf{j}(\mathbf{r})_q$ is defined by
\begin{equation}
\mathbf{j}_q(\mathbf{r}) = -\frac{i}{2} (\mathbf{\nabla-\nabla'})\rho_q(\mathbf{r,r'})\,\, |_{\mathbf{r'=r}} \,\,\,\,\, ,
\end{equation}
where $q$ denotes protons or neutrons. The current vectors are shown in Figs. \ref{fig:15mev} and \ref{fig:12mev} for driven frequencies chosen to excite the GDR and PDR, respectively.

\begin{figure}[h!]
\begin{center} 
\includegraphics[width=9cm]{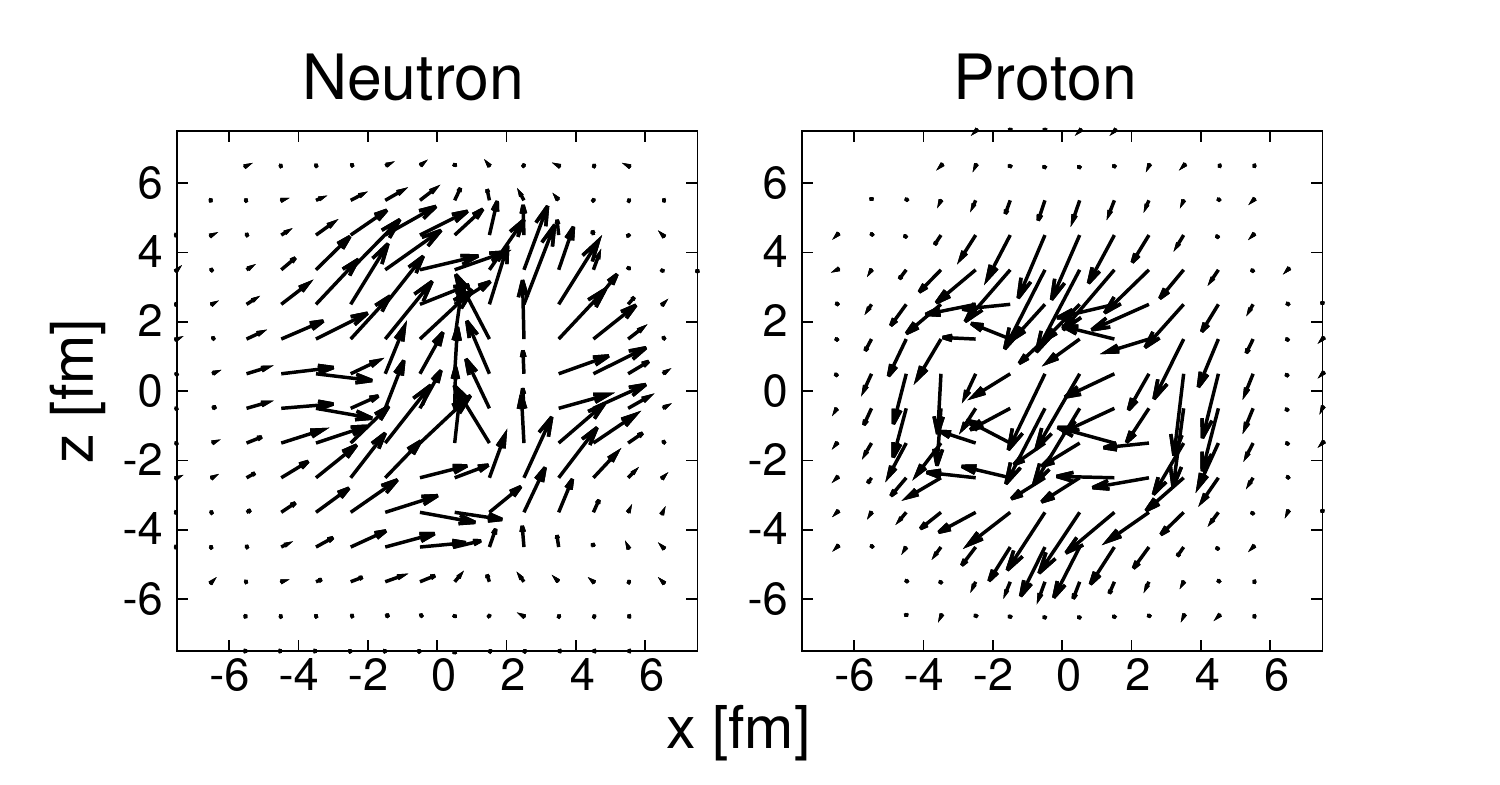} 
\caption{Snapshot in time of normalized proton and neutron current vectors for a frequency corresponding to the GDR. The vectors have been normalised to a length of 2 fm.}
\label{fig:15mev}
\end{center}
\end{figure}
\begin{figure}[h!]
\begin{center} 
\includegraphics[width=9cm]{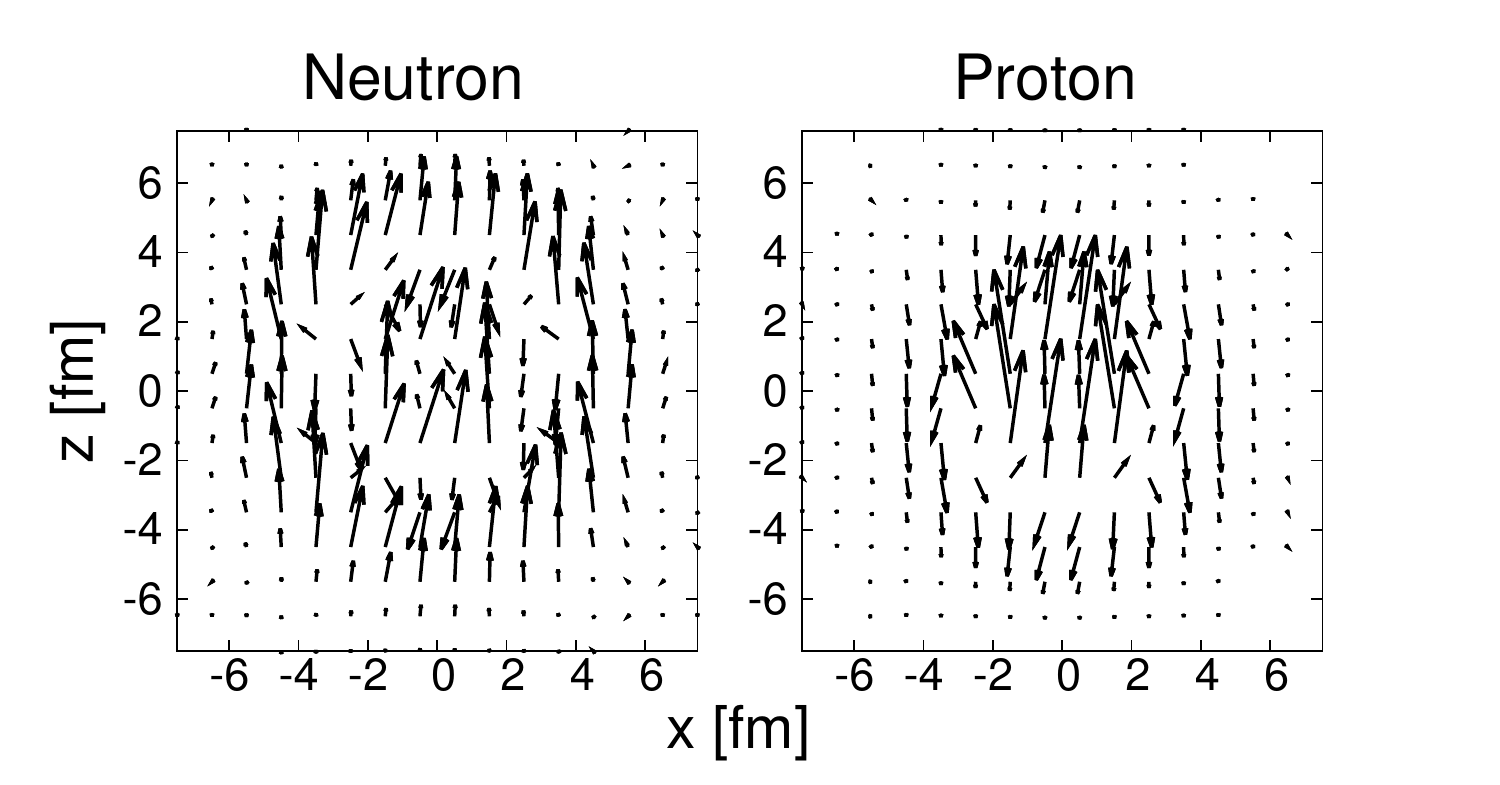} 
\caption{Snapshot in time of normalized current vectors for the nucleus driven at a frequency corresponding to the pygmy enhancement upon the low energy tail of the GDR.}
\label{fig:12mev}
\end{center}
\end{figure}

For the GDR, the Goldhaber-Teller envision is demonstrated once again, showing the flow of protons to be opposite in direction to that of the neutrons. 

For the PDR, a significantly different behaviour can be seen. A core-skin type behaviour is apparent from the vectors, with the neutron currents contributing significantly to a skin and core region. However, the currents do not display a strongly collective core oscillating out of phase with the skin; the currents in the core and skin regions seem largely in phase with one another. The significantly contributing proton currents seem centralized in the core of the nucleus, in phase with the neutron core currents. However, the surrounding current vectors point in different directions to those in the core. These observations imply that the simple picture of the PDR as a proton-neutron core vibrating out of phase with a neutron skin may not be fully valid in the case of a nucleus without an extreme excess of neutrons.

As an alternate approach to isolate the pygmy mode from the GDR, we consider modifying the dipole operator to give the SHF ground state an instantaneous boost exciting only valence skin orbitals against the bulk of the orbitals making up the core. This will allow us to test whether this is a good description of the PDR, and to which extent it coincides with the usual core-skin picture. This method of splitting the dipole operator into a sum of a core and skin part has been used previously within a harmonic oscillator shell model to analyse the collective nature of the PDR \cite{Bar12}. The approach of using exotic dipole operators can also be employed to examine different aspects of the dipole response, for example toroidal dipole modes \cite{Urb12}. 

We define an exotic dipole boost operator by
\begin{equation}
\hat{\textbf{D}}^\prime = \frac{A-\sum v^2}{A}\sum_{i=1}^{\eta}\textbf{r}_{skin} - \frac{\sum v^2}{A}\sum_{j=1}^{\zeta}\textbf{r}_{core}\,\,\,\,\, ,
\label{skin_core}
\end{equation}
where $\eta$ is the number of skin orbitals, $\zeta$ the number of core orbitals, and $\sum v^2$ corresponds to the summed BCS occupations of the skin orbitals in the calculated ground state. The expectation value of $\hat{\textbf{D}}^\prime$ can be tracked as before, and it could be expected that even with the instantaneous boost, only the core-skin vibrational mode will be excited. We test the classical picture of the PDR as a neutron skin vibrating against a core by exciting the valence 1$g_{9/2}$ neutron orbital against all other protons and neutrons. 

\begin{figure}[h!]
\begin{center} 
\includegraphics[width=9cm]{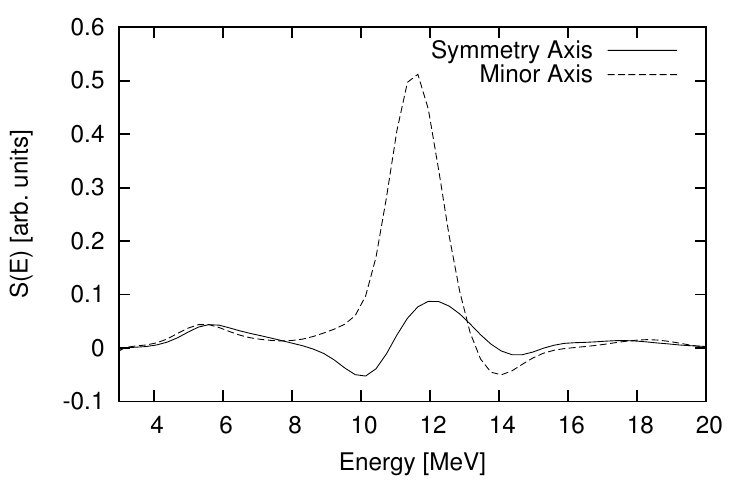} 
\caption{Strength function from exotic dipole boost. The Skyrme interaction was NRAPR. The peak at 12 MeV corresponds to the low lying enhancement upon the tail of the GDR in Fig. \ref{fig:tdhf_resp}.}
\label{fig:cors}
\end{center}
\end{figure}

Shown in Fig. \ref{fig:cors} is the strength function for the operator $\hat{\textbf{D}}^\prime$. In comparison to Fig. \ref{fig:tdhf_resp}, the peak corresponds to the region that was attributed to the PDR upon the tail of the GDR. No significant strength at energies higher than this have been excited, despite the fact that the boost is applied over all frequencies. Due to the fact that the cross section formula in Eq. \ref{strength_calc} is derived from linear response theory using the $E1$ operator $\mathbf{\hat{D}}$ \cite{Rin80}, the response function for this exotic boost operator does not necessarily correspond directly to $\sigma(E)$. Here, the interest lies mostly in the fact that the only observed response is in the region attributed to the PDR. %

\begin{figure}[h!]
\begin{center} 
\includegraphics[width=9cm]{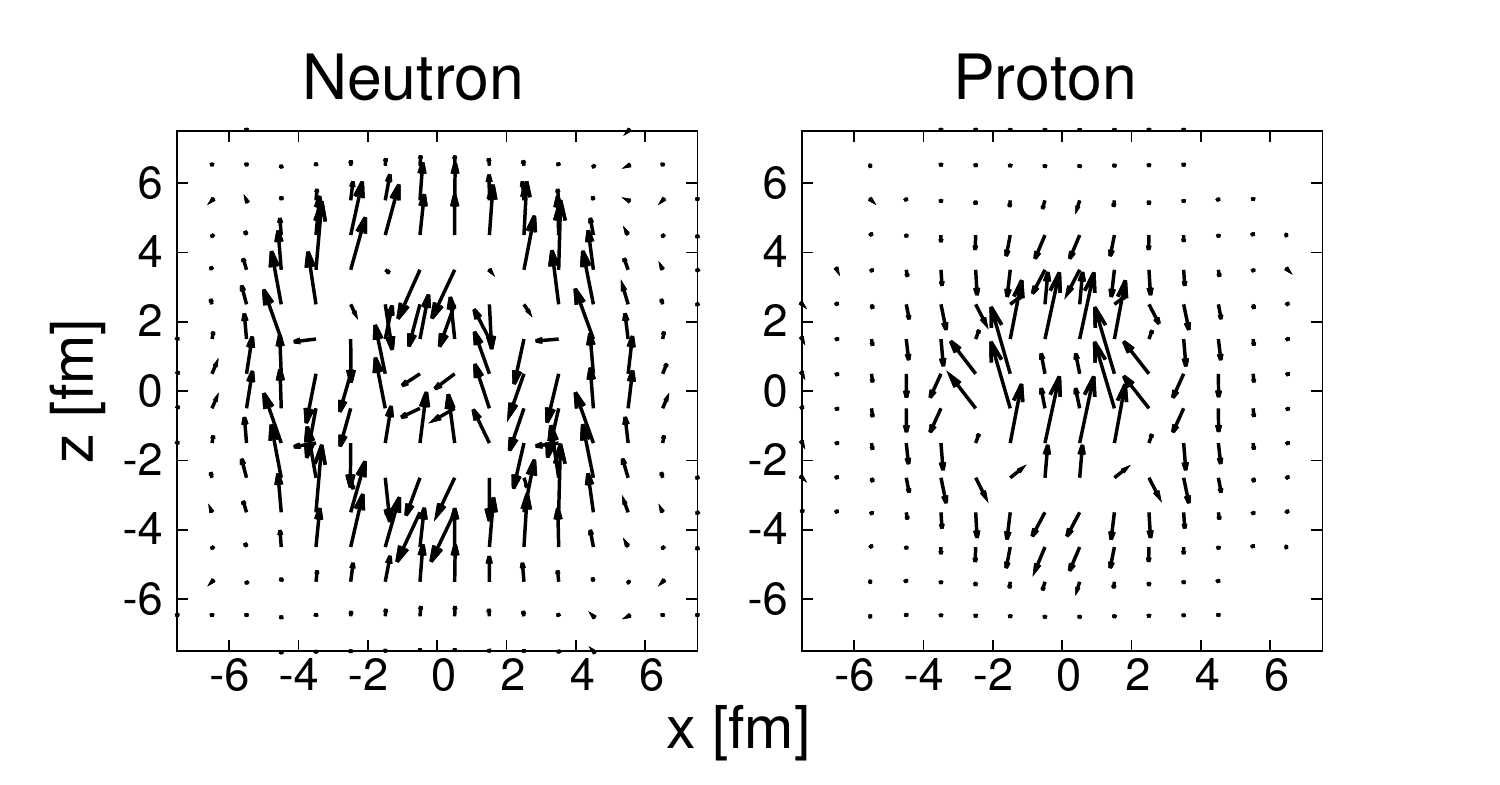} 
\caption{Snapshot in time of the current vectors when using the instantaneous boost using $\hat{\textbf{D}}^\prime$. The 1$g_{9/2}$ neutron orbitals were excited as the skin against the core composed of all the other single particle wavefunctions. }
\label{fig:12curr}
\end{center}
\end{figure}

We can be confident that the response function corresponds to the PDR by analyzing the corresponding current vectors. Fig. \ref{fig:12curr} shows the vectors for a time snapshot when the initial excitation is provided by $\hat{\textbf{D}}^\prime$. The figure displays the same behaviour as seen in Fig. \ref{fig:12mev}, where the nucleus was driven at a fixed excitation frequency corresponding to the PDR. Since the initial boost only consisted of separating the valence neutron orbitals from all others, the fact that the proton vectors also exhibit the same behaviour as seen in Fig. \ref{fig:12curr} shows that after the initial perturbation the nucleus couples to the same pygmy dipole vibrational mode which we observe when exciting the nucleus at a fixed frequency.

This clearly demonstrates the potential of using exotic dipole operators to effectively isolate different parts of the collective isovector dipole response of nuclei. Further theoretical research along this direction is being developed at the moment.

\section{Conclusion}
\label{eight}
Using the high resolution NRF technique, the dipole response of $^{76}$Se in the energy range 4 to 9 MeV has been analyzed through a series of experiments at the HI$\gamma$S and DHIPS facilities. 

The directly observed summed dipole excitation strengths attributed to individual states in the region are 0.132(29)~e$^{2}$fm$^{2}$, and 0.641(109)~$\mu_{N}^{2}$, for electric and magnetic spin $J=1$ states, respectively. The measured excitation strengths above 6 MeV underestimate the actual $E1$ strength due to fragmentation effects which cannot be individually accounted for.
Considering the transitions from low lying 2$^+$ states to the ground state, we have presented evidence that due to branching decays from resonantly excited states, the total scattering cross section deduced from ground state decays only is a significant underestimation of the total.

Once branching decays are accounted for, the deduced total photon scattering cross section seems to connect to previous data at higher energies. Factoring in these branching decays allows an enhancement, which may be attributed to a pygmy resonance, to be seen when fitting previous cross section data to a generalized Lorentzian form. However, no such enhancement is observed when the GDR tail is extrapolated using a standard Lorentzian form.

A 3D TDHF framework has been employed to describe the collective $E1$ vibrational modes of $^{76}$Se. A consistent underestimation of the GDR width and strength, and the location of the PDR enhancement, suggest that further considerations, such as accounting for beyond mean-field and time-dependent pairing effects, are required to fully describe experimental results. We demonstrate a novel technique to isolate the PDR from the GDR within our calculations without having to force the excitation energy of the nucleus.

\section{Acknowledgments}

This work was supported by US DOE grant numbers DE-FG02-91ER-40609 (Yale), DE-FG02-97ER41042 (NCU), DE-FG02-97ER41033 (Duke), US National Science Foundation grant number PHY-0956310 (Kentucky), and UK STFC grant numbers ST/J500768/1, ST/J00051/1 and ST/I005528/1 (Surrey). It was further supported by the Helmholtz Alliance Program of the Helmholtz Association (HA216/EMMI).
The authors are grateful to the operators at the HI$\gamma$S facility for the outstanding quality of the photon beams provided. P. Goddard gratefully acknowledges helpful advice and input from T.~Ahn. 

\bibliography{76se_ref}

\begin{thebibliography}{92}%
\makeatletter
\providecommand \@ifxundefined [1]{%
 \@ifx{#1\undefined}
}%
\providecommand \@ifnum [1]{%
 \ifnum #1\expandafter \@firstoftwo
 \else \expandafter \@secondoftwo
 \fi
}%
\providecommand \@ifx [1]{%
 \ifx #1\expandafter \@firstoftwo
 \else \expandafter \@secondoftwo
 \fi
}%
\providecommand \natexlab [1]{#1}%
\providecommand \enquote  [1]{``#1''}%
\providecommand \bibnamefont  [1]{#1}%
\providecommand \bibfnamefont [1]{#1}%
\providecommand \citenamefont [1]{#1}%
\providecommand \href@noop [0]{\@secondoftwo}%
\providecommand \href [0]{\begingroup \@sanitize@url \@href}%
\providecommand \@href[1]{\@@startlink{#1}\@@href}%
\providecommand \@@href[1]{\endgroup#1\@@endlink}%
\providecommand \@sanitize@url [0]{\catcode `\\12\catcode `\$12\catcode
  `\&12\catcode `\#12\catcode `\^12\catcode `\_12\catcode `\%12\relax}%
\providecommand \@@startlink[1]{}%
\providecommand \@@endlink[0]{}%
\providecommand \url  [0]{\begingroup\@sanitize@url \@url }%
\providecommand \@url [1]{\endgroup\@href {#1}{\urlprefix }}%
\providecommand \urlprefix  [0]{URL }%
\providecommand \Eprint [0]{\href }%
\providecommand \doibase [0]{http://dx.doi.org/}%
\providecommand \selectlanguage [0]{\@gobble}%
\providecommand \bibinfo  [0]{\@secondoftwo}%
\providecommand \bibfield  [0]{\@secondoftwo}%
\providecommand \translation [1]{[#1]}%
\providecommand \BibitemOpen [0]{}%
\providecommand \bibitemStop [0]{}%
\providecommand \bibitemNoStop [0]{.\EOS\space}%
\providecommand \EOS [0]{\spacefactor3000\relax}%
\providecommand \BibitemShut  [1]{\csname bibitem#1\endcsname}%
\let\auto@bib@innerbib\@empty
\bibitem [{\citenamefont {Kneissl}\ \emph {et~al.}(1996)\citenamefont
  {Kneissl}, \citenamefont {Pitz},\ and\ \citenamefont {Zilges}}]{Kne96}%
  \BibitemOpen
  \bibfield  {author} {\bibinfo {author} {\bibfnamefont {U.}~\bibnamefont
  {Kneissl}}, \bibinfo {author} {\bibfnamefont {H.}~\bibnamefont {Pitz}}, \
  and\ \bibinfo {author} {\bibfnamefont {A.}~\bibnamefont {Zilges}},\ }\href
  {\doibase 10.1016/0146-6410(96)00055-5} {\bibfield  {journal} {\bibinfo
  {journal} {Prog. Part. Nucl. Phys.}\ }\textbf {\bibinfo {volume} {37}},\
  \bibinfo {pages} {349 } (\bibinfo {year} {1996})}\BibitemShut {NoStop}%
\bibitem [{\citenamefont {Herzberg}\ \emph {et~al.}(1997)\citenamefont
  {Herzberg}, \citenamefont {von Brentano}, \citenamefont {Eberth},
  \citenamefont {Enders}, \citenamefont {Fischer}, \citenamefont {Huxel},
  \citenamefont {Klemme}, \citenamefont {von Neumann-Cosel}, \citenamefont
  {Nicolay}, \citenamefont {Pietralla}, \citenamefont {Ponomarev},
  \citenamefont {Reif}, \citenamefont {Richter}, \citenamefont {Schlegel},
  \citenamefont {Schwengner}, \citenamefont {Skoda}, \citenamefont {Thomas},
  \citenamefont {Wiedenhover}, \citenamefont {Winter},\ and\ \citenamefont
  {Zilges}}]{Herz97}%
  \BibitemOpen
  \bibfield  {author} {\bibinfo {author} {\bibfnamefont {R.-D.}\ \bibnamefont
  {Herzberg}}, \bibinfo {author} {\bibfnamefont {P.}~\bibnamefont {von
  Brentano}}, \bibinfo {author} {\bibfnamefont {J.}~\bibnamefont {Eberth}},
  \bibinfo {author} {\bibfnamefont {J.}~\bibnamefont {Enders}}, \bibinfo
  {author} {\bibfnamefont {R.}~\bibnamefont {Fischer}}, \bibinfo {author}
  {\bibfnamefont {N.}~\bibnamefont {Huxel}}, \bibinfo {author} {\bibfnamefont
  {T.}~\bibnamefont {Klemme}}, \bibinfo {author} {\bibfnamefont
  {P.}~\bibnamefont {von Neumann-Cosel}}, \bibinfo {author} {\bibfnamefont
  {N.}~\bibnamefont {Nicolay}}, \bibinfo {author} {\bibfnamefont
  {N.}~\bibnamefont {Pietralla}}, \bibinfo {author} {\bibfnamefont
  {V.}~\bibnamefont {Ponomarev}}, \bibinfo {author} {\bibfnamefont
  {J.}~\bibnamefont {Reif}}, \bibinfo {author} {\bibfnamefont {A.}~\bibnamefont
  {Richter}}, \bibinfo {author} {\bibfnamefont {C.}~\bibnamefont {Schlegel}},
  \bibinfo {author} {\bibfnamefont {R.}~\bibnamefont {Schwengner}}, \bibinfo
  {author} {\bibfnamefont {S.}~\bibnamefont {Skoda}}, \bibinfo {author}
  {\bibfnamefont {H.}~\bibnamefont {Thomas}}, \bibinfo {author} {\bibfnamefont
  {I.}~\bibnamefont {Wiedenhover}}, \bibinfo {author} {\bibfnamefont
  {G.}~\bibnamefont {Winter}}, \ and\ \bibinfo {author} {\bibfnamefont
  {A.}~\bibnamefont {Zilges}},\ }\href {\doibase 10.1016/S0370-2693(96)01374-3}
  {\bibfield  {journal} {\bibinfo  {journal} {Phys. Lett. B}\ }\textbf
  {\bibinfo {volume} {390}},\ \bibinfo {pages} {49} (\bibinfo {year}
  {1997})}\BibitemShut {NoStop}%
\bibitem [{\citenamefont {Herzberg}\ \emph {et~al.}(1999)\citenamefont
  {Herzberg}, \citenamefont {Fransen}, \citenamefont {von Brentano},
  \citenamefont {Eberth}, \citenamefont {Enders}, \citenamefont {Fitzler},
  \citenamefont {K\"aubler}, \citenamefont {Kaiser}, \citenamefont {von
  Neumann-Cosel}, \citenamefont {Pietralla}, \citenamefont {Ponomarev},
  \citenamefont {Prade}, \citenamefont {Richter}, \citenamefont {Schnare},
  \citenamefont {Schwengner}, \citenamefont {Skoda}, \citenamefont {Thomas},
  \citenamefont {Tiesler}, \citenamefont {Weisshaar},\ and\ \citenamefont
  {Wiedenh\"over}}]{Herz99}%
  \BibitemOpen
  \bibfield  {author} {\bibinfo {author} {\bibfnamefont {R.-D.}\ \bibnamefont
  {Herzberg}}, \bibinfo {author} {\bibfnamefont {C.}~\bibnamefont {Fransen}},
  \bibinfo {author} {\bibfnamefont {P.}~\bibnamefont {von Brentano}}, \bibinfo
  {author} {\bibfnamefont {J.}~\bibnamefont {Eberth}}, \bibinfo {author}
  {\bibfnamefont {J.}~\bibnamefont {Enders}}, \bibinfo {author} {\bibfnamefont
  {A.}~\bibnamefont {Fitzler}}, \bibinfo {author} {\bibfnamefont
  {L.}~\bibnamefont {K\"aubler}}, \bibinfo {author} {\bibfnamefont
  {H.}~\bibnamefont {Kaiser}}, \bibinfo {author} {\bibfnamefont
  {P.}~\bibnamefont {von Neumann-Cosel}}, \bibinfo {author} {\bibfnamefont
  {N.}~\bibnamefont {Pietralla}}, \bibinfo {author} {\bibfnamefont {V.~Y.}\
  \bibnamefont {Ponomarev}}, \bibinfo {author} {\bibfnamefont {H.}~\bibnamefont
  {Prade}}, \bibinfo {author} {\bibfnamefont {A.}~\bibnamefont {Richter}},
  \bibinfo {author} {\bibfnamefont {H.}~\bibnamefont {Schnare}}, \bibinfo
  {author} {\bibfnamefont {R.}~\bibnamefont {Schwengner}}, \bibinfo {author}
  {\bibfnamefont {S.}~\bibnamefont {Skoda}}, \bibinfo {author} {\bibfnamefont
  {H.~G.}\ \bibnamefont {Thomas}}, \bibinfo {author} {\bibfnamefont
  {H.}~\bibnamefont {Tiesler}}, \bibinfo {author} {\bibfnamefont
  {D.}~\bibnamefont {Weisshaar}}, \ and\ \bibinfo {author} {\bibfnamefont
  {I.}~\bibnamefont {Wiedenh\"over}},\ }\href {\doibase
  10.1103/PhysRevC.60.051307} {\bibfield  {journal} {\bibinfo  {journal} {Phys.
  Rev. C}\ }\textbf {\bibinfo {volume} {60}},\ \bibinfo {pages} {051307}
  (\bibinfo {year} {1999})}\BibitemShut {NoStop}%
\bibitem [{\citenamefont {Savran}\ \emph {et~al.}(2013)\citenamefont {Savran},
  \citenamefont {Aumann},\ and\ \citenamefont {Zilges}}]{Sav2013}%
  \BibitemOpen
  \bibfield  {author} {\bibinfo {author} {\bibfnamefont {D.}~\bibnamefont
  {Savran}}, \bibinfo {author} {\bibfnamefont {T.}~\bibnamefont {Aumann}}, \
  and\ \bibinfo {author} {\bibfnamefont {A.}~\bibnamefont {Zilges}},\ }\href
  {\doibase 10.1016/j.ppnp.2013.02.003} {\bibfield  {journal} {\bibinfo
  {journal} {Prog. Part. Nucl. Phys.}\ }\textbf {\bibinfo {volume} {70}},\
  \bibinfo {pages} {210 } (\bibinfo {year} {2013})}\BibitemShut {NoStop}%
\bibitem [{\citenamefont {Adrich}\ \emph {et~al.}(2005)\citenamefont {Adrich},
  \citenamefont {Klimkiewicz}, \citenamefont {Fallot}, \citenamefont
  {Boretzky}, \citenamefont {Aumann}, \citenamefont {Cortina-Gil},
  \citenamefont {Pramanik}, \citenamefont {Elze}, \citenamefont {Emling},
  \citenamefont {Geissel}, \citenamefont {Hellstr\"om}, \citenamefont {Jones},
  \citenamefont {Kratz}, \citenamefont {Kulessa}, \citenamefont {Leifels},
  \citenamefont {Nociforo}, \citenamefont {Palit}, \citenamefont {Simon},
  \citenamefont {Sur\'owka}, \citenamefont {S\"ummerer},\ and\ \citenamefont
  {Walu\ifmmode~\acute{s}\else \'{s}\fi{}}}]{Adr05}%
  \BibitemOpen
  \bibfield  {author} {\bibinfo {author} {\bibfnamefont {P.}~\bibnamefont
  {Adrich}}, \bibinfo {author} {\bibfnamefont {A.}~\bibnamefont {Klimkiewicz}},
  \bibinfo {author} {\bibfnamefont {M.}~\bibnamefont {Fallot}}, \bibinfo
  {author} {\bibfnamefont {K.}~\bibnamefont {Boretzky}}, \bibinfo {author}
  {\bibfnamefont {T.}~\bibnamefont {Aumann}}, \bibinfo {author} {\bibfnamefont
  {D.}~\bibnamefont {Cortina-Gil}}, \bibinfo {author} {\bibfnamefont {U.~D.}\
  \bibnamefont {Pramanik}}, \bibinfo {author} {\bibfnamefont {T.~W.}\
  \bibnamefont {Elze}}, \bibinfo {author} {\bibfnamefont {H.}~\bibnamefont
  {Emling}}, \bibinfo {author} {\bibfnamefont {H.}~\bibnamefont {Geissel}},
  \bibinfo {author} {\bibfnamefont {M.}~\bibnamefont {Hellstr\"om}}, \bibinfo
  {author} {\bibfnamefont {K.~L.}\ \bibnamefont {Jones}}, \bibinfo {author}
  {\bibfnamefont {J.~V.}\ \bibnamefont {Kratz}}, \bibinfo {author}
  {\bibfnamefont {R.}~\bibnamefont {Kulessa}}, \bibinfo {author} {\bibfnamefont
  {Y.}~\bibnamefont {Leifels}}, \bibinfo {author} {\bibfnamefont
  {C.}~\bibnamefont {Nociforo}}, \bibinfo {author} {\bibfnamefont
  {R.}~\bibnamefont {Palit}}, \bibinfo {author} {\bibfnamefont
  {H.}~\bibnamefont {Simon}}, \bibinfo {author} {\bibfnamefont
  {G.}~\bibnamefont {Sur\'owka}}, \bibinfo {author} {\bibfnamefont
  {K.}~\bibnamefont {S\"ummerer}}, \ and\ \bibinfo {author} {\bibfnamefont
  {W.}~\bibnamefont {Walu\ifmmode~\acute{s}\else \'{s}\fi{}}} (\bibinfo
  {collaboration} {LAND-FRS Collaboration}),\ }\href {\doibase
  10.1103/PhysRevLett.95.132501} {\bibfield  {journal} {\bibinfo  {journal}
  {Phys. Rev. Lett.}\ }\textbf {\bibinfo {volume} {95}},\ \bibinfo {pages}
  {132501} (\bibinfo {year} {2005})}\BibitemShut {NoStop}%
\bibitem [{\citenamefont {Volz}\ \emph {et~al.}(2006)\citenamefont {Volz},
  \citenamefont {Tsoneva}, \citenamefont {Babilon}, \citenamefont {Elvers},
  \citenamefont {Hasper}, \citenamefont {Herzberg}, \citenamefont {Lenske},
  \citenamefont {Lindenberg}, \citenamefont {Savran},\ and\ \citenamefont
  {Zilges}}]{Vol06}%
  \BibitemOpen
  \bibfield  {author} {\bibinfo {author} {\bibfnamefont {S.}~\bibnamefont
  {Volz}}, \bibinfo {author} {\bibfnamefont {N.}~\bibnamefont {Tsoneva}},
  \bibinfo {author} {\bibfnamefont {M.}~\bibnamefont {Babilon}}, \bibinfo
  {author} {\bibfnamefont {M.}~\bibnamefont {Elvers}}, \bibinfo {author}
  {\bibfnamefont {J.}~\bibnamefont {Hasper}}, \bibinfo {author} {\bibfnamefont
  {R.-D.}\ \bibnamefont {Herzberg}}, \bibinfo {author} {\bibfnamefont
  {H.}~\bibnamefont {Lenske}}, \bibinfo {author} {\bibfnamefont
  {K.}~\bibnamefont {Lindenberg}}, \bibinfo {author} {\bibfnamefont
  {D.}~\bibnamefont {Savran}}, \ and\ \bibinfo {author} {\bibfnamefont
  {A.}~\bibnamefont {Zilges}},\ }\href {\doibase
  10.1016/j.nuclphysa.2006.08.012} {\bibfield  {journal} {\bibinfo  {journal}
  {Nucl. Phys. A}\ }\textbf {\bibinfo {volume} {779}},\ \bibinfo {pages} {1 }
  (\bibinfo {year} {2006})}\BibitemShut {NoStop}%
\bibitem [{\citenamefont {Tsoneva}\ and\ \citenamefont {Lenske}(2008)}]{Tso08}%
  \BibitemOpen
  \bibfield  {author} {\bibinfo {author} {\bibfnamefont {N.}~\bibnamefont
  {Tsoneva}}\ and\ \bibinfo {author} {\bibfnamefont {H.}~\bibnamefont
  {Lenske}},\ }\href {\doibase 10.1103/PhysRevC.77.024321} {\bibfield
  {journal} {\bibinfo  {journal} {Phys. Rev. C}\ }\textbf {\bibinfo {volume}
  {77}},\ \bibinfo {pages} {024321} (\bibinfo {year} {2008})}\BibitemShut
  {NoStop}%
\bibitem [{\citenamefont {Mohan}\ \emph {et~al.}(1971)\citenamefont {Mohan},
  \citenamefont {Danos},\ and\ \citenamefont {Biedenharn}}]{Moh71}%
  \BibitemOpen
  \bibfield  {author} {\bibinfo {author} {\bibfnamefont {R.}~\bibnamefont
  {Mohan}}, \bibinfo {author} {\bibfnamefont {M.}~\bibnamefont {Danos}}, \ and\
  \bibinfo {author} {\bibfnamefont {L.~C.}\ \bibnamefont {Biedenharn}},\ }\href
  {\doibase 10.1103/PhysRevC.3.1740} {\bibfield  {journal} {\bibinfo  {journal}
  {Phys. Rev. C}\ }\textbf {\bibinfo {volume} {3}},\ \bibinfo {pages} {1740}
  (\bibinfo {year} {1971})}\BibitemShut {NoStop}%
\bibitem [{\citenamefont {Van~Isacker}\ \emph {et~al.}(1992)\citenamefont
  {Van~Isacker}, \citenamefont {Nagarajan},\ and\ \citenamefont
  {Warner}}]{Van92}%
  \BibitemOpen
  \bibfield  {author} {\bibinfo {author} {\bibfnamefont {P.}~\bibnamefont
  {Van~Isacker}}, \bibinfo {author} {\bibfnamefont {M.~A.}\ \bibnamefont
  {Nagarajan}}, \ and\ \bibinfo {author} {\bibfnamefont {D.~D.}\ \bibnamefont
  {Warner}},\ }\href {\doibase 10.1103/PhysRevC.45.R13} {\bibfield  {journal}
  {\bibinfo  {journal} {Phys. Rev. C}\ }\textbf {\bibinfo {volume} {45}},\
  \bibinfo {pages} {R13} (\bibinfo {year} {1992})}\BibitemShut {NoStop}%
\bibitem [{\citenamefont {Paar}\ \emph {et~al.}(2007)\citenamefont {Paar},
  \citenamefont {Vretenar}, \citenamefont {Khan},\ and\ \citenamefont
  {Col\`o}}]{Paa07}%
  \BibitemOpen
  \bibfield  {author} {\bibinfo {author} {\bibfnamefont {N.}~\bibnamefont
  {Paar}}, \bibinfo {author} {\bibfnamefont {D.}~\bibnamefont {Vretenar}},
  \bibinfo {author} {\bibfnamefont {E.}~\bibnamefont {Khan}}, \ and\ \bibinfo
  {author} {\bibfnamefont {G.}~\bibnamefont {Col\`o}},\ }\href {\doibase
  10.1088/0034-4885/70/5/R02} {\bibfield  {journal} {\bibinfo  {journal} {Rep.
  Prog. Phys.}\ }\textbf {\bibinfo {volume} {70}},\ \bibinfo {pages} {691}
  (\bibinfo {year} {2007})}\BibitemShut {NoStop}%
\bibitem [{\citenamefont {Pe\~na Arteaga}\ \emph {et~al.}(2009)\citenamefont
  {Pe\~na Arteaga}, \citenamefont {Khan},\ and\ \citenamefont {Ring}}]{Art09}%
  \BibitemOpen
  \bibfield  {author} {\bibinfo {author} {\bibfnamefont {D.}~\bibnamefont
  {Pe\~na Arteaga}}, \bibinfo {author} {\bibfnamefont {E.}~\bibnamefont
  {Khan}}, \ and\ \bibinfo {author} {\bibfnamefont {P.}~\bibnamefont {Ring}},\
  }\href {\doibase 10.1103/PhysRevC.79.034311} {\bibfield  {journal} {\bibinfo
  {journal} {Phys. Rev. C}\ }\textbf {\bibinfo {volume} {79}},\ \bibinfo
  {pages} {034311} (\bibinfo {year} {2009})}\BibitemShut {NoStop}%
\bibitem [{\citenamefont {Paar}\ \emph {et~al.}(2005)\citenamefont {Paar},
  \citenamefont {Vretenar},\ and\ \citenamefont {Ring}}]{Paa09}%
  \BibitemOpen
  \bibfield  {author} {\bibinfo {author} {\bibfnamefont {N.}~\bibnamefont
  {Paar}}, \bibinfo {author} {\bibfnamefont {D.}~\bibnamefont {Vretenar}}, \
  and\ \bibinfo {author} {\bibfnamefont {P.}~\bibnamefont {Ring}},\ }\href
  {\doibase 10.1103/PhysRevLett.94.182501} {\bibfield  {journal} {\bibinfo
  {journal} {Phys. Rev. Lett.}\ }\textbf {\bibinfo {volume} {94}},\ \bibinfo
  {pages} {182501} (\bibinfo {year} {2005})}\BibitemShut {NoStop}%
\bibitem [{\citenamefont {Goriely}(1998)}]{Gor98}%
  \BibitemOpen
  \bibfield  {author} {\bibinfo {author} {\bibfnamefont {S.}~\bibnamefont
  {Goriely}},\ }\href {\doibase 10.1016/S0370-2693(98)00907-1} {\bibfield
  {journal} {\bibinfo  {journal} {Phys. Lett. B}\ }\textbf {\bibinfo {volume}
  {436}},\ \bibinfo {pages} {10 } (\bibinfo {year} {1998})}\BibitemShut
  {NoStop}%
\bibitem [{\citenamefont {Savran}\ \emph {et~al.}(2008)\citenamefont {Savran},
  \citenamefont {Fritzsche}, \citenamefont {Hasper}, \citenamefont
  {Lindenberg}, \citenamefont {M\"uller}, \citenamefont {Ponomarev},
  \citenamefont {Sonnabend},\ and\ \citenamefont {Zilges}}]{Sav08}%
  \BibitemOpen
  \bibfield  {author} {\bibinfo {author} {\bibfnamefont {D.}~\bibnamefont
  {Savran}}, \bibinfo {author} {\bibfnamefont {M.}~\bibnamefont {Fritzsche}},
  \bibinfo {author} {\bibfnamefont {J.}~\bibnamefont {Hasper}}, \bibinfo
  {author} {\bibfnamefont {K.}~\bibnamefont {Lindenberg}}, \bibinfo {author}
  {\bibfnamefont {S.}~\bibnamefont {M\"uller}}, \bibinfo {author}
  {\bibfnamefont {V.~Y.}\ \bibnamefont {Ponomarev}}, \bibinfo {author}
  {\bibfnamefont {K.}~\bibnamefont {Sonnabend}}, \ and\ \bibinfo {author}
  {\bibfnamefont {A.}~\bibnamefont {Zilges}},\ }\href {\doibase
  10.1103/PhysRevLett.100.232501} {\bibfield  {journal} {\bibinfo  {journal}
  {Phys. Rev. Lett.}\ }\textbf {\bibinfo {volume} {100}},\ \bibinfo {pages}
  {232501} (\bibinfo {year} {2008})}\BibitemShut {NoStop}%
\bibitem [{\citenamefont {Poltoratska}\ \emph {et~al.}(2012)\citenamefont
  {Poltoratska}, \citenamefont {von Neumann-Cosel}, \citenamefont {Tamii},
  \citenamefont {Adachi}, \citenamefont {Bertulani}, \citenamefont {Carter},
  \citenamefont {Dozono}, \citenamefont {Fujita}, \citenamefont {Fujita},
  \citenamefont {Fujita}, \citenamefont {Hatanaka}, \citenamefont {Itoh},
  \citenamefont {Kawabata}, \citenamefont {Kalmykov}, \citenamefont
  {Krumbholz}, \citenamefont {Litvinova}, \citenamefont {Matsubara},
  \citenamefont {Nakanishi}, \citenamefont {Neveling}, \citenamefont {Okamura},
  \citenamefont {Ong}, \citenamefont {\"Ozel-Tashenov}, \citenamefont
  {Ponomarev}, \citenamefont {Richter}, \citenamefont {Rubio}, \citenamefont
  {Sakaguchi}, \citenamefont {Sakemi}, \citenamefont {Sasamoto}, \citenamefont
  {Shimbara}, \citenamefont {Shimizu}, \citenamefont {Smit}, \citenamefont
  {Suzuki}, \citenamefont {Tameshige}, \citenamefont {Wambach}, \citenamefont
  {Yosoi},\ and\ \citenamefont {Zenihiro}}]{Pol12}%
  \BibitemOpen
  \bibfield  {author} {\bibinfo {author} {\bibfnamefont {I.}~\bibnamefont
  {Poltoratska}}, \bibinfo {author} {\bibfnamefont {P.}~\bibnamefont {von
  Neumann-Cosel}}, \bibinfo {author} {\bibfnamefont {A.}~\bibnamefont {Tamii}},
  \bibinfo {author} {\bibfnamefont {T.}~\bibnamefont {Adachi}}, \bibinfo
  {author} {\bibfnamefont {C.~A.}\ \bibnamefont {Bertulani}}, \bibinfo {author}
  {\bibfnamefont {J.}~\bibnamefont {Carter}}, \bibinfo {author} {\bibfnamefont
  {M.}~\bibnamefont {Dozono}}, \bibinfo {author} {\bibfnamefont
  {H.}~\bibnamefont {Fujita}}, \bibinfo {author} {\bibfnamefont
  {K.}~\bibnamefont {Fujita}}, \bibinfo {author} {\bibfnamefont
  {Y.}~\bibnamefont {Fujita}}, \bibinfo {author} {\bibfnamefont
  {K.}~\bibnamefont {Hatanaka}}, \bibinfo {author} {\bibfnamefont
  {M.}~\bibnamefont {Itoh}}, \bibinfo {author} {\bibfnamefont {T.}~\bibnamefont
  {Kawabata}}, \bibinfo {author} {\bibfnamefont {Y.}~\bibnamefont {Kalmykov}},
  \bibinfo {author} {\bibfnamefont {A.~M.}\ \bibnamefont {Krumbholz}}, \bibinfo
  {author} {\bibfnamefont {E.}~\bibnamefont {Litvinova}}, \bibinfo {author}
  {\bibfnamefont {H.}~\bibnamefont {Matsubara}}, \bibinfo {author}
  {\bibfnamefont {K.}~\bibnamefont {Nakanishi}}, \bibinfo {author}
  {\bibfnamefont {R.}~\bibnamefont {Neveling}}, \bibinfo {author}
  {\bibfnamefont {H.}~\bibnamefont {Okamura}}, \bibinfo {author} {\bibfnamefont
  {H.~J.}\ \bibnamefont {Ong}}, \bibinfo {author} {\bibfnamefont
  {B.}~\bibnamefont {\"Ozel-Tashenov}}, \bibinfo {author} {\bibfnamefont
  {V.~Y.}\ \bibnamefont {Ponomarev}}, \bibinfo {author} {\bibfnamefont
  {A.}~\bibnamefont {Richter}}, \bibinfo {author} {\bibfnamefont
  {B.}~\bibnamefont {Rubio}}, \bibinfo {author} {\bibfnamefont
  {H.}~\bibnamefont {Sakaguchi}}, \bibinfo {author} {\bibfnamefont
  {Y.}~\bibnamefont {Sakemi}}, \bibinfo {author} {\bibfnamefont
  {Y.}~\bibnamefont {Sasamoto}}, \bibinfo {author} {\bibfnamefont
  {Y.}~\bibnamefont {Shimbara}}, \bibinfo {author} {\bibfnamefont
  {Y.}~\bibnamefont {Shimizu}}, \bibinfo {author} {\bibfnamefont {F.~D.}\
  \bibnamefont {Smit}}, \bibinfo {author} {\bibfnamefont {T.}~\bibnamefont
  {Suzuki}}, \bibinfo {author} {\bibfnamefont {Y.}~\bibnamefont {Tameshige}},
  \bibinfo {author} {\bibfnamefont {J.}~\bibnamefont {Wambach}}, \bibinfo
  {author} {\bibfnamefont {M.}~\bibnamefont {Yosoi}}, \ and\ \bibinfo {author}
  {\bibfnamefont {J.}~\bibnamefont {Zenihiro}},\ }\href {\doibase
  10.1103/PhysRevC.85.041304} {\bibfield  {journal} {\bibinfo  {journal} {Phys.
  Rev. C}\ }\textbf {\bibinfo {volume} {85}},\ \bibinfo {pages} {041304}
  (\bibinfo {year} {2012})}\BibitemShut {NoStop}%
\bibitem [{\citenamefont {Savran}\ \emph {et~al.}(2006)\citenamefont {Savran},
  \citenamefont {Babilon}, \citenamefont {van~den Berg}, \citenamefont
  {Harakeh}, \citenamefont {Hasper}, \citenamefont {Matic}, \citenamefont
  {W\"ortche},\ and\ \citenamefont {Zilges}}]{Sav06}%
  \BibitemOpen
  \bibfield  {author} {\bibinfo {author} {\bibfnamefont {D.}~\bibnamefont
  {Savran}}, \bibinfo {author} {\bibfnamefont {M.}~\bibnamefont {Babilon}},
  \bibinfo {author} {\bibfnamefont {A.~M.}\ \bibnamefont {van~den Berg}},
  \bibinfo {author} {\bibfnamefont {M.~N.}\ \bibnamefont {Harakeh}}, \bibinfo
  {author} {\bibfnamefont {J.}~\bibnamefont {Hasper}}, \bibinfo {author}
  {\bibfnamefont {A.}~\bibnamefont {Matic}}, \bibinfo {author} {\bibfnamefont
  {H.~J.}\ \bibnamefont {W\"ortche}}, \ and\ \bibinfo {author} {\bibfnamefont
  {A.}~\bibnamefont {Zilges}},\ }\href {\doibase 10.1103/PhysRevLett.97.172502}
  {\bibfield  {journal} {\bibinfo  {journal} {Phys. Rev. Lett.}\ }\textbf
  {\bibinfo {volume} {97}},\ \bibinfo {pages} {172502} (\bibinfo {year}
  {2006})}\BibitemShut {NoStop}%
\bibitem [{\citenamefont {Tonchev}\ \emph {et~al.}(2010)\citenamefont
  {Tonchev}, \citenamefont {Hammond}, \citenamefont {Kelley}, \citenamefont
  {Kwan}, \citenamefont {Lenske}, \citenamefont {Rusev}, \citenamefont
  {Tornow},\ and\ \citenamefont {Tsoneva}}]{Ton10}%
  \BibitemOpen
  \bibfield  {author} {\bibinfo {author} {\bibfnamefont {A.~P.}\ \bibnamefont
  {Tonchev}}, \bibinfo {author} {\bibfnamefont {S.~L.}\ \bibnamefont
  {Hammond}}, \bibinfo {author} {\bibfnamefont {J.~H.}\ \bibnamefont {Kelley}},
  \bibinfo {author} {\bibfnamefont {E.}~\bibnamefont {Kwan}}, \bibinfo {author}
  {\bibfnamefont {H.}~\bibnamefont {Lenske}}, \bibinfo {author} {\bibfnamefont
  {G.}~\bibnamefont {Rusev}}, \bibinfo {author} {\bibfnamefont
  {W.}~\bibnamefont {Tornow}}, \ and\ \bibinfo {author} {\bibfnamefont
  {N.}~\bibnamefont {Tsoneva}},\ }\href {\doibase
  10.1103/PhysRevLett.104.072501} {\bibfield  {journal} {\bibinfo  {journal}
  {Phys. Rev. Lett.}\ }\textbf {\bibinfo {volume} {104}},\ \bibinfo {pages}
  {072501} (\bibinfo {year} {2010})}\BibitemShut {NoStop}%
\bibitem [{\citenamefont {Govaert}\ \emph {et~al.}(1998)\citenamefont
  {Govaert}, \citenamefont {Bauwens}, \citenamefont {Bryssinck}, \citenamefont
  {De~Frenne}, \citenamefont {Jacobs}, \citenamefont {Mondelaers},
  \citenamefont {Govor},\ and\ \citenamefont {Ponomarev}}]{Gov98}%
  \BibitemOpen
  \bibfield  {author} {\bibinfo {author} {\bibfnamefont {K.}~\bibnamefont
  {Govaert}}, \bibinfo {author} {\bibfnamefont {F.}~\bibnamefont {Bauwens}},
  \bibinfo {author} {\bibfnamefont {J.}~\bibnamefont {Bryssinck}}, \bibinfo
  {author} {\bibfnamefont {D.}~\bibnamefont {De~Frenne}}, \bibinfo {author}
  {\bibfnamefont {E.}~\bibnamefont {Jacobs}}, \bibinfo {author} {\bibfnamefont
  {W.}~\bibnamefont {Mondelaers}}, \bibinfo {author} {\bibfnamefont
  {L.}~\bibnamefont {Govor}}, \ and\ \bibinfo {author} {\bibfnamefont {V.~Y.}\
  \bibnamefont {Ponomarev}},\ }\href {\doibase 10.1103/PhysRevC.57.2229}
  {\bibfield  {journal} {\bibinfo  {journal} {Phys. Rev. C}\ }\textbf {\bibinfo
  {volume} {57}},\ \bibinfo {pages} {2229} (\bibinfo {year}
  {1998})}\BibitemShut {NoStop}%
\bibitem [{\citenamefont {Schwengner}\ \emph {et~al.}(2007)\citenamefont
  {Schwengner}, \citenamefont {Rusev}, \citenamefont {Benouaret}, \citenamefont
  {Beyer}, \citenamefont {Erhard}, \citenamefont {Grosse}, \citenamefont
  {Junghans}, \citenamefont {Klug}, \citenamefont {Kosev}, \citenamefont
  {Kostov}, \citenamefont {Nair}, \citenamefont {Nankov}, \citenamefont
  {Schilling},\ and\ \citenamefont {Wagner}}]{Sch07}%
  \BibitemOpen
  \bibfield  {author} {\bibinfo {author} {\bibfnamefont {R.}~\bibnamefont
  {Schwengner}}, \bibinfo {author} {\bibfnamefont {G.}~\bibnamefont {Rusev}},
  \bibinfo {author} {\bibfnamefont {N.}~\bibnamefont {Benouaret}}, \bibinfo
  {author} {\bibfnamefont {R.}~\bibnamefont {Beyer}}, \bibinfo {author}
  {\bibfnamefont {M.}~\bibnamefont {Erhard}}, \bibinfo {author} {\bibfnamefont
  {E.}~\bibnamefont {Grosse}}, \bibinfo {author} {\bibfnamefont {A.~R.}\
  \bibnamefont {Junghans}}, \bibinfo {author} {\bibfnamefont {J.}~\bibnamefont
  {Klug}}, \bibinfo {author} {\bibfnamefont {K.}~\bibnamefont {Kosev}},
  \bibinfo {author} {\bibfnamefont {L.}~\bibnamefont {Kostov}}, \bibinfo
  {author} {\bibfnamefont {C.}~\bibnamefont {Nair}}, \bibinfo {author}
  {\bibfnamefont {N.}~\bibnamefont {Nankov}}, \bibinfo {author} {\bibfnamefont
  {K.~D.}\ \bibnamefont {Schilling}}, \ and\ \bibinfo {author} {\bibfnamefont
  {A.}~\bibnamefont {Wagner}},\ }\href {\doibase 10.1103/PhysRevC.76.034321}
  {\bibfield  {journal} {\bibinfo  {journal} {Phys. Rev. C}\ }\textbf {\bibinfo
  {volume} {76}},\ \bibinfo {pages} {034321} (\bibinfo {year}
  {2007})}\BibitemShut {NoStop}%
\bibitem [{\citenamefont {Rusev}\ \emph {et~al.}(2008)\citenamefont {Rusev},
  \citenamefont {Schwengner}, \citenamefont {D\"onau}, \citenamefont {Erhard},
  \citenamefont {Grosse}, \citenamefont {Junghans}, \citenamefont {Kosev},
  \citenamefont {Schilling}, \citenamefont {Wagner}, \citenamefont {Be\ifmmode
  \check{c}\else \v{c}\fi{}v\'a\ifmmode~\check{r}\else \v{r}\fi{}},\ and\
  \citenamefont {Krti\ifmmode~\check{c}\else \v{c}\fi{}ka}}]{Rusev08}%
  \BibitemOpen
  \bibfield  {author} {\bibinfo {author} {\bibfnamefont {G.}~\bibnamefont
  {Rusev}}, \bibinfo {author} {\bibfnamefont {R.}~\bibnamefont {Schwengner}},
  \bibinfo {author} {\bibfnamefont {F.}~\bibnamefont {D\"onau}}, \bibinfo
  {author} {\bibfnamefont {M.}~\bibnamefont {Erhard}}, \bibinfo {author}
  {\bibfnamefont {E.}~\bibnamefont {Grosse}}, \bibinfo {author} {\bibfnamefont
  {A.~R.}\ \bibnamefont {Junghans}}, \bibinfo {author} {\bibfnamefont
  {K.}~\bibnamefont {Kosev}}, \bibinfo {author} {\bibfnamefont {K.~D.}\
  \bibnamefont {Schilling}}, \bibinfo {author} {\bibfnamefont {A.}~\bibnamefont
  {Wagner}}, \bibinfo {author} {\bibfnamefont {F.}~\bibnamefont {Be\ifmmode
  \check{c}\else \v{c}\fi{}v\'a\ifmmode~\check{r}\else \v{r}\fi{}}}, \ and\
  \bibinfo {author} {\bibfnamefont {M.}~\bibnamefont
  {Krti\ifmmode~\check{c}\else \v{c}\fi{}ka}},\ }\href {\doibase
  10.1103/PhysRevC.77.064321} {\bibfield  {journal} {\bibinfo  {journal} {Phys.
  Rev. C}\ }\textbf {\bibinfo {volume} {77}},\ \bibinfo {pages} {064321}
  (\bibinfo {year} {2008})}\BibitemShut {NoStop}%
\bibitem [{\citenamefont {Wieland}\ \emph {et~al.}(2009)\citenamefont
  {Wieland}, \citenamefont {Bracco}, \citenamefont {Camera}, \citenamefont
  {Benzoni}, \citenamefont {Blasi}, \citenamefont {Brambilla}, \citenamefont
  {Crespi}, \citenamefont {Leoni}, \citenamefont {Million}, \citenamefont
  {Nicolini}, \citenamefont {Maj}, \citenamefont {Bednarczyk}, \citenamefont
  {Grebosz}, \citenamefont {Kmiecik}, \citenamefont {Meczynski}, \citenamefont
  {Styczen}, \citenamefont {Aumann}, \citenamefont {Banu}, \citenamefont
  {Beck}, \citenamefont {Becker}, \citenamefont {Caceres}, \citenamefont
  {Doornenbal}, \citenamefont {Emling}, \citenamefont {Gerl}, \citenamefont
  {Geissel}, \citenamefont {Gorska}, \citenamefont {Kavatsyuk}, \citenamefont
  {Kavatsyuk}, \citenamefont {Kojouharov}, \citenamefont {Kurz}, \citenamefont
  {Lozeva}, \citenamefont {Saito}, \citenamefont {Saito}, \citenamefont
  {Schaffner}, \citenamefont {Wollersheim}, \citenamefont {Jolie},
  \citenamefont {Reiter}, \citenamefont {Warr}, \citenamefont {deAngelis},
  \citenamefont {Gadea}, \citenamefont {Napoli}, \citenamefont {Lenzi},
  \citenamefont {Lunardi}, \citenamefont {Balabanski}, \citenamefont
  {LoBianco}, \citenamefont {Petrache}, \citenamefont {Saltarelli},
  \citenamefont {Castoldi}, \citenamefont {Zucchiatti}, \citenamefont
  {Walker},\ and\ \citenamefont {B\"urger}}]{Wie09}%
  \BibitemOpen
  \bibfield  {author} {\bibinfo {author} {\bibfnamefont {O.}~\bibnamefont
  {Wieland}}, \bibinfo {author} {\bibfnamefont {A.}~\bibnamefont {Bracco}},
  \bibinfo {author} {\bibfnamefont {F.}~\bibnamefont {Camera}}, \bibinfo
  {author} {\bibfnamefont {G.}~\bibnamefont {Benzoni}}, \bibinfo {author}
  {\bibfnamefont {N.}~\bibnamefont {Blasi}}, \bibinfo {author} {\bibfnamefont
  {S.}~\bibnamefont {Brambilla}}, \bibinfo {author} {\bibfnamefont {F.~C.~L.}\
  \bibnamefont {Crespi}}, \bibinfo {author} {\bibfnamefont {S.}~\bibnamefont
  {Leoni}}, \bibinfo {author} {\bibfnamefont {B.}~\bibnamefont {Million}},
  \bibinfo {author} {\bibfnamefont {R.}~\bibnamefont {Nicolini}}, \bibinfo
  {author} {\bibfnamefont {A.}~\bibnamefont {Maj}}, \bibinfo {author}
  {\bibfnamefont {P.}~\bibnamefont {Bednarczyk}}, \bibinfo {author}
  {\bibfnamefont {J.}~\bibnamefont {Grebosz}}, \bibinfo {author} {\bibfnamefont
  {M.}~\bibnamefont {Kmiecik}}, \bibinfo {author} {\bibfnamefont
  {W.}~\bibnamefont {Meczynski}}, \bibinfo {author} {\bibfnamefont
  {J.}~\bibnamefont {Styczen}}, \bibinfo {author} {\bibfnamefont
  {T.}~\bibnamefont {Aumann}}, \bibinfo {author} {\bibfnamefont
  {A.}~\bibnamefont {Banu}}, \bibinfo {author} {\bibfnamefont {T.}~\bibnamefont
  {Beck}}, \bibinfo {author} {\bibfnamefont {F.}~\bibnamefont {Becker}},
  \bibinfo {author} {\bibfnamefont {L.}~\bibnamefont {Caceres}}, \bibinfo
  {author} {\bibfnamefont {P.}~\bibnamefont {Doornenbal}}, \bibinfo {author}
  {\bibfnamefont {H.}~\bibnamefont {Emling}}, \bibinfo {author} {\bibfnamefont
  {J.}~\bibnamefont {Gerl}}, \bibinfo {author} {\bibfnamefont {H.}~\bibnamefont
  {Geissel}}, \bibinfo {author} {\bibfnamefont {M.}~\bibnamefont {Gorska}},
  \bibinfo {author} {\bibfnamefont {O.}~\bibnamefont {Kavatsyuk}}, \bibinfo
  {author} {\bibfnamefont {M.}~\bibnamefont {Kavatsyuk}}, \bibinfo {author}
  {\bibfnamefont {I.}~\bibnamefont {Kojouharov}}, \bibinfo {author}
  {\bibfnamefont {N.}~\bibnamefont {Kurz}}, \bibinfo {author} {\bibfnamefont
  {R.}~\bibnamefont {Lozeva}}, \bibinfo {author} {\bibfnamefont
  {N.}~\bibnamefont {Saito}}, \bibinfo {author} {\bibfnamefont
  {T.}~\bibnamefont {Saito}}, \bibinfo {author} {\bibfnamefont
  {H.}~\bibnamefont {Schaffner}}, \bibinfo {author} {\bibfnamefont {H.~J.}\
  \bibnamefont {Wollersheim}}, \bibinfo {author} {\bibfnamefont
  {J.}~\bibnamefont {Jolie}}, \bibinfo {author} {\bibfnamefont
  {P.}~\bibnamefont {Reiter}}, \bibinfo {author} {\bibfnamefont
  {N.}~\bibnamefont {Warr}}, \bibinfo {author} {\bibfnamefont {G.}~\bibnamefont
  {deAngelis}}, \bibinfo {author} {\bibfnamefont {A.}~\bibnamefont {Gadea}},
  \bibinfo {author} {\bibfnamefont {D.}~\bibnamefont {Napoli}}, \bibinfo
  {author} {\bibfnamefont {S.}~\bibnamefont {Lenzi}}, \bibinfo {author}
  {\bibfnamefont {S.}~\bibnamefont {Lunardi}}, \bibinfo {author} {\bibfnamefont
  {D.}~\bibnamefont {Balabanski}}, \bibinfo {author} {\bibfnamefont
  {G.}~\bibnamefont {LoBianco}}, \bibinfo {author} {\bibfnamefont
  {C.}~\bibnamefont {Petrache}}, \bibinfo {author} {\bibfnamefont
  {A.}~\bibnamefont {Saltarelli}}, \bibinfo {author} {\bibfnamefont
  {M.}~\bibnamefont {Castoldi}}, \bibinfo {author} {\bibfnamefont
  {A.}~\bibnamefont {Zucchiatti}}, \bibinfo {author} {\bibfnamefont
  {J.}~\bibnamefont {Walker}}, \ and\ \bibinfo {author} {\bibfnamefont
  {A.}~\bibnamefont {B\"urger}},\ }\href {\doibase
  10.1103/PhysRevLett.102.092502} {\bibfield  {journal} {\bibinfo  {journal}
  {Phys. Rev. Lett.}\ }\textbf {\bibinfo {volume} {102}},\ \bibinfo {pages}
  {092502} (\bibinfo {year} {2009})}\BibitemShut {NoStop}%
\bibitem [{\citenamefont {Endres}\ \emph {et~al.}(2010)\citenamefont {Endres},
  \citenamefont {Litvinova}, \citenamefont {Savran}, \citenamefont {Butler},
  \citenamefont {Harakeh}, \citenamefont {Harissopulos}, \citenamefont
  {Herzberg}, \citenamefont {Kr\"ucken}, \citenamefont {Lagoyannis},
  \citenamefont {Pietralla}, \citenamefont {Ponomarev}, \citenamefont
  {Popescu}, \citenamefont {Ring}, \citenamefont {Scheck}, \citenamefont
  {Sonnabend}, \citenamefont {Stoica}, \citenamefont {W\"ortche},\ and\
  \citenamefont {Zilges}}]{End10}%
  \BibitemOpen
  \bibfield  {author} {\bibinfo {author} {\bibfnamefont {J.}~\bibnamefont
  {Endres}}, \bibinfo {author} {\bibfnamefont {E.}~\bibnamefont {Litvinova}},
  \bibinfo {author} {\bibfnamefont {D.}~\bibnamefont {Savran}}, \bibinfo
  {author} {\bibfnamefont {P.~A.}\ \bibnamefont {Butler}}, \bibinfo {author}
  {\bibfnamefont {M.~N.}\ \bibnamefont {Harakeh}}, \bibinfo {author}
  {\bibfnamefont {S.}~\bibnamefont {Harissopulos}}, \bibinfo {author}
  {\bibfnamefont {R.-D.}\ \bibnamefont {Herzberg}}, \bibinfo {author}
  {\bibfnamefont {R.}~\bibnamefont {Kr\"ucken}}, \bibinfo {author}
  {\bibfnamefont {A.}~\bibnamefont {Lagoyannis}}, \bibinfo {author}
  {\bibfnamefont {N.}~\bibnamefont {Pietralla}}, \bibinfo {author}
  {\bibfnamefont {V.~Y.}\ \bibnamefont {Ponomarev}}, \bibinfo {author}
  {\bibfnamefont {L.}~\bibnamefont {Popescu}}, \bibinfo {author} {\bibfnamefont
  {P.}~\bibnamefont {Ring}}, \bibinfo {author} {\bibfnamefont {M.}~\bibnamefont
  {Scheck}}, \bibinfo {author} {\bibfnamefont {K.}~\bibnamefont {Sonnabend}},
  \bibinfo {author} {\bibfnamefont {V.~I.}\ \bibnamefont {Stoica}}, \bibinfo
  {author} {\bibfnamefont {H.~J.}\ \bibnamefont {W\"ortche}}, \ and\ \bibinfo
  {author} {\bibfnamefont {A.}~\bibnamefont {Zilges}},\ }\href {\doibase
  10.1103/PhysRevLett.105.212503} {\bibfield  {journal} {\bibinfo  {journal}
  {Phys. Rev. Lett.}\ }\textbf {\bibinfo {volume} {105}},\ \bibinfo {pages}
  {212503} (\bibinfo {year} {2010})}\BibitemShut {NoStop}%
\bibitem [{\citenamefont {Raman}\ \emph {et~al.}(2001)\citenamefont {Raman},
  \citenamefont {Nestor~Jr.},\ and\ \citenamefont {Tikkanen}}]{Ram01}%
  \BibitemOpen
  \bibfield  {author} {\bibinfo {author} {\bibfnamefont {S.}~\bibnamefont
  {Raman}}, \bibinfo {author} {\bibfnamefont {C.}~\bibnamefont {Nestor~Jr.}}, \
  and\ \bibinfo {author} {\bibfnamefont {P.}~\bibnamefont {Tikkanen}},\ }\href
  {\doibase 10.1006/adnd.2001.0858} {\bibfield  {journal} {\bibinfo  {journal}
  {At. Data Nucl. Data Tables}\ }\textbf {\bibinfo {volume} {78}},\ \bibinfo
  {pages} {1 } (\bibinfo {year} {2001})}\BibitemShut {NoStop}%
\bibitem [{\citenamefont {Carlos}\ \emph {et~al.}(1976)\citenamefont {Carlos},
  \citenamefont {Beil}, \citenamefont {Berg\`re}, \citenamefont {Fagot},
  \citenamefont {Lepr\^etre}, \citenamefont {Veyssi\`ere},\ and\ \citenamefont
  {Solodukhov}}]{Carl76}%
  \BibitemOpen
  \bibfield  {author} {\bibinfo {author} {\bibfnamefont {P.}~\bibnamefont
  {Carlos}}, \bibinfo {author} {\bibfnamefont {H.}~\bibnamefont {Beil}},
  \bibinfo {author} {\bibfnamefont {R.}~\bibnamefont {Berg\`re}}, \bibinfo
  {author} {\bibfnamefont {J.}~\bibnamefont {Fagot}}, \bibinfo {author}
  {\bibfnamefont {A.}~\bibnamefont {Lepr\^etre}}, \bibinfo {author}
  {\bibfnamefont {A.}~\bibnamefont {Veyssi\`ere}}, \ and\ \bibinfo {author}
  {\bibfnamefont {G.}~\bibnamefont {Solodukhov}},\ }\href {\doibase
  10.1016/0375-9474(76)90012-9} {\bibfield  {journal} {\bibinfo  {journal}
  {Nucl. Phys. A}\ }\textbf {\bibinfo {volume} {258}},\ \bibinfo {pages} {365 }
  (\bibinfo {year} {1976})}\BibitemShut {NoStop}%
\bibitem [{\citenamefont {Nathan}\ and\ \citenamefont {Moreh}(1980)}]{Mor80}%
  \BibitemOpen
  \bibfield  {author} {\bibinfo {author} {\bibfnamefont {A.}~\bibnamefont
  {Nathan}}\ and\ \bibinfo {author} {\bibfnamefont {R.}~\bibnamefont {Moreh}},\
  }\href {\doibase 10.1016/0370-2693(80)90657-7} {\bibfield  {journal}
  {\bibinfo  {journal} {Phys. Lett. B}\ }\textbf {\bibinfo {volume} {91}},\
  \bibinfo {pages} {38 } (\bibinfo {year} {1980})}\BibitemShut {NoStop}%
\bibitem [{\citenamefont {Elliott}\ and\ \citenamefont {Vogel}(2002)}]{Ell02}%
  \BibitemOpen
  \bibfield  {author} {\bibinfo {author} {\bibfnamefont {S.}~\bibnamefont
  {Elliott}}\ and\ \bibinfo {author} {\bibfnamefont {P.}~\bibnamefont
  {Vogel}},\ }\href {\doibase 10.1146/annurev.nucl.52.050102.090641} {\bibfield
   {journal} {\bibinfo  {journal} {Ann. Rev. Nucl. Part. Sci.}\ }\textbf
  {\bibinfo {volume} {52}},\ \bibinfo {pages} {115} (\bibinfo {year}
  {2002})}\BibitemShut {NoStop}%
\bibitem [{\citenamefont {Schiffer}\ \emph {et~al.}(2008)\citenamefont
  {Schiffer}, \citenamefont {Freeman}, \citenamefont {Clark}, \citenamefont
  {Deibel}, \citenamefont {Fitzpatrick}, \citenamefont {Gros}, \citenamefont
  {Heinz}, \citenamefont {Hirata}, \citenamefont {Jiang}, \citenamefont {Kay},
  \citenamefont {Parikh}, \citenamefont {Parker}, \citenamefont {Rehm},
  \citenamefont {Villari}, \citenamefont {Werner},\ and\ \citenamefont
  {Wrede}}]{Sch08}%
  \BibitemOpen
  \bibfield  {author} {\bibinfo {author} {\bibfnamefont {J.~P.}\ \bibnamefont
  {Schiffer}}, \bibinfo {author} {\bibfnamefont {S.~J.}\ \bibnamefont
  {Freeman}}, \bibinfo {author} {\bibfnamefont {J.~A.}\ \bibnamefont {Clark}},
  \bibinfo {author} {\bibfnamefont {C.}~\bibnamefont {Deibel}}, \bibinfo
  {author} {\bibfnamefont {C.~R.}\ \bibnamefont {Fitzpatrick}}, \bibinfo
  {author} {\bibfnamefont {S.}~\bibnamefont {Gros}}, \bibinfo {author}
  {\bibfnamefont {A.}~\bibnamefont {Heinz}}, \bibinfo {author} {\bibfnamefont
  {D.}~\bibnamefont {Hirata}}, \bibinfo {author} {\bibfnamefont {C.~L.}\
  \bibnamefont {Jiang}}, \bibinfo {author} {\bibfnamefont {B.~P.}\ \bibnamefont
  {Kay}}, \bibinfo {author} {\bibfnamefont {A.}~\bibnamefont {Parikh}},
  \bibinfo {author} {\bibfnamefont {P.~D.}\ \bibnamefont {Parker}}, \bibinfo
  {author} {\bibfnamefont {K.~E.}\ \bibnamefont {Rehm}}, \bibinfo {author}
  {\bibfnamefont {A.~C.~C.}\ \bibnamefont {Villari}}, \bibinfo {author}
  {\bibfnamefont {V.}~\bibnamefont {Werner}}, \ and\ \bibinfo {author}
  {\bibfnamefont {C.}~\bibnamefont {Wrede}},\ }\href {\doibase
  10.1103/PhysRevLett.100.112501} {\bibfield  {journal} {\bibinfo  {journal}
  {Phys. Rev. Lett.}\ }\textbf {\bibinfo {volume} {100}},\ \bibinfo {pages}
  {112501} (\bibinfo {year} {2008})}\BibitemShut {NoStop}%
\bibitem [{\citenamefont {Freeman}\ \emph {et~al.}(2007)\citenamefont
  {Freeman}, \citenamefont {Schiffer}, \citenamefont {Villari}, \citenamefont
  {Clark}, \citenamefont {Deibel}, \citenamefont {Gros}, \citenamefont {Heinz},
  \citenamefont {Hirata}, \citenamefont {Jiang}, \citenamefont {Kay},
  \citenamefont {Parikh}, \citenamefont {Parker}, \citenamefont {Qian},
  \citenamefont {Rehm}, \citenamefont {Tang}, \citenamefont {Werner},\ and\
  \citenamefont {Wrede}}]{Fre07}%
  \BibitemOpen
  \bibfield  {author} {\bibinfo {author} {\bibfnamefont {S.~J.}\ \bibnamefont
  {Freeman}}, \bibinfo {author} {\bibfnamefont {J.~P.}\ \bibnamefont
  {Schiffer}}, \bibinfo {author} {\bibfnamefont {A.~C.~C.}\ \bibnamefont
  {Villari}}, \bibinfo {author} {\bibfnamefont {J.~A.}\ \bibnamefont {Clark}},
  \bibinfo {author} {\bibfnamefont {C.}~\bibnamefont {Deibel}}, \bibinfo
  {author} {\bibfnamefont {S.}~\bibnamefont {Gros}}, \bibinfo {author}
  {\bibfnamefont {A.}~\bibnamefont {Heinz}}, \bibinfo {author} {\bibfnamefont
  {D.}~\bibnamefont {Hirata}}, \bibinfo {author} {\bibfnamefont {C.~L.}\
  \bibnamefont {Jiang}}, \bibinfo {author} {\bibfnamefont {B.~P.}\ \bibnamefont
  {Kay}}, \bibinfo {author} {\bibfnamefont {A.}~\bibnamefont {Parikh}},
  \bibinfo {author} {\bibfnamefont {P.~D.}\ \bibnamefont {Parker}}, \bibinfo
  {author} {\bibfnamefont {J.}~\bibnamefont {Qian}}, \bibinfo {author}
  {\bibfnamefont {K.~E.}\ \bibnamefont {Rehm}}, \bibinfo {author}
  {\bibfnamefont {X.~D.}\ \bibnamefont {Tang}}, \bibinfo {author}
  {\bibfnamefont {V.}~\bibnamefont {Werner}}, \ and\ \bibinfo {author}
  {\bibfnamefont {C.}~\bibnamefont {Wrede}},\ }\href {\doibase
  10.1103/PhysRevC.75.051301} {\bibfield  {journal} {\bibinfo  {journal} {Phys.
  Rev. C}\ }\textbf {\bibinfo {volume} {75}},\ \bibinfo {pages} {051301}
  (\bibinfo {year} {2007})}\BibitemShut {NoStop}%
\bibitem [{\citenamefont {Klapdor-Kleingrothaus}\ \emph
  {et~al.}(2004)\citenamefont {Klapdor-Kleingrothaus}, \citenamefont
  {Krivosheina}, \citenamefont {Dietz},\ and\ \citenamefont
  {Chkvorets}}]{Klap04}%
  \BibitemOpen
  \bibfield  {author} {\bibinfo {author} {\bibfnamefont {H.}~\bibnamefont
  {Klapdor-Kleingrothaus}}, \bibinfo {author} {\bibfnamefont {I.}~\bibnamefont
  {Krivosheina}}, \bibinfo {author} {\bibfnamefont {A.}~\bibnamefont {Dietz}},
  \ and\ \bibinfo {author} {\bibfnamefont {O.}~\bibnamefont {Chkvorets}},\
  }\href {\doibase 10.1016/j.physletb.2004.02.025} {\bibfield  {journal}
  {\bibinfo  {journal} {Phys. Lett. B}\ }\textbf {\bibinfo {volume} {586}},\
  \bibinfo {pages} {198 } (\bibinfo {year} {2004})}\BibitemShut {NoStop}%
\bibitem [{\citenamefont {Brine}\ \emph {et~al.}(2006)\citenamefont {Brine},
  \citenamefont {Stevenson}, \citenamefont {Maruhn},\ and\ \citenamefont
  {Reinhard}}]{Bri08}%
  \BibitemOpen
  \bibfield  {author} {\bibinfo {author} {\bibfnamefont {M.~P.}\ \bibnamefont
  {Brine}}, \bibinfo {author} {\bibfnamefont {P.~D.}\ \bibnamefont
  {Stevenson}}, \bibinfo {author} {\bibfnamefont {J.~A.}\ \bibnamefont
  {Maruhn}}, \ and\ \bibinfo {author} {\bibfnamefont {P.-G.}\ \bibnamefont
  {Reinhard}},\ }\href {\doibase 10.1142/S0218301306005009} {\bibfield
  {journal} {\bibinfo  {journal} {Int. J. Mod. Phys. E}\ }\textbf {\bibinfo
  {volume} {15}},\ \bibinfo {pages} {1417} (\bibinfo {year}
  {2006})}\BibitemShut {NoStop}%
\bibitem [{\citenamefont {Stevenson}\ and\ \citenamefont
  {Fracasso}(2010)}]{Stev10}%
  \BibitemOpen
  \bibfield  {author} {\bibinfo {author} {\bibfnamefont {P.~D.}\ \bibnamefont
  {Stevenson}}\ and\ \bibinfo {author} {\bibfnamefont {S.}~\bibnamefont
  {Fracasso}},\ }\href {http://stacks.iop.org/0954-3899/37/i=6/a=064030}
  {\bibfield  {journal} {\bibinfo  {journal} {J. Phys. G: Nucl. Part. Phys.}\
  }\textbf {\bibinfo {volume} {37}},\ \bibinfo {pages} {064030} (\bibinfo
  {year} {2010})}\BibitemShut {NoStop}%
\bibitem [{\citenamefont {Nakatsukasa}\ \emph {et~al.}(2007)\citenamefont
  {Nakatsukasa}, \citenamefont {Inakura},\ and\ \citenamefont
  {Yabana}}]{Ina09}%
  \BibitemOpen
  \bibfield  {author} {\bibinfo {author} {\bibfnamefont {T.}~\bibnamefont
  {Nakatsukasa}}, \bibinfo {author} {\bibfnamefont {T.}~\bibnamefont
  {Inakura}}, \ and\ \bibinfo {author} {\bibfnamefont {K.}~\bibnamefont
  {Yabana}},\ }\href {\doibase 10.1103/PhysRevC.76.024318} {\bibfield
  {journal} {\bibinfo  {journal} {Phys. Rev. C}\ }\textbf {\bibinfo {volume}
  {76}},\ \bibinfo {pages} {024318} (\bibinfo {year} {2007})}\BibitemShut
  {NoStop}%
\bibitem [{\citenamefont {Avogadro}\ and\ \citenamefont
  {Nakatsukasa}(2011)}]{Avo11}%
  \BibitemOpen
  \bibfield  {author} {\bibinfo {author} {\bibfnamefont {P.}~\bibnamefont
  {Avogadro}}\ and\ \bibinfo {author} {\bibfnamefont {T.}~\bibnamefont
  {Nakatsukasa}},\ }\href {\doibase 10.1103/PhysRevC.84.014314} {\bibfield
  {journal} {\bibinfo  {journal} {Phys. Rev. C}\ }\textbf {\bibinfo {volume}
  {84}},\ \bibinfo {pages} {014314} (\bibinfo {year} {2011})}\BibitemShut
  {NoStop}%
\bibitem [{\citenamefont {Toivanen}\ and\ \citenamefont
  {Suhonen}(1997)}]{Suh97}%
  \BibitemOpen
  \bibfield  {author} {\bibinfo {author} {\bibfnamefont {J.}~\bibnamefont
  {Toivanen}}\ and\ \bibinfo {author} {\bibfnamefont {J.}~\bibnamefont
  {Suhonen}},\ }\href {\doibase 10.1103/PhysRevC.55.2314} {\bibfield  {journal}
  {\bibinfo  {journal} {Phys. Rev. C}\ }\textbf {\bibinfo {volume} {55}},\
  \bibinfo {pages} {2314} (\bibinfo {year} {1997})}\BibitemShut {NoStop}%
\bibitem [{\citenamefont {Cooper}\ \emph {et~al.}(2012)\citenamefont {Cooper},
  \citenamefont {Reichel}, \citenamefont {Werner}, \citenamefont {Bettermann},
  \citenamefont {Alikhani}, \citenamefont {Aslanidou}, \citenamefont {Bauer},
  \citenamefont {Coquard}, \citenamefont {Fritzsche}, \citenamefont
  {Fritzsche}, \citenamefont {Glorius}, \citenamefont {Goddard}, \citenamefont
  {M\"oller}, \citenamefont {Pietralla}, \citenamefont {Reese}, \citenamefont
  {Romig}, \citenamefont {Savran}, \citenamefont {Schnorrenberger},
  \citenamefont {Siebenh\"uhner}, \citenamefont {Simon}, \citenamefont
  {Sonnabend}, \citenamefont {Smith}, \citenamefont {Walz}, \citenamefont
  {Yates}, \citenamefont {Yevetska},\ and\ \citenamefont {Zweidinger}}]{Coo10}%
  \BibitemOpen
  \bibfield  {author} {\bibinfo {author} {\bibfnamefont {N.}~\bibnamefont
  {Cooper}}, \bibinfo {author} {\bibfnamefont {F.}~\bibnamefont {Reichel}},
  \bibinfo {author} {\bibfnamefont {V.}~\bibnamefont {Werner}}, \bibinfo
  {author} {\bibfnamefont {L.}~\bibnamefont {Bettermann}}, \bibinfo {author}
  {\bibfnamefont {B.}~\bibnamefont {Alikhani}}, \bibinfo {author}
  {\bibfnamefont {S.}~\bibnamefont {Aslanidou}}, \bibinfo {author}
  {\bibfnamefont {C.}~\bibnamefont {Bauer}}, \bibinfo {author} {\bibfnamefont
  {L.}~\bibnamefont {Coquard}}, \bibinfo {author} {\bibfnamefont
  {M.}~\bibnamefont {Fritzsche}}, \bibinfo {author} {\bibfnamefont
  {Y.}~\bibnamefont {Fritzsche}}, \bibinfo {author} {\bibfnamefont
  {J.}~\bibnamefont {Glorius}}, \bibinfo {author} {\bibfnamefont {P.~M.}\
  \bibnamefont {Goddard}}, \bibinfo {author} {\bibfnamefont {T.}~\bibnamefont
  {M\"oller}}, \bibinfo {author} {\bibfnamefont {N.}~\bibnamefont {Pietralla}},
  \bibinfo {author} {\bibfnamefont {M.}~\bibnamefont {Reese}}, \bibinfo
  {author} {\bibfnamefont {C.}~\bibnamefont {Romig}}, \bibinfo {author}
  {\bibfnamefont {D.}~\bibnamefont {Savran}}, \bibinfo {author} {\bibfnamefont
  {L.}~\bibnamefont {Schnorrenberger}}, \bibinfo {author} {\bibfnamefont
  {F.}~\bibnamefont {Siebenh\"uhner}}, \bibinfo {author} {\bibfnamefont
  {V.~V.}\ \bibnamefont {Simon}}, \bibinfo {author} {\bibfnamefont
  {K.}~\bibnamefont {Sonnabend}}, \bibinfo {author} {\bibfnamefont {M.~K.}\
  \bibnamefont {Smith}}, \bibinfo {author} {\bibfnamefont {C.}~\bibnamefont
  {Walz}}, \bibinfo {author} {\bibfnamefont {S.~W.}\ \bibnamefont {Yates}},
  \bibinfo {author} {\bibfnamefont {O.}~\bibnamefont {Yevetska}}, \ and\
  \bibinfo {author} {\bibfnamefont {M.}~\bibnamefont {Zweidinger}},\ }\href
  {\doibase 10.1103/PhysRevC.86.034313} {\bibfield  {journal} {\bibinfo
  {journal} {Phys. Rev. C}\ }\textbf {\bibinfo {volume} {86}},\ \bibinfo
  {pages} {034313} (\bibinfo {year} {2012})}\BibitemShut {NoStop}%
\bibitem [{\citenamefont {Sonnabend}\ \emph {et~al.}(2011)\citenamefont
  {Sonnabend}, \citenamefont {Savran}, \citenamefont {Beller}, \citenamefont
  {BÃ¼ssing}, \citenamefont {Constantinescu}, \citenamefont {Elvers},
  \citenamefont {Endres}, \citenamefont {Fritzsche}, \citenamefont {Glorius},
  \citenamefont {Hasper}, \citenamefont {Isaak}, \citenamefont {LÃ¶her},
  \citenamefont {MÃ¼ller}, \citenamefont {Pietralla}, \citenamefont {Romig},
  \citenamefont {Sauerwein}, \citenamefont {Schnorrenberger}, \citenamefont
  {WÃ¤lzlein}, \citenamefont {Zilges},\ and\ \citenamefont
  {Zweidinger}}]{Son11}%
  \BibitemOpen
  \bibfield  {author} {\bibinfo {author} {\bibfnamefont {K.}~\bibnamefont
  {Sonnabend}}, \bibinfo {author} {\bibfnamefont {D.}~\bibnamefont {Savran}},
  \bibinfo {author} {\bibfnamefont {J.}~\bibnamefont {Beller}}, \bibinfo
  {author} {\bibfnamefont {M.}~\bibnamefont {BÃ¼ssing}}, \bibinfo {author}
  {\bibfnamefont {A.}~\bibnamefont {Constantinescu}}, \bibinfo {author}
  {\bibfnamefont {M.}~\bibnamefont {Elvers}}, \bibinfo {author} {\bibfnamefont
  {J.}~\bibnamefont {Endres}}, \bibinfo {author} {\bibfnamefont
  {M.}~\bibnamefont {Fritzsche}}, \bibinfo {author} {\bibfnamefont
  {J.}~\bibnamefont {Glorius}}, \bibinfo {author} {\bibfnamefont
  {J.}~\bibnamefont {Hasper}}, \bibinfo {author} {\bibfnamefont
  {J.}~\bibnamefont {Isaak}}, \bibinfo {author} {\bibfnamefont
  {B.}~\bibnamefont {LÃ¶her}}, \bibinfo {author} {\bibfnamefont
  {S.}~\bibnamefont {MÃ¼ller}}, \bibinfo {author} {\bibfnamefont
  {N.}~\bibnamefont {Pietralla}}, \bibinfo {author} {\bibfnamefont
  {C.}~\bibnamefont {Romig}}, \bibinfo {author} {\bibfnamefont
  {A.}~\bibnamefont {Sauerwein}}, \bibinfo {author} {\bibfnamefont
  {L.}~\bibnamefont {Schnorrenberger}}, \bibinfo {author} {\bibfnamefont
  {C.}~\bibnamefont {WÃ¤lzlein}}, \bibinfo {author} {\bibfnamefont
  {A.}~\bibnamefont {Zilges}}, \ and\ \bibinfo {author} {\bibfnamefont
  {M.}~\bibnamefont {Zweidinger}},\ }\href {\doibase
  10.1016/j.nima.2011.02.107} {\bibfield  {journal} {\bibinfo  {journal} {Nucl.
  Instrum. Methods A}\ }\textbf {\bibinfo {volume} {640}},\ \bibinfo {pages} {6
  } (\bibinfo {year} {2011})}\BibitemShut {NoStop}%
\bibitem [{\citenamefont {Rusev}\ \emph {et~al.}(2013)\citenamefont {Rusev},
  \citenamefont {Tsoneva}, \citenamefont {D\"onau}, \citenamefont {Frauendorf},
  \citenamefont {Schwengner}, \citenamefont {Tonchev}, \citenamefont {Adekola},
  \citenamefont {Hammond}, \citenamefont {Kelley}, \citenamefont {Kwan},
  \citenamefont {Lenske}, \citenamefont {Tornow},\ and\ \citenamefont
  {Wagner}}]{Gen13}%
  \BibitemOpen
  \bibfield  {author} {\bibinfo {author} {\bibfnamefont {G.}~\bibnamefont
  {Rusev}}, \bibinfo {author} {\bibfnamefont {N.}~\bibnamefont {Tsoneva}},
  \bibinfo {author} {\bibfnamefont {F.}~\bibnamefont {D\"onau}}, \bibinfo
  {author} {\bibfnamefont {S.}~\bibnamefont {Frauendorf}}, \bibinfo {author}
  {\bibfnamefont {R.}~\bibnamefont {Schwengner}}, \bibinfo {author}
  {\bibfnamefont {A.~P.}\ \bibnamefont {Tonchev}}, \bibinfo {author}
  {\bibfnamefont {A.~S.}\ \bibnamefont {Adekola}}, \bibinfo {author}
  {\bibfnamefont {S.~L.}\ \bibnamefont {Hammond}}, \bibinfo {author}
  {\bibfnamefont {J.~H.}\ \bibnamefont {Kelley}}, \bibinfo {author}
  {\bibfnamefont {E.}~\bibnamefont {Kwan}}, \bibinfo {author} {\bibfnamefont
  {H.}~\bibnamefont {Lenske}}, \bibinfo {author} {\bibfnamefont
  {W.}~\bibnamefont {Tornow}}, \ and\ \bibinfo {author} {\bibfnamefont
  {A.}~\bibnamefont {Wagner}},\ }\href {\doibase
  10.1103/PhysRevLett.110.022503} {\bibfield  {journal} {\bibinfo  {journal}
  {Phys. Rev. Lett.}\ }\textbf {\bibinfo {volume} {110}},\ \bibinfo {pages}
  {022503} (\bibinfo {year} {2013})}\BibitemShut {NoStop}%
\bibitem [{\citenamefont {Pietralla}\ \emph {et~al.}(2001)\citenamefont
  {Pietralla}, \citenamefont {Berant}, \citenamefont {Litvinenko},
  \citenamefont {Hartman}, \citenamefont {Mikhailov}, \citenamefont {Pinayev},
  \citenamefont {Swift}, \citenamefont {Ahmed}, \citenamefont {Kelley},
  \citenamefont {Nelson}, \citenamefont {Prior}, \citenamefont {Sabourov},
  \citenamefont {Tonchev},\ and\ \citenamefont {Weller}}]{Pie02}%
  \BibitemOpen
  \bibfield  {author} {\bibinfo {author} {\bibfnamefont {N.}~\bibnamefont
  {Pietralla}}, \bibinfo {author} {\bibfnamefont {Z.}~\bibnamefont {Berant}},
  \bibinfo {author} {\bibfnamefont {V.~N.}\ \bibnamefont {Litvinenko}},
  \bibinfo {author} {\bibfnamefont {S.}~\bibnamefont {Hartman}}, \bibinfo
  {author} {\bibfnamefont {F.~F.}\ \bibnamefont {Mikhailov}}, \bibinfo {author}
  {\bibfnamefont {I.~V.}\ \bibnamefont {Pinayev}}, \bibinfo {author}
  {\bibfnamefont {G.}~\bibnamefont {Swift}}, \bibinfo {author} {\bibfnamefont
  {M.~W.}\ \bibnamefont {Ahmed}}, \bibinfo {author} {\bibfnamefont {J.~H.}\
  \bibnamefont {Kelley}}, \bibinfo {author} {\bibfnamefont {S.~O.}\
  \bibnamefont {Nelson}}, \bibinfo {author} {\bibfnamefont {R.}~\bibnamefont
  {Prior}}, \bibinfo {author} {\bibfnamefont {K.}~\bibnamefont {Sabourov}},
  \bibinfo {author} {\bibfnamefont {A.~P.}\ \bibnamefont {Tonchev}}, \ and\
  \bibinfo {author} {\bibfnamefont {H.~R.}\ \bibnamefont {Weller}},\ }\href
  {\doibase 10.1103/PhysRevLett.88.012502} {\bibfield  {journal} {\bibinfo
  {journal} {Phys. Rev. Lett.}\ }\textbf {\bibinfo {volume} {88}},\ \bibinfo
  {pages} {012502} (\bibinfo {year} {2001})}\BibitemShut {NoStop}%
\bibitem [{\citenamefont {Krane}\ \emph {et~al.}(1973)\citenamefont {Krane},
  \citenamefont {Steffen},\ and\ \citenamefont {Wheeler}}]{Kra73}%
  \BibitemOpen
  \bibfield  {author} {\bibinfo {author} {\bibfnamefont {K.}~\bibnamefont
  {Krane}}, \bibinfo {author} {\bibfnamefont {R.}~\bibnamefont {Steffen}}, \
  and\ \bibinfo {author} {\bibfnamefont {R.}~\bibnamefont {Wheeler}},\ }\href
  {\doibase 10.1016/S0092-640X(73)80016-6} {\bibfield  {journal} {\bibinfo
  {journal} {At. Data Nucl. Data Tables}\ }\textbf {\bibinfo {volume} {11}},\
  \bibinfo {pages} {351 } (\bibinfo {year} {1973})}\BibitemShut {NoStop}%
\bibitem [{\citenamefont {Fagg}\ and\ \citenamefont {Hanna}(1959)}]{Fag59}%
  \BibitemOpen
  \bibfield  {author} {\bibinfo {author} {\bibfnamefont {L.~W.}\ \bibnamefont
  {Fagg}}\ and\ \bibinfo {author} {\bibfnamefont {S.~S.}\ \bibnamefont
  {Hanna}},\ }\href {\doibase 10.1103/RevModPhys.31.711} {\bibfield  {journal}
  {\bibinfo  {journal} {Rev. Mod. Phys.}\ }\textbf {\bibinfo {volume} {31}},\
  \bibinfo {pages} {711} (\bibinfo {year} {1959})}\BibitemShut {NoStop}%
\bibitem [{\citenamefont {Pietralla}\ \emph {et~al.}(2003)\citenamefont
  {Pietralla}, \citenamefont {Ahmed}, \citenamefont {Fransen}, \citenamefont
  {Litvinenko}, \citenamefont {Tonchev},\ and\ \citenamefont {Weller}}]{Pie03}%
  \BibitemOpen
  \bibfield  {author} {\bibinfo {author} {\bibfnamefont {N.}~\bibnamefont
  {Pietralla}}, \bibinfo {author} {\bibfnamefont {M.~W.}\ \bibnamefont
  {Ahmed}}, \bibinfo {author} {\bibfnamefont {C.}~\bibnamefont {Fransen}},
  \bibinfo {author} {\bibfnamefont {V.~N.}\ \bibnamefont {Litvinenko}},
  \bibinfo {author} {\bibfnamefont {A.~P.}\ \bibnamefont {Tonchev}}, \ and\
  \bibinfo {author} {\bibfnamefont {H.~R.}\ \bibnamefont {Weller}},\ }\href
  {\doibase 10.1063/1.1556666} {\bibfield  {journal} {\bibinfo  {journal} {AIP
  Conf. Proc.}\ }\textbf {\bibinfo {volume} {656}},\ \bibinfo {pages} {365}
  (\bibinfo {year} {2003})}\BibitemShut {NoStop}%
\bibitem [{\citenamefont {Werner}(2004)}]{Wer04}%
  \BibitemOpen
  \bibfield  {author} {\bibinfo {author} {\bibfnamefont {V.}~\bibnamefont
  {Werner}},\ }\emph {\bibinfo {title} {Proton-neutron symmetry at the limits
  of collectivity}},\ \href@noop {} {Ph.D. thesis},\ \bibinfo  {school}
  {Universit\"at zu K\"oln} (\bibinfo {year} {2004})\BibitemShut {NoStop}%
\bibitem [{\citenamefont {Biedenharn}\ and\ \citenamefont
  {Rose}(1953)}]{Bie53}%
  \BibitemOpen
  \bibfield  {author} {\bibinfo {author} {\bibfnamefont {L.~C.}\ \bibnamefont
  {Biedenharn}}\ and\ \bibinfo {author} {\bibfnamefont {M.~E.}\ \bibnamefont
  {Rose}},\ }\href {\doibase 10.1103/RevModPhys.25.729} {\bibfield  {journal}
  {\bibinfo  {journal} {Rev. Mod. Phys.}\ }\textbf {\bibinfo {volume} {25}},\
  \bibinfo {pages} {729} (\bibinfo {year} {1953})}\BibitemShut {NoStop}%
\bibitem [{\citenamefont {Carman}\ \emph {et~al.}(1996)\citenamefont {Carman},
  \citenamefont {Litveninko}, \citenamefont {Madey}, \citenamefont {Neuman},
  \citenamefont {Norum}, \citenamefont {O'Shea}, \citenamefont {Roberson},
  \citenamefont {Scarlett}, \citenamefont {Schreiber},\ and\ \citenamefont
  {Weller}}]{Car96}%
  \BibitemOpen
  \bibfield  {author} {\bibinfo {author} {\bibfnamefont {T.~S.}\ \bibnamefont
  {Carman}}, \bibinfo {author} {\bibfnamefont {V.}~\bibnamefont {Litveninko}},
  \bibinfo {author} {\bibfnamefont {J.}~\bibnamefont {Madey}}, \bibinfo
  {author} {\bibfnamefont {C.}~\bibnamefont {Neuman}}, \bibinfo {author}
  {\bibfnamefont {B.}~\bibnamefont {Norum}}, \bibinfo {author} {\bibfnamefont
  {P.~G.}\ \bibnamefont {O'Shea}}, \bibinfo {author} {\bibfnamefont {N.~R.}\
  \bibnamefont {Roberson}}, \bibinfo {author} {\bibfnamefont {C.~Y.}\
  \bibnamefont {Scarlett}}, \bibinfo {author} {\bibfnamefont {E.}~\bibnamefont
  {Schreiber}}, \ and\ \bibinfo {author} {\bibfnamefont {H.~R.}\ \bibnamefont
  {Weller}},\ }\href {\doibase 10.1016/0168-9002(95)01486-1} {\bibfield
  {journal} {\bibinfo  {journal} {Nucl. Instrum. Methods A}\ }\textbf {\bibinfo
  {volume} {378}},\ \bibinfo {pages} {1 } (\bibinfo {year} {1996})}\BibitemShut
  {NoStop}%
\bibitem [{\citenamefont {Litvinenko}\ \emph {et~al.}(1997)\citenamefont
  {Litvinenko}, \citenamefont {Burnham}, \citenamefont {Emamian}, \citenamefont
  {Hower}, \citenamefont {Madey}, \citenamefont {Morcombe}, \citenamefont
  {O'Shea}, \citenamefont {Park}, \citenamefont {Sachtschale}, \citenamefont
  {Straub}, \citenamefont {Swift}, \citenamefont {Wang}, \citenamefont {Wu},
  \citenamefont {Canon}, \citenamefont {Howell}, \citenamefont {Roberson},
  \citenamefont {Schreiber}, \citenamefont {Spraker}, \citenamefont {Tornow},
  \citenamefont {Weller}, \citenamefont {Pinayev}, \citenamefont {Gavrilov},
  \citenamefont {Fedotov}, \citenamefont {Kulipanov}, \citenamefont {Kurkin},
  \citenamefont {Mikhailov}, \citenamefont {Popik}, \citenamefont {Skrinsky},
  \citenamefont {Vinokurov}, \citenamefont {Norum}, \citenamefont {Lumpkin},\
  and\ \citenamefont {Yang}}]{Lit97}%
  \BibitemOpen
  \bibfield  {author} {\bibinfo {author} {\bibfnamefont {V.~N.}\ \bibnamefont
  {Litvinenko}}, \bibinfo {author} {\bibfnamefont {B.}~\bibnamefont {Burnham}},
  \bibinfo {author} {\bibfnamefont {M.}~\bibnamefont {Emamian}}, \bibinfo
  {author} {\bibfnamefont {N.}~\bibnamefont {Hower}}, \bibinfo {author}
  {\bibfnamefont {J.~M.~J.}\ \bibnamefont {Madey}}, \bibinfo {author}
  {\bibfnamefont {P.}~\bibnamefont {Morcombe}}, \bibinfo {author}
  {\bibfnamefont {P.~G.}\ \bibnamefont {O'Shea}}, \bibinfo {author}
  {\bibfnamefont {S.~H.}\ \bibnamefont {Park}}, \bibinfo {author}
  {\bibfnamefont {R.}~\bibnamefont {Sachtschale}}, \bibinfo {author}
  {\bibfnamefont {K.~D.}\ \bibnamefont {Straub}}, \bibinfo {author}
  {\bibfnamefont {G.}~\bibnamefont {Swift}}, \bibinfo {author} {\bibfnamefont
  {P.}~\bibnamefont {Wang}}, \bibinfo {author} {\bibfnamefont {Y.}~\bibnamefont
  {Wu}}, \bibinfo {author} {\bibfnamefont {R.~S.}\ \bibnamefont {Canon}},
  \bibinfo {author} {\bibfnamefont {C.~R.}\ \bibnamefont {Howell}}, \bibinfo
  {author} {\bibfnamefont {N.~R.}\ \bibnamefont {Roberson}}, \bibinfo {author}
  {\bibfnamefont {E.~C.}\ \bibnamefont {Schreiber}}, \bibinfo {author}
  {\bibfnamefont {M.}~\bibnamefont {Spraker}}, \bibinfo {author} {\bibfnamefont
  {W.}~\bibnamefont {Tornow}}, \bibinfo {author} {\bibfnamefont {H.~R.}\
  \bibnamefont {Weller}}, \bibinfo {author} {\bibfnamefont {I.~V.}\
  \bibnamefont {Pinayev}}, \bibinfo {author} {\bibfnamefont {N.~G.}\
  \bibnamefont {Gavrilov}}, \bibinfo {author} {\bibfnamefont {M.~G.}\
  \bibnamefont {Fedotov}}, \bibinfo {author} {\bibfnamefont {G.~N.}\
  \bibnamefont {Kulipanov}}, \bibinfo {author} {\bibfnamefont {G.~Y.}\
  \bibnamefont {Kurkin}}, \bibinfo {author} {\bibfnamefont {S.~F.}\
  \bibnamefont {Mikhailov}}, \bibinfo {author} {\bibfnamefont {V.~M.}\
  \bibnamefont {Popik}}, \bibinfo {author} {\bibfnamefont {A.~N.}\ \bibnamefont
  {Skrinsky}}, \bibinfo {author} {\bibfnamefont {N.~A.}\ \bibnamefont
  {Vinokurov}}, \bibinfo {author} {\bibfnamefont {B.~E.}\ \bibnamefont
  {Norum}}, \bibinfo {author} {\bibfnamefont {A.}~\bibnamefont {Lumpkin}}, \
  and\ \bibinfo {author} {\bibfnamefont {B.}~\bibnamefont {Yang}},\ }\href
  {\doibase 10.1103/PhysRevLett.78.4569} {\bibfield  {journal} {\bibinfo
  {journal} {Phys. Rev. Lett.}\ }\textbf {\bibinfo {volume} {78}},\ \bibinfo
  {pages} {4569} (\bibinfo {year} {1997})}\BibitemShut {NoStop}%
\bibitem [{\citenamefont {Weller}\ \emph {et~al.}(2009)\citenamefont {Weller},
  \citenamefont {Ahmed}, \citenamefont {Gao}, \citenamefont {Tornow},
  \citenamefont {Wu}, \citenamefont {Gai},\ and\ \citenamefont
  {Miskimen}}]{Wel08}%
  \BibitemOpen
  \bibfield  {author} {\bibinfo {author} {\bibfnamefont {H.~R.}\ \bibnamefont
  {Weller}}, \bibinfo {author} {\bibfnamefont {M.~W.}\ \bibnamefont {Ahmed}},
  \bibinfo {author} {\bibfnamefont {H.}~\bibnamefont {Gao}}, \bibinfo {author}
  {\bibfnamefont {W.}~\bibnamefont {Tornow}}, \bibinfo {author} {\bibfnamefont
  {Y.~K.}\ \bibnamefont {Wu}}, \bibinfo {author} {\bibfnamefont
  {M.}~\bibnamefont {Gai}}, \ and\ \bibinfo {author} {\bibfnamefont
  {R.}~\bibnamefont {Miskimen}},\ }\href {\doibase 10.1016/j.ppnp.2008.07.001}
  {\bibfield  {journal} {\bibinfo  {journal} {Prog. Part. Nucl. Phys.}\
  }\textbf {\bibinfo {volume} {62}},\ \bibinfo {pages} {257 } (\bibinfo {year}
  {2009})}\BibitemShut {NoStop}%
\bibitem [{\citenamefont {Hammond}\ \emph {et~al.}(2012)\citenamefont
  {Hammond}, \citenamefont {Adekola}, \citenamefont {Angell}, \citenamefont
  {Karwowski}, \citenamefont {Kwan}, \citenamefont {Rusev}, \citenamefont
  {Tonchev}, \citenamefont {Tornow}, \citenamefont {Howell},\ and\
  \citenamefont {Kelley}}]{Ham12}%
  \BibitemOpen
  \bibfield  {author} {\bibinfo {author} {\bibfnamefont {S.~L.}\ \bibnamefont
  {Hammond}}, \bibinfo {author} {\bibfnamefont {A.~S.}\ \bibnamefont
  {Adekola}}, \bibinfo {author} {\bibfnamefont {C.~T.}\ \bibnamefont {Angell}},
  \bibinfo {author} {\bibfnamefont {H.~J.}\ \bibnamefont {Karwowski}}, \bibinfo
  {author} {\bibfnamefont {E.}~\bibnamefont {Kwan}}, \bibinfo {author}
  {\bibfnamefont {G.}~\bibnamefont {Rusev}}, \bibinfo {author} {\bibfnamefont
  {A.~P.}\ \bibnamefont {Tonchev}}, \bibinfo {author} {\bibfnamefont
  {W.}~\bibnamefont {Tornow}}, \bibinfo {author} {\bibfnamefont {C.~R.}\
  \bibnamefont {Howell}}, \ and\ \bibinfo {author} {\bibfnamefont {J.~H.}\
  \bibnamefont {Kelley}},\ }\href {\doibase 10.1103/PhysRevC.85.044302}
  {\bibfield  {journal} {\bibinfo  {journal} {Phys. Rev. C}\ }\textbf {\bibinfo
  {volume} {85}},\ \bibinfo {pages} {044302} (\bibinfo {year}
  {2012})}\BibitemShut {NoStop}%
\bibitem [{\citenamefont {Agostinelli}\ \emph {et~al.}(2003)\citenamefont
  {Agostinelli}, \citenamefont {Allison}, \citenamefont {Amako}, \citenamefont
  {Apostolakis}, \citenamefont {Araujo}, \citenamefont {Arce}, \citenamefont
  {Asai}, \citenamefont {Axen}, \citenamefont {Banerjee}, \citenamefont
  {Barrand}, \citenamefont {Behner}, \citenamefont {Bellagamba}, \citenamefont
  {Boudreau}, \citenamefont {Broglia}, \citenamefont {Brunengo}, \citenamefont
  {Burkhardt}, \citenamefont {Chauvie}, \citenamefont {Chuma}, \citenamefont
  {Chytracek}, \citenamefont {Cooperman}, \citenamefont {Cosmo}, \citenamefont
  {Degtyarenko}, \citenamefont {Dell'Acqua}, \citenamefont {Depaola},
  \citenamefont {Dietrich}, \citenamefont {Enami}, \citenamefont {Feliciello},
  \citenamefont {Ferguson}, \citenamefont {Fesefeldt}, \citenamefont {Folger},
  \citenamefont {Foppiano}, \citenamefont {Forti}, \citenamefont {Garelli},
  \citenamefont {Giani}, \citenamefont {Giannitrapani}, \citenamefont {Gibin},
  \citenamefont {Cadenas}, \citenamefont {Gonz\`alez}, \citenamefont {Abril},
  \citenamefont {Greeniaus}, \citenamefont {Greiner}, \citenamefont {Grichine},
  \citenamefont {Grossheim}, \citenamefont {Guatelli}, \citenamefont
  {Gumplinger}, \citenamefont {Hamatsu}, \citenamefont {Hashimoto},
  \citenamefont {Hasui}, \citenamefont {Heikkinen}, \citenamefont {Howard},
  \citenamefont {Ivanchenko}, \citenamefont {Johnson}, \citenamefont {Jones},
  \citenamefont {Kallenbach}, \citenamefont {Kanaya}, \citenamefont {Kawabata},
  \citenamefont {Kawabata}, \citenamefont {Kawaguti}, \citenamefont {Kelner},
  \citenamefont {Kent}, \citenamefont {Kimura}, \citenamefont {Kodama},
  \citenamefont {Kokoulin}, \citenamefont {Kossov}, \citenamefont {Kurashige},
  \citenamefont {Lamanna}, \citenamefont {Lamp\`en}, \citenamefont {Lara},
  \citenamefont {Lefebure}, \citenamefont {Lei}, \citenamefont {Liendl},
  \citenamefont {Lockman}, \citenamefont {Longo}, \citenamefont {Magni},
  \citenamefont {Maire}, \citenamefont {Medernach}, \citenamefont {Minamimoto},
  \citenamefont {de~Freitas}, \citenamefont {Morita}, \citenamefont {Murakami},
  \citenamefont {Nagamatu}, \citenamefont {Nartallo}, \citenamefont {Nieminen},
  \citenamefont {Nishimura}, \citenamefont {Ohtsubo}, \citenamefont {Okamura},
  \citenamefont {O'Neale}, \citenamefont {Oohata}, \citenamefont {Paech},
  \citenamefont {Perl}, \citenamefont {Pfeiffer}, \citenamefont {Pia},
  \citenamefont {Ranjard}, \citenamefont {Rybin}, \citenamefont {Sadilov},
  \citenamefont {Salvo}, \citenamefont {Santin}, \citenamefont {Sasaki},
  \citenamefont {Savvas}, \citenamefont {Sawada}, \citenamefont {Scherer},
  \citenamefont {Sei}, \citenamefont {Sirotenko}, \citenamefont {Smith},
  \citenamefont {Starkov}, \citenamefont {Stoecker}, \citenamefont {Sulkimo},
  \citenamefont {Takahata}, \citenamefont {Tanaka}, \citenamefont {Tcherniaev},
  \citenamefont {Tehrani}, \citenamefont {Tropeano}, \citenamefont {Truscott},
  \citenamefont {Uno}, \citenamefont {Urban}, \citenamefont {Urban},
  \citenamefont {Verderi}, \citenamefont {Walkden}, \citenamefont {Wander},
  \citenamefont {Weber}, \citenamefont {Wellisch}, \citenamefont {Wenaus},
  \citenamefont {Williams}, \citenamefont {Wright}, \citenamefont {Yamada},
  \citenamefont {Yoshida},\ and\ \citenamefont {Zschiesche}}]{Ago03}%
  \BibitemOpen
  \bibfield  {author} {\bibinfo {author} {\bibfnamefont {S.}~\bibnamefont
  {Agostinelli}}, \bibinfo {author} {\bibfnamefont {J.}~\bibnamefont
  {Allison}}, \bibinfo {author} {\bibfnamefont {K.}~\bibnamefont {Amako}},
  \bibinfo {author} {\bibfnamefont {J.}~\bibnamefont {Apostolakis}}, \bibinfo
  {author} {\bibfnamefont {H.}~\bibnamefont {Araujo}}, \bibinfo {author}
  {\bibfnamefont {P.}~\bibnamefont {Arce}}, \bibinfo {author} {\bibfnamefont
  {M.}~\bibnamefont {Asai}}, \bibinfo {author} {\bibfnamefont {D.}~\bibnamefont
  {Axen}}, \bibinfo {author} {\bibfnamefont {S.}~\bibnamefont {Banerjee}},
  \bibinfo {author} {\bibfnamefont {G.}~\bibnamefont {Barrand}}, \bibinfo
  {author} {\bibfnamefont {F.}~\bibnamefont {Behner}}, \bibinfo {author}
  {\bibfnamefont {L.}~\bibnamefont {Bellagamba}}, \bibinfo {author}
  {\bibfnamefont {J.}~\bibnamefont {Boudreau}}, \bibinfo {author}
  {\bibfnamefont {L.}~\bibnamefont {Broglia}}, \bibinfo {author} {\bibfnamefont
  {A.}~\bibnamefont {Brunengo}}, \bibinfo {author} {\bibfnamefont
  {H.}~\bibnamefont {Burkhardt}}, \bibinfo {author} {\bibfnamefont
  {S.}~\bibnamefont {Chauvie}}, \bibinfo {author} {\bibfnamefont
  {J.}~\bibnamefont {Chuma}}, \bibinfo {author} {\bibfnamefont
  {R.}~\bibnamefont {Chytracek}}, \bibinfo {author} {\bibfnamefont
  {G.}~\bibnamefont {Cooperman}}, \bibinfo {author} {\bibfnamefont
  {G.}~\bibnamefont {Cosmo}}, \bibinfo {author} {\bibfnamefont
  {P.}~\bibnamefont {Degtyarenko}}, \bibinfo {author} {\bibfnamefont
  {A.}~\bibnamefont {Dell'Acqua}}, \bibinfo {author} {\bibfnamefont
  {G.}~\bibnamefont {Depaola}}, \bibinfo {author} {\bibfnamefont
  {D.}~\bibnamefont {Dietrich}}, \bibinfo {author} {\bibfnamefont
  {R.}~\bibnamefont {Enami}}, \bibinfo {author} {\bibfnamefont
  {A.}~\bibnamefont {Feliciello}}, \bibinfo {author} {\bibfnamefont
  {C.}~\bibnamefont {Ferguson}}, \bibinfo {author} {\bibfnamefont
  {H.}~\bibnamefont {Fesefeldt}}, \bibinfo {author} {\bibfnamefont
  {G.}~\bibnamefont {Folger}}, \bibinfo {author} {\bibfnamefont
  {F.}~\bibnamefont {Foppiano}}, \bibinfo {author} {\bibfnamefont
  {A.}~\bibnamefont {Forti}}, \bibinfo {author} {\bibfnamefont
  {S.}~\bibnamefont {Garelli}}, \bibinfo {author} {\bibfnamefont
  {S.}~\bibnamefont {Giani}}, \bibinfo {author} {\bibfnamefont
  {R.}~\bibnamefont {Giannitrapani}}, \bibinfo {author} {\bibfnamefont
  {D.}~\bibnamefont {Gibin}}, \bibinfo {author} {\bibfnamefont {J.~G.}\
  \bibnamefont {Cadenas}}, \bibinfo {author} {\bibfnamefont {I.}~\bibnamefont
  {Gonz\`alez}}, \bibinfo {author} {\bibfnamefont {G.~G.}\ \bibnamefont
  {Abril}}, \bibinfo {author} {\bibfnamefont {G.}~\bibnamefont {Greeniaus}},
  \bibinfo {author} {\bibfnamefont {W.}~\bibnamefont {Greiner}}, \bibinfo
  {author} {\bibfnamefont {V.}~\bibnamefont {Grichine}}, \bibinfo {author}
  {\bibfnamefont {A.}~\bibnamefont {Grossheim}}, \bibinfo {author}
  {\bibfnamefont {S.}~\bibnamefont {Guatelli}}, \bibinfo {author}
  {\bibfnamefont {P.}~\bibnamefont {Gumplinger}}, \bibinfo {author}
  {\bibfnamefont {R.}~\bibnamefont {Hamatsu}}, \bibinfo {author} {\bibfnamefont
  {K.}~\bibnamefont {Hashimoto}}, \bibinfo {author} {\bibfnamefont
  {H.}~\bibnamefont {Hasui}}, \bibinfo {author} {\bibfnamefont
  {A.}~\bibnamefont {Heikkinen}}, \bibinfo {author} {\bibfnamefont
  {A.}~\bibnamefont {Howard}}, \bibinfo {author} {\bibfnamefont
  {V.}~\bibnamefont {Ivanchenko}}, \bibinfo {author} {\bibfnamefont
  {A.}~\bibnamefont {Johnson}}, \bibinfo {author} {\bibfnamefont
  {F.}~\bibnamefont {Jones}}, \bibinfo {author} {\bibfnamefont
  {J.}~\bibnamefont {Kallenbach}}, \bibinfo {author} {\bibfnamefont
  {N.}~\bibnamefont {Kanaya}}, \bibinfo {author} {\bibfnamefont
  {M.}~\bibnamefont {Kawabata}}, \bibinfo {author} {\bibfnamefont
  {Y.}~\bibnamefont {Kawabata}}, \bibinfo {author} {\bibfnamefont
  {M.}~\bibnamefont {Kawaguti}}, \bibinfo {author} {\bibfnamefont
  {S.}~\bibnamefont {Kelner}}, \bibinfo {author} {\bibfnamefont
  {P.}~\bibnamefont {Kent}}, \bibinfo {author} {\bibfnamefont {A.}~\bibnamefont
  {Kimura}}, \bibinfo {author} {\bibfnamefont {T.}~\bibnamefont {Kodama}},
  \bibinfo {author} {\bibfnamefont {R.}~\bibnamefont {Kokoulin}}, \bibinfo
  {author} {\bibfnamefont {M.}~\bibnamefont {Kossov}}, \bibinfo {author}
  {\bibfnamefont {H.}~\bibnamefont {Kurashige}}, \bibinfo {author}
  {\bibfnamefont {E.}~\bibnamefont {Lamanna}}, \bibinfo {author} {\bibfnamefont
  {T.}~\bibnamefont {Lamp\`en}}, \bibinfo {author} {\bibfnamefont
  {V.}~\bibnamefont {Lara}}, \bibinfo {author} {\bibfnamefont {V.}~\bibnamefont
  {Lefebure}}, \bibinfo {author} {\bibfnamefont {F.}~\bibnamefont {Lei}},
  \bibinfo {author} {\bibfnamefont {M.}~\bibnamefont {Liendl}}, \bibinfo
  {author} {\bibfnamefont {W.}~\bibnamefont {Lockman}}, \bibinfo {author}
  {\bibfnamefont {F.}~\bibnamefont {Longo}}, \bibinfo {author} {\bibfnamefont
  {S.}~\bibnamefont {Magni}}, \bibinfo {author} {\bibfnamefont
  {M.}~\bibnamefont {Maire}}, \bibinfo {author} {\bibfnamefont
  {E.}~\bibnamefont {Medernach}}, \bibinfo {author} {\bibfnamefont
  {K.}~\bibnamefont {Minamimoto}}, \bibinfo {author} {\bibfnamefont {P.~M.}\
  \bibnamefont {de~Freitas}}, \bibinfo {author} {\bibfnamefont
  {Y.}~\bibnamefont {Morita}}, \bibinfo {author} {\bibfnamefont
  {K.}~\bibnamefont {Murakami}}, \bibinfo {author} {\bibfnamefont
  {M.}~\bibnamefont {Nagamatu}}, \bibinfo {author} {\bibfnamefont
  {R.}~\bibnamefont {Nartallo}}, \bibinfo {author} {\bibfnamefont
  {P.}~\bibnamefont {Nieminen}}, \bibinfo {author} {\bibfnamefont
  {T.}~\bibnamefont {Nishimura}}, \bibinfo {author} {\bibfnamefont
  {K.}~\bibnamefont {Ohtsubo}}, \bibinfo {author} {\bibfnamefont
  {M.}~\bibnamefont {Okamura}}, \bibinfo {author} {\bibfnamefont
  {S.}~\bibnamefont {O'Neale}}, \bibinfo {author} {\bibfnamefont
  {Y.}~\bibnamefont {Oohata}}, \bibinfo {author} {\bibfnamefont
  {K.}~\bibnamefont {Paech}}, \bibinfo {author} {\bibfnamefont
  {J.}~\bibnamefont {Perl}}, \bibinfo {author} {\bibfnamefont {A.}~\bibnamefont
  {Pfeiffer}}, \bibinfo {author} {\bibfnamefont {M.}~\bibnamefont {Pia}},
  \bibinfo {author} {\bibfnamefont {F.}~\bibnamefont {Ranjard}}, \bibinfo
  {author} {\bibfnamefont {A.}~\bibnamefont {Rybin}}, \bibinfo {author}
  {\bibfnamefont {S.}~\bibnamefont {Sadilov}}, \bibinfo {author} {\bibfnamefont
  {E.~D.}\ \bibnamefont {Salvo}}, \bibinfo {author} {\bibfnamefont
  {G.}~\bibnamefont {Santin}}, \bibinfo {author} {\bibfnamefont
  {T.}~\bibnamefont {Sasaki}}, \bibinfo {author} {\bibfnamefont
  {N.}~\bibnamefont {Savvas}}, \bibinfo {author} {\bibfnamefont
  {Y.}~\bibnamefont {Sawada}}, \bibinfo {author} {\bibfnamefont
  {S.}~\bibnamefont {Scherer}}, \bibinfo {author} {\bibfnamefont
  {S.}~\bibnamefont {Sei}}, \bibinfo {author} {\bibfnamefont {V.}~\bibnamefont
  {Sirotenko}}, \bibinfo {author} {\bibfnamefont {D.}~\bibnamefont {Smith}},
  \bibinfo {author} {\bibfnamefont {N.}~\bibnamefont {Starkov}}, \bibinfo
  {author} {\bibfnamefont {H.}~\bibnamefont {Stoecker}}, \bibinfo {author}
  {\bibfnamefont {J.}~\bibnamefont {Sulkimo}}, \bibinfo {author} {\bibfnamefont
  {M.}~\bibnamefont {Takahata}}, \bibinfo {author} {\bibfnamefont
  {S.}~\bibnamefont {Tanaka}}, \bibinfo {author} {\bibfnamefont
  {E.}~\bibnamefont {Tcherniaev}}, \bibinfo {author} {\bibfnamefont {E.~S.}\
  \bibnamefont {Tehrani}}, \bibinfo {author} {\bibfnamefont {M.}~\bibnamefont
  {Tropeano}}, \bibinfo {author} {\bibfnamefont {P.}~\bibnamefont {Truscott}},
  \bibinfo {author} {\bibfnamefont {H.}~\bibnamefont {Uno}}, \bibinfo {author}
  {\bibfnamefont {L.}~\bibnamefont {Urban}}, \bibinfo {author} {\bibfnamefont
  {P.}~\bibnamefont {Urban}}, \bibinfo {author} {\bibfnamefont
  {M.}~\bibnamefont {Verderi}}, \bibinfo {author} {\bibfnamefont
  {A.}~\bibnamefont {Walkden}}, \bibinfo {author} {\bibfnamefont
  {W.}~\bibnamefont {Wander}}, \bibinfo {author} {\bibfnamefont
  {H.}~\bibnamefont {Weber}}, \bibinfo {author} {\bibfnamefont
  {J.}~\bibnamefont {Wellisch}}, \bibinfo {author} {\bibfnamefont
  {T.}~\bibnamefont {Wenaus}}, \bibinfo {author} {\bibfnamefont
  {D.}~\bibnamefont {Williams}}, \bibinfo {author} {\bibfnamefont
  {D.}~\bibnamefont {Wright}}, \bibinfo {author} {\bibfnamefont
  {T.}~\bibnamefont {Yamada}}, \bibinfo {author} {\bibfnamefont
  {H.}~\bibnamefont {Yoshida}}, \ and\ \bibinfo {author} {\bibfnamefont
  {D.}~\bibnamefont {Zschiesche}},\ }\href {\doibase
  10.1016/S0168-9002(03)01368-8} {\bibfield  {journal} {\bibinfo  {journal}
  {Nucl. Instrum. Methods A}\ }\textbf {\bibinfo {volume} {506}},\ \bibinfo
  {pages} {250 } (\bibinfo {year} {2003})}\BibitemShut {NoStop}%
\bibitem [{\citenamefont {Pietralla}\ \emph {et~al.}(1994)\citenamefont
  {Pietralla}, \citenamefont {von Brentano}, \citenamefont {Casten},
  \citenamefont {Otsuka},\ and\ \citenamefont {Zamfir}}]{Pie94}%
  \BibitemOpen
  \bibfield  {author} {\bibinfo {author} {\bibfnamefont {N.}~\bibnamefont
  {Pietralla}}, \bibinfo {author} {\bibfnamefont {P.}~\bibnamefont {von
  Brentano}}, \bibinfo {author} {\bibfnamefont {R.~F.}\ \bibnamefont {Casten}},
  \bibinfo {author} {\bibfnamefont {T.}~\bibnamefont {Otsuka}}, \ and\ \bibinfo
  {author} {\bibfnamefont {N.~V.}\ \bibnamefont {Zamfir}},\ }\href {\doibase
  10.1103/PhysRevLett.73.2962} {\bibfield  {journal} {\bibinfo  {journal}
  {Phys. Rev. Lett.}\ }\textbf {\bibinfo {volume} {73}},\ \bibinfo {pages}
  {2962} (\bibinfo {year} {1994})}\BibitemShut {NoStop}%
\bibitem [{\citenamefont {Ajzenberg-Selove}(1990)}]{ndsC}%
  \BibitemOpen
  \bibfield  {author} {\bibinfo {author} {\bibfnamefont {F.}~\bibnamefont
  {Ajzenberg-Selove}},\ }\href {\doibase 10.1016/0375-9474(90)90271-M}
  {\bibfield  {journal} {\bibinfo  {journal} {Nucl. Phys. A}\ }\textbf
  {\bibinfo {volume} {506}},\ \bibinfo {pages} {1 } (\bibinfo {year}
  {1990})}\BibitemShut {NoStop}%
\bibitem [{\citenamefont {Sun}\ \emph {et~al.}(2009{\natexlab{a}})\citenamefont
  {Sun}, \citenamefont {Li}, \citenamefont {Rusev}, \citenamefont {Tonchev},\
  and\ \citenamefont {Wu}}]{Sun09}%
  \BibitemOpen
  \bibfield  {author} {\bibinfo {author} {\bibfnamefont {C.}~\bibnamefont
  {Sun}}, \bibinfo {author} {\bibfnamefont {J.}~\bibnamefont {Li}}, \bibinfo
  {author} {\bibfnamefont {G.}~\bibnamefont {Rusev}}, \bibinfo {author}
  {\bibfnamefont {A.~P.}\ \bibnamefont {Tonchev}}, \ and\ \bibinfo {author}
  {\bibfnamefont {Y.~K.}\ \bibnamefont {Wu}},\ }\href {\doibase
  10.1103/PhysRevSTAB.12.062801} {\bibfield  {journal} {\bibinfo  {journal}
  {Phys. Rev. ST Accel. Beams}\ }\textbf {\bibinfo {volume} {12}},\ \bibinfo
  {pages} {062801} (\bibinfo {year} {2009}{\natexlab{a}})}\BibitemShut
  {NoStop}%
\bibitem [{\citenamefont {Sun}\ \emph {et~al.}(2009{\natexlab{b}})\citenamefont
  {Sun}, \citenamefont {Wu}, \citenamefont {Rusev},\ and\ \citenamefont
  {Tonchev}}]{Sun09i}%
  \BibitemOpen
  \bibfield  {author} {\bibinfo {author} {\bibfnamefont {C.}~\bibnamefont
  {Sun}}, \bibinfo {author} {\bibfnamefont {Y.}~\bibnamefont {Wu}}, \bibinfo
  {author} {\bibfnamefont {G.}~\bibnamefont {Rusev}}, \ and\ \bibinfo {author}
  {\bibfnamefont {A.}~\bibnamefont {Tonchev}},\ }\href {\doibase
  10.1016/j.nima.2009.03.237} {\bibfield  {journal} {\bibinfo  {journal} {Nucl.
  Instr. Meth. Phys. Res. A}\ }\textbf {\bibinfo {volume} {605}},\ \bibinfo
  {pages} {312 } (\bibinfo {year} {2009}{\natexlab{b}})}\BibitemShut {NoStop}%
\bibitem [{\citenamefont {Kwan}\ \emph {et~al.}(2011)\citenamefont {Kwan},
  \citenamefont {Rusev}, \citenamefont {Adekola}, \citenamefont {D\"onau},
  \citenamefont {Hammond}, \citenamefont {Howell}, \citenamefont {Karwowski},
  \citenamefont {Kelley}, \citenamefont {Pedroni}, \citenamefont {Raut},
  \citenamefont {Tonchev},\ and\ \citenamefont {Tornow}}]{Kwa11}%
  \BibitemOpen
  \bibfield  {author} {\bibinfo {author} {\bibfnamefont {E.}~\bibnamefont
  {Kwan}}, \bibinfo {author} {\bibfnamefont {G.}~\bibnamefont {Rusev}},
  \bibinfo {author} {\bibfnamefont {A.~S.}\ \bibnamefont {Adekola}}, \bibinfo
  {author} {\bibfnamefont {F.}~\bibnamefont {D\"onau}}, \bibinfo {author}
  {\bibfnamefont {S.~L.}\ \bibnamefont {Hammond}}, \bibinfo {author}
  {\bibfnamefont {C.~R.}\ \bibnamefont {Howell}}, \bibinfo {author}
  {\bibfnamefont {H.~J.}\ \bibnamefont {Karwowski}}, \bibinfo {author}
  {\bibfnamefont {J.~H.}\ \bibnamefont {Kelley}}, \bibinfo {author}
  {\bibfnamefont {R.~S.}\ \bibnamefont {Pedroni}}, \bibinfo {author}
  {\bibfnamefont {R.}~\bibnamefont {Raut}}, \bibinfo {author} {\bibfnamefont
  {A.~P.}\ \bibnamefont {Tonchev}}, \ and\ \bibinfo {author} {\bibfnamefont
  {W.}~\bibnamefont {Tornow}},\ }\href {\doibase 10.1103/PhysRevC.83.041601}
  {\bibfield  {journal} {\bibinfo  {journal} {Phys. Rev. C}\ }\textbf {\bibinfo
  {volume} {83}},\ \bibinfo {pages} {041601} (\bibinfo {year}
  {2011})}\BibitemShut {NoStop}%
\bibitem [{\citenamefont {Singh}(1995)}]{ndsSe}%
  \BibitemOpen
  \bibfield  {author} {\bibinfo {author} {\bibfnamefont {B.}~\bibnamefont
  {Singh}},\ }\href {\doibase 10.1006/ndsh.1995.1005} {\bibfield  {journal}
  {\bibinfo  {journal} {Nucl. Data Sheets}\ }\textbf {\bibinfo {volume} {74}},\
  \bibinfo {pages} {63 } (\bibinfo {year} {1995})}\BibitemShut {NoStop}%
\bibitem [{\citenamefont {Pietralla}\ \emph {et~al.}(1998)\citenamefont
  {Pietralla}, \citenamefont {von Brentano}, \citenamefont {Herzberg},
  \citenamefont {Kneissl}, \citenamefont {Lo~Iudice}, \citenamefont {Maser},
  \citenamefont {Pitz},\ and\ \citenamefont {Zilges}}]{Pie98}%
  \BibitemOpen
  \bibfield  {author} {\bibinfo {author} {\bibfnamefont {N.}~\bibnamefont
  {Pietralla}}, \bibinfo {author} {\bibfnamefont {P.}~\bibnamefont {von
  Brentano}}, \bibinfo {author} {\bibfnamefont {R.-D.}\ \bibnamefont
  {Herzberg}}, \bibinfo {author} {\bibfnamefont {U.}~\bibnamefont {Kneissl}},
  \bibinfo {author} {\bibfnamefont {N.}~\bibnamefont {Lo~Iudice}}, \bibinfo
  {author} {\bibfnamefont {H.}~\bibnamefont {Maser}}, \bibinfo {author}
  {\bibfnamefont {H.~H.}\ \bibnamefont {Pitz}}, \ and\ \bibinfo {author}
  {\bibfnamefont {A.}~\bibnamefont {Zilges}},\ }\href {\doibase
  10.1103/PhysRevC.58.184} {\bibfield  {journal} {\bibinfo  {journal} {Phys.
  Rev. C}\ }\textbf {\bibinfo {volume} {58}},\ \bibinfo {pages} {184} (\bibinfo
  {year} {1998})}\BibitemShut {NoStop}%
\bibitem [{\citenamefont {Heyde}\ \emph {et~al.}(2010)\citenamefont {Heyde},
  \citenamefont {von Neumann-Cosel},\ and\ \citenamefont {Richter}}]{Hey10}%
  \BibitemOpen
  \bibfield  {author} {\bibinfo {author} {\bibfnamefont {K.}~\bibnamefont
  {Heyde}}, \bibinfo {author} {\bibfnamefont {P.}~\bibnamefont {von
  Neumann-Cosel}}, \ and\ \bibinfo {author} {\bibfnamefont {A.}~\bibnamefont
  {Richter}},\ }\href {\doibase 10.1103/RevModPhys.82.2365} {\bibfield
  {journal} {\bibinfo  {journal} {Rev. Mod. Phys.}\ }\textbf {\bibinfo {volume}
  {82}},\ \bibinfo {pages} {2365} (\bibinfo {year} {2010})}\BibitemShut
  {NoStop}%
\bibitem [{\citenamefont {Hagmann}\ \emph {et~al.}(2009)\citenamefont
  {Hagmann}, \citenamefont {Hall}, \citenamefont {Johnson}, \citenamefont
  {McNabb}, \citenamefont {Kelley}, \citenamefont {Huibregtse}, \citenamefont
  {Kwan}, \citenamefont {Rusev},\ and\ \citenamefont {Tonchev}}]{Hag09}%
  \BibitemOpen
  \bibfield  {author} {\bibinfo {author} {\bibfnamefont {C.~A.}\ \bibnamefont
  {Hagmann}}, \bibinfo {author} {\bibfnamefont {J.~M.}\ \bibnamefont {Hall}},
  \bibinfo {author} {\bibfnamefont {M.~S.}\ \bibnamefont {Johnson}}, \bibinfo
  {author} {\bibfnamefont {D.~P.}\ \bibnamefont {McNabb}}, \bibinfo {author}
  {\bibfnamefont {J.~H.}\ \bibnamefont {Kelley}}, \bibinfo {author}
  {\bibfnamefont {C.}~\bibnamefont {Huibregtse}}, \bibinfo {author}
  {\bibfnamefont {E.}~\bibnamefont {Kwan}}, \bibinfo {author} {\bibfnamefont
  {G.}~\bibnamefont {Rusev}}, \ and\ \bibinfo {author} {\bibfnamefont {A.~P.}\
  \bibnamefont {Tonchev}},\ }\href {\doibase 10.1063/1.3238328} {\bibfield
  {journal} {\bibinfo  {journal} {J. Appl. Phys.}\ }\textbf {\bibinfo {volume}
  {106}},\ \bibinfo {eid} {084901} (\bibinfo {year} {2009})}\BibitemShut
  {NoStop}%
\bibitem [{\citenamefont {Goeke}\ and\ \citenamefont {Speth}(1982)}]{Goe82}%
  \BibitemOpen
  \bibfield  {author} {\bibinfo {author} {\bibfnamefont {K.}~\bibnamefont
  {Goeke}}\ and\ \bibinfo {author} {\bibfnamefont {J.}~\bibnamefont {Speth}},\
  }\href {\doibase 10.1146/annurev.ns.32.120182.000433} {\bibfield  {journal}
  {\bibinfo  {journal} {Ann. Rev. Nucl. Part. Sci.}\ }\textbf {\bibinfo
  {volume} {32}},\ \bibinfo {pages} {65} (\bibinfo {year} {1982})}\BibitemShut
  {NoStop}%
\bibitem [{\citenamefont {Ring}\ and\ \citenamefont {Schuck}(1980)}]{Rin80}%
  \BibitemOpen
  \bibfield  {author} {\bibinfo {author} {\bibfnamefont {P.}~\bibnamefont
  {Ring}}\ and\ \bibinfo {author} {\bibfnamefont {P.}~\bibnamefont {Schuck}},\
  }\href@noop {} {\emph {\bibinfo {title} {The Nuclear Many-Body Problem}}}\
  (\bibinfo  {publisher} {Springer-Verlag},\ \bibinfo {address} {Berlin},\
  \bibinfo {year} {1980})\BibitemShut {NoStop}%
\bibitem [{\citenamefont {Simenel}(2012)}]{Simenel12}%
  \BibitemOpen
  \bibfield  {author} {\bibinfo {author} {\bibfnamefont {C.}~\bibnamefont
  {Simenel}},\ }\href {\doibase 10.1140/epja/i2012-12152-0} {\bibfield
  {journal} {\bibinfo  {journal} {Eur. J. Phys. A}\ }\textbf {\bibinfo {volume}
  {48}},\ \bibinfo {pages} {1} (\bibinfo {year} {2012})}\BibitemShut {NoStop}%
\bibitem [{\citenamefont {Soloviev}(1992)}]{Sol92}%
  \BibitemOpen
  \bibfield  {author} {\bibinfo {author} {\bibfnamefont {V.}~\bibnamefont
  {Soloviev}},\ }\href@noop {} {\emph {\bibinfo {title} {Theory of Atomic
  Nuclei: Quasiparticles and Phonons}}}\ (\bibinfo  {publisher} {IoP
  Publishing},\ \bibinfo {address} {Bristol},\ \bibinfo {year}
  {1992})\BibitemShut {NoStop}%
\bibitem [{\citenamefont {Lacroix}\ \emph {et~al.}(2004)\citenamefont
  {Lacroix}, \citenamefont {Ayik},\ and\ \citenamefont {Chomaz}}]{Lac04}%
  \BibitemOpen
  \bibfield  {author} {\bibinfo {author} {\bibfnamefont {D.}~\bibnamefont
  {Lacroix}}, \bibinfo {author} {\bibfnamefont {S.}~\bibnamefont {Ayik}}, \
  and\ \bibinfo {author} {\bibfnamefont {P.}~\bibnamefont {Chomaz}},\ }\href
  {\doibase 10.1016/j.ppnp.2004.02.002} {\bibfield  {journal} {\bibinfo
  {journal} {Prog. Part. Nucl. Phys.}\ }\textbf {\bibinfo {volume} {52}},\
  \bibinfo {pages} {497 } (\bibinfo {year} {2004})}\BibitemShut {NoStop}%
\bibitem [{\citenamefont {Lacroix}\ \emph {et~al.}(1999)\citenamefont
  {Lacroix}, \citenamefont {Chomaz},\ and\ \citenamefont {Ayik}}]{Lac99}%
  \BibitemOpen
  \bibfield  {author} {\bibinfo {author} {\bibfnamefont {D.}~\bibnamefont
  {Lacroix}}, \bibinfo {author} {\bibfnamefont {P.}~\bibnamefont {Chomaz}}, \
  and\ \bibinfo {author} {\bibfnamefont {S.}~\bibnamefont {Ayik}},\ }\href
  {\doibase 10.1016/S0375-9474(99)00136-0} {\bibfield  {journal} {\bibinfo
  {journal} {Nucl. Phys. A}\ }\textbf {\bibinfo {volume} {651}},\ \bibinfo
  {pages} {369 } (\bibinfo {year} {1999})}\BibitemShut {NoStop}%
\bibitem [{\citenamefont {Reinhard}\ and\ \citenamefont
  {Suraud}(1992)}]{Rein92}%
  \BibitemOpen
  \bibfield  {author} {\bibinfo {author} {\bibfnamefont {P.-G.}\ \bibnamefont
  {Reinhard}}\ and\ \bibinfo {author} {\bibfnamefont {E.}~\bibnamefont
  {Suraud}},\ }\href {\doibase 10.1016/0003-4916(52)90043-2} {\bibfield
  {journal} {\bibinfo  {journal} {Ann. Phys.}\ }\textbf {\bibinfo {volume}
  {216}},\ \bibinfo {pages} {98 } (\bibinfo {year} {1992})}\BibitemShut
  {NoStop}%
\bibitem [{\citenamefont {Lacroix}\ \emph {et~al.}(2013)\citenamefont
  {Lacroix}, \citenamefont {Gambacurta},\ and\ \citenamefont {Ayik}}]{Lac13}%
  \BibitemOpen
  \bibfield  {author} {\bibinfo {author} {\bibfnamefont {D.}~\bibnamefont
  {Lacroix}}, \bibinfo {author} {\bibfnamefont {D.}~\bibnamefont {Gambacurta}},
  \ and\ \bibinfo {author} {\bibfnamefont {S.}~\bibnamefont {Ayik}},\ }\href
  {http://arxiv.org/abs/1303.0748} {} (\bibinfo {year} {2013}),\ \Eprint
  {http://arxiv.org/abs/1303.0748} {arXiv:1303.0748} \BibitemShut {NoStop}%
\bibitem [{\citenamefont {Avez}\ \emph {et~al.}(2008)\citenamefont {Avez},
  \citenamefont {Simenel},\ and\ \citenamefont {Chomaz}}]{Ave08}%
  \BibitemOpen
  \bibfield  {author} {\bibinfo {author} {\bibfnamefont {B.}~\bibnamefont
  {Avez}}, \bibinfo {author} {\bibfnamefont {C.}~\bibnamefont {Simenel}}, \
  and\ \bibinfo {author} {\bibfnamefont {P.}~\bibnamefont {Chomaz}},\ }\href
  {\doibase 10.1103/PhysRevC.78.044318} {\bibfield  {journal} {\bibinfo
  {journal} {Phys. Rev. C}\ }\textbf {\bibinfo {volume} {78}},\ \bibinfo
  {pages} {044318} (\bibinfo {year} {2008})}\BibitemShut {NoStop}%
\bibitem [{\citenamefont {Stetcu}\ \emph {et~al.}(2011)\citenamefont {Stetcu},
  \citenamefont {Bulgac}, \citenamefont {Magierski},\ and\ \citenamefont
  {Roche}}]{Ste11}%
  \BibitemOpen
  \bibfield  {author} {\bibinfo {author} {\bibfnamefont {I.}~\bibnamefont
  {Stetcu}}, \bibinfo {author} {\bibfnamefont {A.}~\bibnamefont {Bulgac}},
  \bibinfo {author} {\bibfnamefont {P.}~\bibnamefont {Magierski}}, \ and\
  \bibinfo {author} {\bibfnamefont {K.~J.}\ \bibnamefont {Roche}},\ }\href
  {\doibase 10.1103/PhysRevC.84.051309} {\bibfield  {journal} {\bibinfo
  {journal} {Phys. Rev. C}\ }\textbf {\bibinfo {volume} {84}},\ \bibinfo
  {pages} {051309} (\bibinfo {year} {2011})}\BibitemShut {NoStop}%
\bibitem [{\citenamefont {Ebata}\ \emph {et~al.}(2010)\citenamefont {Ebata},
  \citenamefont {Nakatsukasa}, \citenamefont {Inakura}, \citenamefont
  {Yoshida}, \citenamefont {Hashimoto},\ and\ \citenamefont {Yabana}}]{Eba10}%
  \BibitemOpen
  \bibfield  {author} {\bibinfo {author} {\bibfnamefont {S.}~\bibnamefont
  {Ebata}}, \bibinfo {author} {\bibfnamefont {T.}~\bibnamefont {Nakatsukasa}},
  \bibinfo {author} {\bibfnamefont {T.}~\bibnamefont {Inakura}}, \bibinfo
  {author} {\bibfnamefont {K.}~\bibnamefont {Yoshida}}, \bibinfo {author}
  {\bibfnamefont {Y.}~\bibnamefont {Hashimoto}}, \ and\ \bibinfo {author}
  {\bibfnamefont {K.}~\bibnamefont {Yabana}},\ }\href {\doibase
  10.1103/PhysRevC.82.034306} {\bibfield  {journal} {\bibinfo  {journal} {Phys.
  Rev. C}\ }\textbf {\bibinfo {volume} {82}},\ \bibinfo {pages} {034306}
  (\bibinfo {year} {2010})}\BibitemShut {NoStop}%
\bibitem [{\citenamefont {Gross}\ \emph {et~al.}(1986)\citenamefont {Gross},
  \citenamefont {Runge},\ and\ \citenamefont {Heinonen}}]{Gro91}%
  \BibitemOpen
  \bibfield  {author} {\bibinfo {author} {\bibfnamefont {E.}~\bibnamefont
  {Gross}}, \bibinfo {author} {\bibfnamefont {E.}~\bibnamefont {Runge}}, \ and\
  \bibinfo {author} {\bibfnamefont {O.}~\bibnamefont {Heinonen}},\ }\href@noop
  {} {\emph {\bibinfo {title} {Many-Particle Theory}}}\ (\bibinfo  {publisher}
  {Adam Hilger},\ \bibinfo {address} {Bristol},\ \bibinfo {year}
  {1986})\BibitemShut {NoStop}%
\bibitem [{\citenamefont {Kohn}\ and\ \citenamefont {Sham}(1965)}]{Koh65}%
  \BibitemOpen
  \bibfield  {author} {\bibinfo {author} {\bibfnamefont {W.}~\bibnamefont
  {Kohn}}\ and\ \bibinfo {author} {\bibfnamefont {L.~J.}\ \bibnamefont
  {Sham}},\ }\href {\doibase 10.1103/PhysRev.140.A1133} {\bibfield  {journal}
  {\bibinfo  {journal} {Phys. Rev.}\ }\textbf {\bibinfo {volume} {140}},\
  \bibinfo {pages} {A1133} (\bibinfo {year} {1965})}\BibitemShut {NoStop}%
\bibitem [{\citenamefont {Vautherin}\ and\ \citenamefont
  {Brink}(1972)}]{Vau72}%
  \BibitemOpen
  \bibfield  {author} {\bibinfo {author} {\bibfnamefont {D.}~\bibnamefont
  {Vautherin}}\ and\ \bibinfo {author} {\bibfnamefont {D.~M.}\ \bibnamefont
  {Brink}},\ }\href {\doibase 10.1103/PhysRevC.5.626} {\bibfield  {journal}
  {\bibinfo  {journal} {Phys. Rev. C}\ }\textbf {\bibinfo {volume} {5}},\
  \bibinfo {pages} {626} (\bibinfo {year} {1972})}\BibitemShut {NoStop}%
\bibitem [{\citenamefont {Skyrme}(1959)}]{Sky59}%
  \BibitemOpen
  \bibfield  {author} {\bibinfo {author} {\bibfnamefont {T.}~\bibnamefont
  {Skyrme}},\ }\href {\doibase 10.1016/0029-5582(58)90345-6} {\bibfield
  {journal} {\bibinfo  {journal} {Nucl. Phys.}\ }\textbf {\bibinfo {volume}
  {9}},\ \bibinfo {pages} {615 } (\bibinfo {year} {1958-1959})}\BibitemShut
  {NoStop}%
\bibitem [{\citenamefont {Engel}\ \emph {et~al.}(1975)\citenamefont {Engel},
  \citenamefont {Brink}, \citenamefont {Goeke}, \citenamefont {Krieger},\ and\
  \citenamefont {Vautherin}}]{Eng74}%
  \BibitemOpen
  \bibfield  {author} {\bibinfo {author} {\bibfnamefont {Y.}~\bibnamefont
  {Engel}}, \bibinfo {author} {\bibfnamefont {D.}~\bibnamefont {Brink}},
  \bibinfo {author} {\bibfnamefont {K.}~\bibnamefont {Goeke}}, \bibinfo
  {author} {\bibfnamefont {S.}~\bibnamefont {Krieger}}, \ and\ \bibinfo
  {author} {\bibfnamefont {D.}~\bibnamefont {Vautherin}},\ }\href {\doibase
  10.1016/0375-9474(75)90184-0} {\bibfield  {journal} {\bibinfo  {journal}
  {Nucl. Phys. A}\ }\textbf {\bibinfo {volume} {249}},\ \bibinfo {pages} {215 }
  (\bibinfo {year} {1975})}\BibitemShut {NoStop}%
\bibitem [{\citenamefont {Dirac}(1930)}]{Dir30}%
  \BibitemOpen
  \bibfield  {author} {\bibinfo {author} {\bibfnamefont {P.~A.~M.}\
  \bibnamefont {Dirac}},\ }\href {\doibase 10.1017/S0305004100016108}
  {\bibfield  {journal} {\bibinfo  {journal} {Math. Proc. Cambridge Phil.
  Soc.}\ }\textbf {\bibinfo {volume} {26}},\ \bibinfo {pages} {376} (\bibinfo
  {year} {1930})}\BibitemShut {NoStop}%
\bibitem [{\citenamefont {Harekeh}\ and\ \citenamefont {van~der
  Woude}(2001)}]{Har01}%
  \BibitemOpen
  \bibfield  {author} {\bibinfo {author} {\bibfnamefont {M.~N.}\ \bibnamefont
  {Harekeh}}\ and\ \bibinfo {author} {\bibfnamefont {A.}~\bibnamefont {van~der
  Woude}},\ }\href@noop {} {\emph {\bibinfo {title} {Giant Resonances}}}\
  (\bibinfo  {publisher} {Claredon Press},\ \bibinfo {address} {Oxford},\
  \bibinfo {year} {2001})\BibitemShut {NoStop}%
\bibitem [{\citenamefont {Maruhn}\ \emph {et~al.}(2005)\citenamefont {Maruhn},
  \citenamefont {Reinhard}, \citenamefont {Stevenson}, \citenamefont {Stone},\
  and\ \citenamefont {Strayer}}]{Mar05}%
  \BibitemOpen
  \bibfield  {author} {\bibinfo {author} {\bibfnamefont {J.~A.}\ \bibnamefont
  {Maruhn}}, \bibinfo {author} {\bibfnamefont {P.~G.}\ \bibnamefont
  {Reinhard}}, \bibinfo {author} {\bibfnamefont {P.~D.}\ \bibnamefont
  {Stevenson}}, \bibinfo {author} {\bibfnamefont {J.~R.}\ \bibnamefont
  {Stone}}, \ and\ \bibinfo {author} {\bibfnamefont {M.~R.}\ \bibnamefont
  {Strayer}},\ }\href {\doibase 10.1103/PhysRevC.71.064328} {\bibfield
  {journal} {\bibinfo  {journal} {Phys. Rev. C}\ }\textbf {\bibinfo {volume}
  {71}},\ \bibinfo {pages} {064328} (\bibinfo {year} {2005})}\BibitemShut
  {NoStop}%
\bibitem [{\citenamefont {Calvayrac}\ \emph {et~al.}(1997)\citenamefont
  {Calvayrac}, \citenamefont {Reinhard},\ and\ \citenamefont {Suraud}}]{Rei97}%
  \BibitemOpen
  \bibfield  {author} {\bibinfo {author} {\bibfnamefont {F.}~\bibnamefont
  {Calvayrac}}, \bibinfo {author} {\bibfnamefont {P.}~\bibnamefont {Reinhard}},
  \ and\ \bibinfo {author} {\bibfnamefont {E.}~\bibnamefont {Suraud}},\ }\href
  {\doibase 10.1006/aphy.1996.5654} {\bibfield  {journal} {\bibinfo  {journal}
  {Ann. Phys.}\ }\textbf {\bibinfo {volume} {255}},\ \bibinfo {pages} {125 }
  (\bibinfo {year} {1997})}\BibitemShut {NoStop}%
\bibitem [{\citenamefont {Reinhard}\ \emph {et~al.}(2006)\citenamefont
  {Reinhard}, \citenamefont {Stevenson}, \citenamefont {Almehed}, \citenamefont
  {Maruhn},\ and\ \citenamefont {Strayer}}]{Rei06}%
  \BibitemOpen
  \bibfield  {author} {\bibinfo {author} {\bibfnamefont {P.-G.}\ \bibnamefont
  {Reinhard}}, \bibinfo {author} {\bibfnamefont {P.~D.}\ \bibnamefont
  {Stevenson}}, \bibinfo {author} {\bibfnamefont {D.}~\bibnamefont {Almehed}},
  \bibinfo {author} {\bibfnamefont {J.~A.}\ \bibnamefont {Maruhn}}, \ and\
  \bibinfo {author} {\bibfnamefont {M.~R.}\ \bibnamefont {Strayer}},\ }\href
  {\doibase 10.1103/PhysRevE.73.036709} {\bibfield  {journal} {\bibinfo
  {journal} {Phys. Rev. E}\ }\textbf {\bibinfo {volume} {73}},\ \bibinfo
  {pages} {036709} (\bibinfo {year} {2006})}\BibitemShut {NoStop}%
\bibitem [{\citenamefont {Bohr}\ and\ \citenamefont {Mottleson}(1975)}]{Boh75}%
  \BibitemOpen
  \bibfield  {author} {\bibinfo {author} {\bibfnamefont {A.}~\bibnamefont
  {Bohr}}\ and\ \bibinfo {author} {\bibfnamefont {B.}~\bibnamefont
  {Mottleson}},\ }\href@noop {} {\emph {\bibinfo {title} {Nuclear
  Structure}}},\ Vol.~\bibinfo {volume} {II}\ (\bibinfo  {publisher} {W. A.
  Benjamin},\ \bibinfo {address} {New York},\ \bibinfo {year}
  {1975})\BibitemShut {NoStop}%
\bibitem [{\citenamefont {Steiner}\ \emph {et~al.}(2005)\citenamefont
  {Steiner}, \citenamefont {Prakash}, \citenamefont {Lattimer},\ and\
  \citenamefont {Ellis}}]{Ste05}%
  \BibitemOpen
  \bibfield  {author} {\bibinfo {author} {\bibfnamefont {A.}~\bibnamefont
  {Steiner}}, \bibinfo {author} {\bibfnamefont {M.}~\bibnamefont {Prakash}},
  \bibinfo {author} {\bibfnamefont {J.}~\bibnamefont {Lattimer}}, \ and\
  \bibinfo {author} {\bibfnamefont {P.}~\bibnamefont {Ellis}},\ }\href
  {\doibase 10.1016/j.physrep.2005.02.004} {\bibfield  {journal} {\bibinfo
  {journal} {Phys. Rep.}\ }\textbf {\bibinfo {volume} {411}},\ \bibinfo {pages}
  {325 } (\bibinfo {year} {2005})}\BibitemShut {NoStop}%
\bibitem [{\citenamefont {Reinhard}\ and\ \citenamefont
  {Flocard}(1995)}]{Flo95}%
  \BibitemOpen
  \bibfield  {author} {\bibinfo {author} {\bibfnamefont {P.-G.}\ \bibnamefont
  {Reinhard}}\ and\ \bibinfo {author} {\bibfnamefont {H.}~\bibnamefont
  {Flocard}},\ }\href {\doibase 10.1016/0375-9474(94)00770-N} {\bibfield
  {journal} {\bibinfo  {journal} {Nucl. Phys. A}\ }\textbf {\bibinfo {volume}
  {584}},\ \bibinfo {pages} {467 } (\bibinfo {year} {1995})}\BibitemShut
  {NoStop}%
\bibitem [{\citenamefont {Chabanat}\ \emph {et~al.}(1995)\citenamefont
  {Chabanat}, \citenamefont {Bonche}, \citenamefont {Haensel}, \citenamefont
  {Meyer},\ and\ \citenamefont {Schaeffer}}]{Chab95}%
  \BibitemOpen
  \bibfield  {author} {\bibinfo {author} {\bibfnamefont {E.}~\bibnamefont
  {Chabanat}}, \bibinfo {author} {\bibfnamefont {P.}~\bibnamefont {Bonche}},
  \bibinfo {author} {\bibfnamefont {P.}~\bibnamefont {Haensel}}, \bibinfo
  {author} {\bibfnamefont {J.}~\bibnamefont {Meyer}}, \ and\ \bibinfo {author}
  {\bibfnamefont {R.}~\bibnamefont {Schaeffer}},\ }\href
  {http://stacks.iop.org/1402-4896/1995/i=T56/a=034} {\bibfield  {journal}
  {\bibinfo  {journal} {Phys. Scr.}\ }\textbf {\bibinfo {volume} {1995}},\
  \bibinfo {pages} {231} (\bibinfo {year} {1995})}\BibitemShut {NoStop}%
\bibitem [{\citenamefont {Dutra}\ \emph {et~al.}(2012)\citenamefont {Dutra},
  \citenamefont {Louren\ifmmode~\mbox{\c{c}}\else \c{c}\fi{}o}, \citenamefont
  {S\'a~Martins}, \citenamefont {Delfino}, \citenamefont {Stone},\ and\
  \citenamefont {Stevenson}}]{Dut12}%
  \BibitemOpen
  \bibfield  {author} {\bibinfo {author} {\bibfnamefont {M.}~\bibnamefont
  {Dutra}}, \bibinfo {author} {\bibfnamefont {O.}~\bibnamefont
  {Louren\ifmmode~\mbox{\c{c}}\else \c{c}\fi{}o}}, \bibinfo {author}
  {\bibfnamefont {J.~S.}\ \bibnamefont {S\'a~Martins}}, \bibinfo {author}
  {\bibfnamefont {A.}~\bibnamefont {Delfino}}, \bibinfo {author} {\bibfnamefont
  {J.~R.}\ \bibnamefont {Stone}}, \ and\ \bibinfo {author} {\bibfnamefont
  {P.~D.}\ \bibnamefont {Stevenson}},\ }\href {\doibase
  10.1103/PhysRevC.85.035201} {\bibfield  {journal} {\bibinfo  {journal} {Phys.
  Rev. C}\ }\textbf {\bibinfo {volume} {85}},\ \bibinfo {pages} {035201}
  (\bibinfo {year} {2012})}\BibitemShut {NoStop}%
\bibitem [{\citenamefont {Roca-Maza}\ \emph {et~al.}(2013)\citenamefont
  {Roca-Maza}, \citenamefont {Brenna}, \citenamefont {Agrawal}, \citenamefont
  {Bortignon}, \citenamefont {Col\`o}, \citenamefont {Cao}, \citenamefont
  {Paar},\ and\ \citenamefont {Vretenar}}]{Roc13}%
  \BibitemOpen
  \bibfield  {author} {\bibinfo {author} {\bibfnamefont {X.}~\bibnamefont
  {Roca-Maza}}, \bibinfo {author} {\bibfnamefont {M.}~\bibnamefont {Brenna}},
  \bibinfo {author} {\bibfnamefont {B.~K.}\ \bibnamefont {Agrawal}}, \bibinfo
  {author} {\bibfnamefont {P.~F.}\ \bibnamefont {Bortignon}}, \bibinfo {author}
  {\bibfnamefont {G.}~\bibnamefont {Col\`o}}, \bibinfo {author} {\bibfnamefont
  {L.-G.}\ \bibnamefont {Cao}}, \bibinfo {author} {\bibfnamefont
  {N.}~\bibnamefont {Paar}}, \ and\ \bibinfo {author} {\bibfnamefont
  {D.}~\bibnamefont {Vretenar}},\ }\href {\doibase 10.1103/PhysRevC.87.034301}
  {\bibfield  {journal} {\bibinfo  {journal} {Phys. Rev. C}\ }\textbf {\bibinfo
  {volume} {87}},\ \bibinfo {pages} {034301} (\bibinfo {year}
  {2013})}\BibitemShut {NoStop}%
\bibitem [{\citenamefont {Goddard}\ \emph {et~al.}(2013)\citenamefont
  {Goddard}, \citenamefont {Stevenson},\ and\ \citenamefont {Rios}}]{God13}%
  \BibitemOpen
  \bibfield  {author} {\bibinfo {author} {\bibfnamefont {P.~M.}\ \bibnamefont
  {Goddard}}, \bibinfo {author} {\bibfnamefont {P.~D.}\ \bibnamefont
  {Stevenson}}, \ and\ \bibinfo {author} {\bibfnamefont {A.}~\bibnamefont
  {Rios}},\ }\href {\doibase 10.1103/PhysRevLett.110.032503} {\bibfield
  {journal} {\bibinfo  {journal} {Phys. Rev. Lett.}\ }\textbf {\bibinfo
  {volume} {110}},\ \bibinfo {pages} {032503} (\bibinfo {year}
  {2013})}\BibitemShut {NoStop}%
\bibitem [{\citenamefont {Audi}\ \emph {et~al.}(2003)\citenamefont {Audi},
  \citenamefont {Bersillon}, \citenamefont {Blachot},\ and\ \citenamefont
  {Wapstra}}]{Aud03}%
  \BibitemOpen
  \bibfield  {author} {\bibinfo {author} {\bibfnamefont {G.}~\bibnamefont
  {Audi}}, \bibinfo {author} {\bibfnamefont {O.}~\bibnamefont {Bersillon}},
  \bibinfo {author} {\bibfnamefont {J.}~\bibnamefont {Blachot}}, \ and\
  \bibinfo {author} {\bibfnamefont {A.}~\bibnamefont {Wapstra}},\ }\href
  {\doibase 10.1016/j.nuclphysa.2003.11.001} {\bibfield  {journal} {\bibinfo
  {journal} {Nucl. Phys. A}\ }\textbf {\bibinfo {volume} {729}},\ \bibinfo
  {pages} {3 } (\bibinfo {year} {2003})}\BibitemShut {NoStop}%
\bibitem [{\citenamefont {Pardi}\ and\ \citenamefont
  {Stevenson}(2013)}]{Par13}%
  \BibitemOpen
  \bibfield  {author} {\bibinfo {author} {\bibfnamefont {C.~I.}\ \bibnamefont
  {Pardi}}\ and\ \bibinfo {author} {\bibfnamefont {P.~D.}\ \bibnamefont
  {Stevenson}},\ }\href {\doibase 10.1103/PhysRevC.87.014330} {\bibfield
  {journal} {\bibinfo  {journal} {Phys. Rev. C}\ }\textbf {\bibinfo {volume}
  {87}},\ \bibinfo {pages} {014330} (\bibinfo {year} {2013})}\BibitemShut
  {NoStop}%
\bibitem [{\citenamefont {Scamps}\ and\ \citenamefont
  {Lacroix}(2013)}]{Scam13}%
  \BibitemOpen
  \bibfield  {author} {\bibinfo {author} {\bibfnamefont {G.}~\bibnamefont
  {Scamps}}\ and\ \bibinfo {author} {\bibfnamefont {D.}~\bibnamefont
  {Lacroix}},\ }\href {/http://arxiv.org/abs/1304.2497} {} (\bibinfo {year}
  {2013}),\ \Eprint {http://arxiv.org/abs/1304.2497} {arXiv:1304.2497}
  \BibitemShut {NoStop}%
\bibitem [{\citenamefont {Lacroix}\ \emph {et~al.}(2001)\citenamefont
  {Lacroix}, \citenamefont {Ayik},\ and\ \citenamefont {Chomaz}}]{Lac01}%
  \BibitemOpen
  \bibfield  {author} {\bibinfo {author} {\bibfnamefont {D.}~\bibnamefont
  {Lacroix}}, \bibinfo {author} {\bibfnamefont {S.}~\bibnamefont {Ayik}}, \
  and\ \bibinfo {author} {\bibfnamefont {P.}~\bibnamefont {Chomaz}},\ }\href
  {\doibase 10.1103/PhysRevC.63.064305} {\bibfield  {journal} {\bibinfo
  {journal} {Phys. Rev. C}\ }\textbf {\bibinfo {volume} {63}},\ \bibinfo
  {pages} {064305} (\bibinfo {year} {2001})}\BibitemShut {NoStop}%
\bibitem [{\citenamefont {Goldhaber}\ and\ \citenamefont
  {Teller}(1948)}]{Gol48}%
  \BibitemOpen
  \bibfield  {author} {\bibinfo {author} {\bibfnamefont {M.}~\bibnamefont
  {Goldhaber}}\ and\ \bibinfo {author} {\bibfnamefont {E.}~\bibnamefont
  {Teller}},\ }\href {\doibase 10.1103/PhysRev.74.1046} {\bibfield  {journal}
  {\bibinfo  {journal} {Phys. Rev.}\ }\textbf {\bibinfo {volume} {74}},\
  \bibinfo {pages} {1046} (\bibinfo {year} {1948})}\BibitemShut {NoStop}%
\bibitem [{\citenamefont {Baran}\ \emph {et~al.}(2012)\citenamefont {Baran},
  \citenamefont {Frecus}, \citenamefont {Colonna},\ and\ \citenamefont
  {Di~Toro}}]{Bar12}%
  \BibitemOpen
  \bibfield  {author} {\bibinfo {author} {\bibfnamefont {V.}~\bibnamefont
  {Baran}}, \bibinfo {author} {\bibfnamefont {B.}~\bibnamefont {Frecus}},
  \bibinfo {author} {\bibfnamefont {M.}~\bibnamefont {Colonna}}, \ and\
  \bibinfo {author} {\bibfnamefont {M.}~\bibnamefont {Di~Toro}},\ }\href
  {\doibase 10.1103/PhysRevC.85.051601} {\bibfield  {journal} {\bibinfo
  {journal} {Phys. Rev. C}\ }\textbf {\bibinfo {volume} {85}},\ \bibinfo
  {pages} {051601} (\bibinfo {year} {2012})}\BibitemShut {NoStop}%
\bibitem [{\citenamefont {Urban}(2012)}]{Urb12}%
  \BibitemOpen
  \bibfield  {author} {\bibinfo {author} {\bibfnamefont {M.}~\bibnamefont
  {Urban}},\ }\href {\doibase 10.1103/PhysRevC.85.034322} {\bibfield  {journal}
  {\bibinfo  {journal} {Phys. Rev. C}\ }\textbf {\bibinfo {volume} {85}},\
  \bibinfo {pages} {034322} (\bibinfo {year} {2012})}\BibitemShut {NoStop}%
\end{thebibliography}%

\end{document}